\documentclass[vecphys]{svmult}

\usepackage{makeidx}         
\usepackage{graphicx}        
\usepackage{multicol}        
\usepackage[bottom]{footmisc}

\makeindex             

\usepackage{bbm}

\newcounter{pr}
\newtheorem{prob}[pr]{{\bf Problem}}
\newenvironment{Problem}{\begin{prob} \rm}{\end{prob}}

\newcounter{sol}
\newtheorem{solu}[sol]{{\bf Solution to problem}}

\newcounter{def}
\newtheorem{mydefinition}[def]{{\bf Definition}}
\newenvironment{Definition}{\begin{mydefinition} \em}{\end{mydefinition}}

\begin{document}

\title*{From Special Geometry to Black Hole Partition Functions}
\author{Thomas Mohaupt}
\institute{
Department of Mathematical Sciences,
University of Liverpool,
Peach Street,
Liverpool L69 7ZL,
United Kingdom.
\texttt{Thomas.Mohaupt@liv.ac.uk}}
%
%
\maketitle


{\bf Abstract:} These notes are based on lectures given 
at the Erwin-Schr\"odinger Institut in Vienna in 2006/2007
and at the 2007 School on Attractor Mechanism in Frascati.
Lecture I reviews special geometry from the superconformal
point of view. Lecture II discusses the black hole attractor
mechanism, the underlying variational principle and 
black hole partition functions. 
Lecture III applies the formalism introduced in the previous
lectures to large and small
BPS black holes in $N=4$ supergravity.
Lecture IV is devoted to the microscopic desription of these
black holes 
in $N=4$ string compactifications. The lecture notes include
problems which allow the readers to develop
some of the key ideas by themselves. Appendix A reviews special 
geometry from the mathematical point of view. Appendix B 
provides the necessary background in modular forms needed
for understanding S-duality and  string state counting.

\section{Introduction}

Recent years have witnessed a renewed interest in the detailed
study of supersymmetric black holes in string theory. This 
has been triggered by the work of H.~Ooguri, A.~Strominger
and C.~Vafa \cite{Ooguri:2004zv}, who introduced the so-called
mixed partition function for supersymmetric black holes, and who formulated
an intriguing conjecture about its relation to the partition function
of the topological string. The ability to test these ideas
in a highly non-trivial way relies on two previous developments, which
have been unfolding over the last decade. The first is that
string theory provides models of black holes at the fundamental
or `microscopic' level, where microstates
can be identified and counted with high precision, at least
for supersymmetric black holes 
\cite{StromingerVafa,Maldacena:1997de,Vafa:1997gr}. 
The second development is that one can handle subleading
contributions to the thermodynamical or `macroscopic'
black hole entropy. The macroscopic description of black holes
is provided by solutions to the equations of motion of effective,
four-dimensional supergravity theories, which approximate the
underlying string theory at length scales which are large
compared to the string, Planck and compactification scale. In this framework 
subleading contributions manifest themselves as higher derivative
terms in the effective action. For a 
particular class of higher derivative terms in $N=2$ supergravity,
which are usually referred to as `$R^2$-terms', it is possible
to construct exact near-horizon asymptotic solutions and to compute 
the black hole entropy to high 
precision \cite{LopesCardoso:1998wt,LopesCardoso:2000qm}. 
The subleading corrections
to the macroscopic entropy agree with the subleading contributions
to the microscopic entropy, provided that the area law for the
entropy is replaced by Wald's generalized formula, which applies
to any diffeomorphism invariant Lagrangian \cite{Wald}.

The main tools which make it possible to handle the $R^2$-terms
are the superconformal calculus, which allows the
off-shell construction of $N=2$ supergravity coupled to vector 
multiplets, and the so-called special geometry,
which highly constrains the vector multiplet couplings.
The reason for this simplification
is that scalars and gauge fields sit in the
same supermultiplet, so that the electric-magnetic duality 
of the gauge fields imprints itself on the whole multiplet. As a result
the complicated structure of the theory, including an
infinite class of higher derivative terms, becomes manageable and
transparent, once all quantities are organised such that they 
transform as functions or vectors under the symplectic 
transformations which implement electric-magnetic duality.
This is particularly important
if the $N=2$ supergravity theory is the effective field theory 
of a string compactification, because string dualities form a 
subset of these symplectic transformations.

In these lectures we give a detailed account of the whole
story, starting from the construction of $N=2$ supergravity, proceeding 
to the definition of black hole partition functions, and ending with   
microscopic state counting. In more detail,
the first lecture is devoted to special geometry, the superconformal
calculus and the construction of $N=2$ supergravity with vector 
multiplets, including the $R^2$-terms. The essential concept
of gauge equivalence is explained using non-supersymmetric toy examples.
When reviewing the  construction of $N=2$ supergravity we focus on the
emergence of special geometry and stress the central role of symplectic
covariance. Appendix A, which gives an account of special geometry from the
mathematical point of view, provides an additional perspective 
on the subject. Lecture II starts by reviewing the concept of 
BPS or supersymmetric states and solitons. Its main point is the
black hole variational principle, which underlies the black hole
attractor equations. Based on this, conjectures
about the relation between the macroscopically defined 
black hole free energy and the microscopically defined black hole 
partition functions are formulated. We do not only discuss
how $R^2$-terms enter into this, but also give a detailed discussion
of the crucial role played by the so-called non-holomorphic corrections,
which are essential for making physical quantities, such as the black hole 
entropy, duality invariant. 

The second half of the lectures is devoted to tests of 
the conjectures formulated in Lecture II. 
For concreteness and simplicity, I only discuss the simplest
string compactification with $N=4$ supersymmetry, namely
the compactification of the heterotic string on $T^6$. After 
explaining how the $N=2$ formalism can be used to analyse
$N=4$ theories, we will see that $N=4$ black holes are governed
by a simplified, reduced variational principle for the dilaton.
There are two different types of supersymmetric black holes 
in $N=4$ compactifications, called `large' and `small'
black holes, and we summarize the results on the entropy for
both of them.

With Lecture IV we turn to the microscopic
side of the story. While the counting of $\frac{1}{2}$-BPS
states, corresponding to small black holes,
is explained in full detail, we also give an outline of how
this generalises to $\frac{1}{4}$-BPS states, corresponding to
large black holes. With the state degeneracy at hand, 
the corresponding black hole partition functions can be 
computed and confronted with the predictions made on the 
basis of the macroscopically defined free energy. We give
a critical discussion of the results and point out which
open problems need to be addressed in the future. 
While Appendix A reviews K\"ahler and special K\"ahler
geometry from the mathematical point of view, Appendix B
collects some background material on modular forms.

The selection of the material and the presentation 
are based on two principles. The first is to give a pedagogical
account, which should be accessible to students, postdocs, 
and researchers working in other fields. The second is to 
present this field from the perspective which I found useful
in my own work. For this reason various topics which are 
relevant or related to the subject are not covered in detail, in particular
the topological string, precision state counting for other 
$N=4$ compactifications and for $N=2$ compactifications, and 
the whole field of non-supersymmetric extremal black holes.
But this should not be a problem, 
given that these topics are already covered by other 
excellent recent reviews and lectures notes. See in particular
\cite{Sen:2007qy} for an extensive review of the entropy function
formalism and non-supersymmetric black holes, and \cite{Pioline:2006ni}
for a review emphasizing the role of the topological string.
The selection of references follows the same principles. I have not
tried to give a complete account, but to select those 
references which I believe are most useful for the reader. 
The references are usually given in paragraphs entitled
`Further reading and references' at the end of sections or
subsections.

At the ends of Lectures I and IV I formulate 
exercises which should be instructive for beginners. 
The solutions of these exercises are available
upon request. In addition, some  further exercises
are suggested within the lectures.

\section{Lecture I: Special Geometry \label{LI}}

Our first topic is the so-called special geometry which 
governs the couplings of ${N}=2$ supergravity 
with vector multiplets. We start with a review of
the St\"uckelberg mechanism for gravity, explain 
how this can be generalized to the gauge equivalence
between gravity and a gauge theory of the conformal 
group, and then sketch how this can be used to 
construct ${N}=2$ supergravity in the framework
of the superconformal tensor calculus.

\subsection{Gauge equivalence and  the St\"uckelberg mechanism for 
gravity \label{LI:1}}

The Einstein-Hilbert action
\begin{equation}
S[g] = - \frac{1}{2 \kappa^2} \int d^n x \sqrt{-g} R  
\label{EHaction}
\end{equation}
is not invariant under local dilatations
\begin{equation}
\delta g_{\mu \nu} = - 2 \Lambda(x) g_{\mu \nu} \;.
\end{equation}
However, we can enforce local dilatation invariance at the 
expense of introducing a `compensator'. Let $\phi(x)$ be
a scalar field, which transforms as 
\begin{equation}
\delta \phi = \frac{1}{2} (n-2) \Lambda \phi \;.
\end{equation}
Then the action
\begin{equation}
\tilde{S}[g,\phi] = - \int d^n x \sqrt{-g}
\left( \phi^2 R - 4 \frac{n-1}{n-2} \partial_\mu \phi 
\partial^\mu \phi \right) \;
\label{EHactionDil}
\end{equation}
is invariant under local 
dilatations. If we impose the `dilatational gauge'
\begin{equation}
\phi(x) = a = \mbox{const.} \;,
\end{equation}
we obtain the gauge fixed action
\begin{equation}
\tilde{S}_{\rm g.f.} = - a^2 \int d^n x \sqrt{-g} R \;.
\end{equation}
This is proportional to the Einstein-Hilbert action
(\ref{EHaction}), and becomes equal to it if we choose
the constant $a$ to satisfy $a^2=\frac{1}{2\kappa^2}$. 

The actions $S[g]$ and $\tilde{S}[g,\phi]$ are said to
be `gauge equivalent'. We can go from $S[g]$ to 
$\tilde{S}[g,\phi]$ by adding the compensator $\phi$,
while we get from $\tilde{S}[g,\phi]$ to $S[g]$ by 
gauge fixing the additional local scale symmetry. Both
theories are equivalent, because the extra degree of
freedom $\phi$ is balanced by the additional symmetry.

There is an alternative view of the relation between 
$S[g]$ and $\tilde{S}[g,\phi]$. If we perform the field redefinition
\begin{equation}
g_{\mu \nu} = \phi^{{(n-2)}/4} \tilde{g}_{\mu \nu} \;,
\label{split_trace}
\end{equation}
then 
\begin{equation}
S[g] = \tilde{S}[\tilde{g},\phi] \;.
\end{equation}
Conversely, starting from $\tilde{S}[\tilde{g},\phi]$,
we can remove $\phi$ by a field-dependent gauge transformation 
with parameter $\exp (\Lambda) = \frac{b}{\phi}$, where $b=\mbox{const.}$
The field redefinition (\ref{split_trace}) decomposes the metric
into its trace (a scalar) and its traceless part
(associated with the
graviton). This is analogous
to the St\"uckelberg mechanism for a massive vector field, 
which decomposes the vector field into a
massless vector (the transverse part) 
and a scalar (the longitudinal part), and which makes
the action invariant under $U(1)$ gauge transformations.\\[2ex]
We conclude with some further remarks:
\begin{enumerate}
\item
The same procedure can be applied in the presence of matter. 
The compensator field has to be added in such a way that
it compensates for the transformation of matter fields
under dilatations. Derivatives need to 
be covariantized with respect to dilatations (we will see
how this works in section \ref{GravityAsGaugeTheory}.)
\item
It is possible to write down a dilatation invariant action for
gravity, which only involves the metric and its derivatives,
but this action is quadratic rather than linear in the
curvature:\footnote{In contrast to other formulae in this subsection,
the following formula refers specifically to $n=4$ dimensions.}
\begin{equation}
S[g] = \int d^4 x \sqrt{-g} \left(  R_{\mu \nu} R^{\mu \nu} - \frac{1}{3}
R^2 \right) \;.
\end{equation}
This actions contains terms with up to four derivatives. These and other
higher derivative terms  typically occur when quantum or stringy 
corrections to the Einstein-Hilbert action are taken into account.
\item
When looking at $\tilde{S}[g,\phi]$, one sees that the kinetic
term for the scalar $\phi$ has the `wrong' sign, meaning that
the kinetic energy is not positive definite. This signals that
$\phi$ is not a matter field, but a compensator. 
\end{enumerate}

\subsection{Gravity as a constrained gauge theory of the 
conformal group \label{GravityAsGaugeTheory}}

Let us 
recall some standard concepts of gauge theory. Given 
a reductive\footnote{A direct sum of simple and abelian Lie 
algebras.} Lie algebra with generators $X_A$ and relations
$[X_A,X_B]=f_{AB}^C X_C$, we define a Lie algebra valued 
gauge field (connection)
\begin{equation}
h_\mu = h_\mu^A X_A \;.
\end{equation}
The corresponding covariante derivative (frequently also  
called the connection)
is
\begin{equation}
D_\mu = \partial_\mu - i h_\mu \;,
\end{equation}
where it is understood that $h_\mu$ operates on the representation
of the field on which $D_\mu$ operates. The field strength (curvature)
is
\begin{equation}
R_{\mu \nu}^A = 2 \partial_{[\mu} h_{\nu]}^A + 2 h_{[\mu}^B h_{\nu]}^C 
f^A_{BC} 
\;.
\end{equation}

We now specialize to the conformal group, which is generated
by translations $P^a$, Lorentz transformations $M^{ab}$,
dilatations $D$ and special conformal 
transformations $K^a$. Here $a,b =0,1,2,3$ are internal indices.
We denote the
corresponding gauge fields (with hindsight) by 
$e_{\mu}^a, \omega_{\mu}^{ab}, b_\mu, f_\mu^a$, where $\mu$
is a space-time index. The corresponding field strength
are denoted $R(P)_{\mu \nu}^a$, $R(M)_{\mu \nu}^{ab}$,
$R(D)_{\mu \nu}$, $R(K)_{\mu \nu}^a$.

So far the conformal transformations have been treated as
internal symmetries,
acting as gauge transformations
at each point of space-time, but not acting on space-time. 
The set-up is precisely as in any standard gauge theory, except that
our gauge group is not compact and wouldn't lead to a
unitary Yang-Mills-type theory.    

But now the so-called conventional constraints are imposed, which
enforce that the local translations are identified with 
diffeomorphisms of space-time, while the local Lorentz transformations
become Lorentz transformations of local frames.
\begin{enumerate}
\item
The first constraint is 
\begin{equation}
R(P)_{\mu \nu}^a =0 \;.
\end{equation}
It can be shown that this implies that local translations act 
as space-time diffeomorphisms, modulo gauge transformations. 
As a result, the M-connection $\omega_{\mu}^{ab}$ becomes 
a dependent field, and can be expressed in terms of the P-connection
$e_\mu^a$ and the D-connection $b_\mu$:
\begin{eqnarray}
\omega_\mu^{ab} &=& \omega(e)_\mu^{ab} - 2 e_\mu^{[a} e^{b]\nu} b_\nu \;,\\
\omega(e)_{\mu b}^{\;\;\;\;c} &=& \frac{1}{2} e_\mu^{\;\;a}
( - \Omega_{ab}^{\;\;\;\;c} + \Omega_{b\;\;a}^{\;\;c} + \Omega^c_{\;\;
ab} ) \label{Solve_omega}\\
\Omega_{ab}^c &=& e_a^\mu e_b^\nu \left(
\partial_\mu e_\nu^c - \partial_\nu e_\mu^c \right) \;,\\
\mbox{where} \;\;\;e_\mu^a e_a^\nu &=& \delta_\mu^\nu \;.\nonumber
\end{eqnarray}
\item
The second constraint imposes `Ricci-flatness' on the M-curvature:
\begin{equation}
e^\nu_b R(M)^{ab}_{\mu \nu} = 0\;.
\end{equation}
This constraint allows to solve for the K-connection:
\begin{equation}
f_\mu^a = \frac{1}{2} e^{\nu a} \left( R_{\mu \nu} - \frac{1}{6}
R g_{\mu \nu} \right) \;,
\label{SolveK}
\end{equation}
where 
\begin{equation}
R_{\mu \nu}^{ab} := R(\omega)_{\mu \nu}^{ab} :=
2 \partial_{[\mu} \omega_{\nu]}^{ab} - 2
\omega_{[\mu}^{ac} \omega_{\nu]}^{db}\eta_{cd}
\label{Romega}
\end{equation}
is the part of the M-curvature which does not involve the K-connection:
\begin{equation}
R(M)_{\mu \nu}^{ab} = R(\omega)_{\mu \nu}^{ab} - 4 f_{[\mu}^{[a} e_{\nu]}^{b]}
\;.
\end{equation}
\end{enumerate}
By inspection of (\ref{Solve_omega}) and (\ref{Romega}) we can
identify $\omega(e)_{\mu}^{ab}$ with the spin connections, 
$R_{\mu \nu}^{ab}$ with the space-time curvature, $e_{\mu}^a$
with the vielbein and $\Omega_{ab}^{\;\;\;\;c}$ with the anholonomity
coefficients.\footnote{The anholonomity coefficients measure the deviation
of a given frame (choice of basis of tangent space at each point)
from a coordinate frame (choice of basis corresponding to the tangent
vector fields of a coordinate system).} 
While $\omega_\mu^{ab}$ and $f_{\mu}^a$
are now dependent quantities, the D-connection $b_\mu$ is still
an independent field. However, it can be shown that 
$b_\mu$ can be gauged away 
using K-transformations, and the vielbein $e_\mu^a$ remains
as the only independent physical field. Thus we have matched the
field content of gravity. To obtain the Einstein-Hilbert action,
we start from the conformally invariant action for a scalar field
$\phi$:
\begin{equation}
S = - \int d^4 x e \phi (D_c)^2 \phi \;,
\label{ActionConfGrav}
\end{equation}
where $(D_c)^2 = D_\mu D^\mu$ is the conformal
D'Alambert operator. In the K-gauge $b_\mu=0$ this becomes
\begin{equation}
S = \int d^4 x e \left( \partial_\mu \phi \partial^\mu \phi - \frac{1}{6}
R \phi^2 \right) \;.
\end{equation}
As in our discussion of the St\"uckelberg mechanism, we can 
now impose the D-gauge $\phi=\phi_0=\mbox{const.}$ to obtain
the Einstein-Hilbert action. Observe that the kinetic term
for $\phi$ has again the `wrong' sign, indicating that this
field is a compensator. Note that the Einstein-Hilbert action is
obtained from a conformal matter action, and not 
from a Yang-Mills-type action with Lagrangian
$\sim (R(M)_{\mu \nu}^{ab})^2$. As we have seen already in the discussion 
of the St\"uckelberg mechanism, such actions are higher order
in derivatives, and become interesting once we want to 
include higher order corrections to the Einstein-Hilbert action.

\subsection{Rigid $N=2$ vector multiplets}

Before we can adapt the method of the previous section 
to the case of ${N}=2$ supergravity, we need
to review rigidly supersymmetric $N=2$ vector multiplets.
An $N=2$ off-shell vector multiplet has the following
components:
\begin{equation}
\left( X, \lambda_i, A_\mu | Y_{ij} \right) \;.
\end{equation}
$X$ is a complex scalar and $\lambda_i$ is a doublet of 
Weyl spinors. The ${N}=2$ supersymmetry algebra
has the R-symmetry group $SU(2) \times U(1)$, and 
the index $i=1,2$ belongs to the fundamental representation 
of $SU(2)$. $A_\mu$ is a gauge field, and $Y_{ij}$ is an 
$SU(2)$-triplet ($Y_{ij} = Y_{ji}$) of scalars, which
is subject to the reality constraint 
$\overline{Y}^{ij} = Y_{ij}$\footnote{$SU(2)$ indices are raised 
and lowered with the invariant tenor $\varepsilon_{ij}
=-\varepsilon_{ji}$.}. All together there
are 8 bosonic and 8 fermionic degrees of freedom.

If we build an action with abelian gauge symmetry,
then the gauge field $A_\mu$ will only enter through
its field strength $F_{\mu \nu}=2\partial_{[\mu}A_{\nu]}$, which is part of
a so-called restricted\footnote{While a general chiral 
$N=2$ chiral multiplet has $16+16$ components, a 
restricted chiral multiplet is 
obtained by imposing additional 
conditions and has only $8+8$ (independent) components.
Moreover, the anti-selfdual tensor field $F^-_{\mu \nu}$ of a 
restricted chiral multiplet is 
subject to a Bianchi identity, which allows to interpret it as a
field strength.} chiral ${\cal N}=2$ multiplet
\begin{equation}
\underline{X} = 
\left(X, \lambda_i, F^-_{\mu \nu}, \ldots | Y_{ij}, \ldots \right) \;,
\end{equation}
where the omitted fields are dependent. $F^{-}_{\mu \nu}$ is
the anti-selfdual part of the field strength $F_{\mu \nu}$. 
The selfdual part $F^+_{\mu \nu}$
resides in the complex conjugate of the above multiplet, together
with the complex conjugate scalar $\overline{X}$ and fermions of the 
opposite chirality. 

We take an arbitrary number $n+1$ of such multiplets and label 
them by $I=0,1,\ldots, n$. The general Lagrangian is given by
a chiral integral over $N=2$ superspace,
\begin{equation}
{\cal L}^{\rm rigid} = \int d^4 \theta F(\underline{X}^I) + \mbox{c.c.} \;,
\label{LvmRigidSuper}
\end{equation}
where $F(\underline{X}^I)$ is a function which depends arbitrarily on 
the restricted chiral superfields $\underline{X}^I$ but not on their
complex conjugates. Restricting the superfield 
$F(\underline{X}^I)$ to its lowest component, 
we obtain a holomorphic function $F(X^I)$ of the
scalar fields, called the prepotential. The bosonic part of the
resulting component Lagrangian is given by the highest component of
the same superfield and reads
\begin{equation}
{\cal L}^{\rm rigid} = 
i(\partial_\mu F_I \partial^\mu \overline{X}^I -
\partial_\mu \overline{F}_I \partial^\mu {X}^I )
+ \frac{i}{4} F_{IJ} F^{-I}_{\mu \nu} F^{-J|\mu \nu}
- \frac{i}{4} \overline{F}_{IJ} F^{+I}_{\mu \nu} F^{+J|\mu \nu} \;.
\label{LvmRigid}
\end{equation}
Here $\overline{X}^I$ is the complex conjugate of $X^I$, etc., and
\begin{equation}
F_I = \frac{\partial F}{\partial X^I} \;,\;\;\;
F_{IJ}=\frac{\partial^2 F}{\partial X^I \partial X^J } \;,\;\;\mbox{etc.}
\end{equation}
The equations of motion for the gauge fields are:\footnote{As an 
additional exercise, convince yourself that you get the Maxwell equations 
if the gauge couplings are constant.}
\begin{eqnarray}
\partial_\mu \left(
G^{-|\mu \nu}_I - G^{+|\mu \nu}_I  \right) &=& 0 \;,\label{ELeqs}\\
\partial_\mu \left(
F^{-|\mu \nu}_I - F^{+|\mu \nu}_I  \right) &=& 0 \;. \label{Bianchi}
\end{eqnarray}
Equations (\ref{ELeqs}) are the Euler-Lagrange equations resulting
from variations of the gauge fields $A^I_\mu$. We formulated them using the
dual gauge fields
\begin{equation}
G^{\pm | \mu \nu}_I := 2i \frac{\partial {\cal L}}{\partial 
F^{I\pm}_{\mu \nu}} \;.
\end{equation}
Equations (\ref{Bianchi}) are the corresponding Bianchi identities. 
The combined
set of field equations is invariant under linear transformations of 
the $2n+2$ field strength $(F^{I\pm}, G^\pm_I)^T$. Since the
dual field strength are dependent quantities, we would like
to interpret the rotated set of field equations as the Euler-Lagrange
equations and Bianchi identites of a `dual' Lagrangian. Up to 
rescalings of the field strength, this restricts the linear transformations 
to the symplectic group $Sp(2n+2,\mathbbm{R})$. These symplectic
rotations generalize the electric-magnetic duality transformations
of Maxwell theory.\footnote{To see this more clearly, take 
$F_{IJ}$ to be constant and restrict yourself to one single gauge field.
The resulting $Sp(2,\mathbbm{R}) \simeq SL(2,\mathbbm{R})$ mixes 
the field strength with its Hodge dual.}

Since $G^{I-}_{\mu \nu} \propto F_{IJ} F^{J-}_{\mu \nu}$, the
gauge couplings $F_{IJ}$ must transform fractionally linearly:
\begin{equation}
\mathbbm{F} \rightarrow (W + V \mathbbm{F}) (U + Z \mathbbm{F})^{-1} \;,
\end{equation}
where $\mathbbm{F} = (F_{IJ})$ and 
\begin{equation}
\left( \begin{array}{cc}
U & Z \\ W & V \\
\end{array} \right) \in Sp(2n+2,\mathbbm{R}) \;.
\end{equation}
This transformation must be induced by a symplectic
rotation of the scalars. This is the case
if $(X^I, F_I)^T$ transforms 
linearly, with the same matrix as the field strength. 

Quantities which transform linearly, such as the field strength
$(F^{I\pm}_{\mu \nu}, G^{\pm}_{I|\mu \nu})^T$ and the scalars $(X^I, F_I)^T$ are 
called symplectic vectors. A function $f(X)$ is called a 
symplectic function if 
\begin{equation}
f(X) = \tilde{f}(\tilde{X}) \;.
\end{equation}
Note that the prepotential $F(X)$ is {\em not}
a symplectic function, but transforms in a rather complicated way.
However, we can easily construct examples of symplectic functions,
by contracting symplectic vectors. The following symplectic
functions will occur in the following:
\begin{eqnarray}
K &=& i \left( X^I \overline{F}_I - F_I \overline{X}^I 
\right) \;, \label{KaehlerPotentialRigid}\\
{\cal F}_{\mu \nu}^- &=& X^I G^{-}_{I|\mu \nu} - F_I F^{I-}_{\mu \nu} \;.
\end{eqnarray}

The scalar part of the action (\ref{LvmRigid}) can be rewritten 
as follows:
\begin{equation}
{\cal L}^{\rm rigid}_{\rm scalar} = - N_{IJ} 
\partial_\mu X^I \partial^\mu X^J \;,
\end{equation}
where
\begin{equation}
N_{IJ} = -i \left( F_{IJ} - \overline{F}_{IJ} \right)= \frac{\partial^2
K}{\partial X^I \partial \overline{X}^J} \;.
\label{RigidMetric}
\end{equation}
$N_{IJ}$ can be interpreted as a Riemannian metric on 
the target manifold of the scalars $X^I$, which we denote
$M$. In fact,
$N_{IJ}$ is a K\"ahler metric with K\"ahler potential
(\ref{KaehlerPotentialRigid}). Thus the scalar manifold 
$M$ is a K\"ahler manifold. Moreover, 
$M$  is a non-generic K\"ahler manifold, 
because its K\"ahler
potential can be expressed in terms of the holomorphic
prepotential $F(X^I)$. Such manifolds are called `affine 
special K\"ahler manifolds.' 

An intrinsic definition of affine special K\"ahler manifolds
can be given in terms of the so-called
special connection $\nabla$ (which is different from the
Levi-Civita connection of the metric $N_{IJ}$). This is 
explained in appendix \ref{AppA}.
Equivalently, an affine special K\"ahler manifold 
can be chararacterised (locally) by the existence
of a so-called K\"ahlerian  Lagrangian immersion 
\begin{equation}
\Phi: M \rightarrow T^*\mathbbm{C}^{n+1} 
\simeq \mathbbm{C}^{2n+2} \;.
\end{equation}
In this construction the special K\"ahler metric
of $M$ is obtained by pulling
back a flat K\"ahler metric from $T^*\mathbbm{C}^{n+1}$.
In other words, all specific properties
of $M$ 
are encoded in the immersion $\Phi$. Since the immersion
is Lagrangian, it has a generating function, which
is nothing but the prepotential: $\Phi=dF$.
The immersed manifold $M$ is
(generically) the graph of a map $X^I \rightarrow W_I = F_I (X)$,
where $(X^I, W_I)$ are symplectic coordinates on 
$T^*\mathbbm{C}^{n+1}$. Along the 
immersed manifold, half of the coordinates of $T^*\mathbbm{C}^{n+1}$
become functions of the other half: the $X^I$ are coordinates
on $M$ while the $W_I$ can be expressed in 
terms of the $X^I$ using the prepotential as 
$W_I = \frac{\partial F}{\partial X^I}$. 
We refer the interested reader to appendix \ref{AppA}
for more details on the mathematical aspects of this construction.

\subsection{Rigid superconformal vector multiplets}

The superconformal calculus provides a systematic way 
to obtain the Lagrangian of $N=2$ Poincar\'e supergravity
by exploiting its gauge equivalence with $N=2$ conformal
supergravity. This proceeds in the following steps:
\begin{enumerate}
\item
Construct the general Lagrangian for rigid superconformal
vector multiplets.
\item
Gauge the superconformal group to obtain conformal supergravity.
\item
Gauge fix the additional transformations to obtain
Poincar\'e supergravity.
\end{enumerate}
One can use the gauge equivalence to study
Poincar\'e supergravity in terms of conformal supergravity,
which is useful because one can maintain manifest symplectic
covariance. In practice one might gauge fix some transformations,
while keeping others intact, or use gauge invariant quantities.

As a first step, we need to discuss the additional constraints
resulting from rigid $N=2$ superconformal invariance. Besides the
conformal generators $P^a$, $M^{ab}$, $D$, $K^a$, the $N=2$ superconformal
algebra contains the generators $A$ and $V^\Lambda$ of 
the $U(1)\times SU(2)$ R-symmetry, the supersymmetry generators
$Q$ and the special supersymmetry generators $S$. Note that
the superconformal algebra has a second set of supersymmetry 
transformations which balances the additional bosonic symmetry
transformations. 

The dilatations and chiral $U(1)$ transformations naturally combine
into complex scale transformations. The scalars have scaling weight
$w=1$ and $U(1)$ charge $c=-1$:
\begin{equation}
X^I \rightarrow \lambda X^I \;,\;\;\;
\lambda = |\lambda| e^{-i\phi} \in \mathbbm{C}^* \;.
\end{equation}
Scale invariance of the action requires that the prepotential is
homogenous of degree 2:
\begin{equation}
F(\lambda X^I) = \lambda^2 F(X^I) \;.
\end{equation}
Geometrically, this implies that the scalar manifold
$M$ of  rigid superconformal 
vector multiplets is 
a complex cone. Such manifolds are called `conical affine
special K\"ahler manifolds'.

\subsection{$N=2$ conformal supergravity}

The construction of $N=2$ supergravity now proceeds along
the lines of the $N=0$ example given in section \ref{GravityAsGaugeTheory}.
Starting from (\ref{LvmRigidSuper}), one needs to covariantize
all derivatives with respect to superconformal transformations.
The corresponding gauge fields are:
$e_\mu^a$ (Translations),
$\omega_\mu^{ab}$ (Lorentz transformations),
$b_\mu$ (Dilatations),
$f_\mu^a$ (special conformal transformations),
$A_\mu$ (chiral $U(1)$ transformations),
${\cal V}_{\mu i}^j$ ($SU(2)$ transformations),
$\psi^i_\mu$ (supersymmetry transformations) and
$\phi_\mu^i$ (special supersymmetry transformations).

As in section \ref{GravityAsGaugeTheory} one needs to
impose constraints, which then allow to solve for some
of the gauge fields. The remaining, independent gauge fields
belong to the Weyl multiplet,
\begin{equation}
\left( e_\mu^a, \psi_\mu^i, b_\mu, A_\mu, {\cal V}_{\mu i}^j 
| T^-_{ab}, \chi^i, D \right) \;,
\end{equation}
together with the auxiliary fields $T^-_{ab}$ (anti-selfdual tensor),
$\chi^i$ (spinor doublet) and $D$ (scalar).
The only physical degrees of freedom 
contributed to Poincar\'e supergravity from this multiplet are
the graviton $e_\mu^a$ and the two gravitini $\psi_\mu^i$. 
The other connections can be gauged away or
become dependent fields upon gauge fixing.

While covariantization of (\ref{LvmRigidSuper}) with respect
to superconformal transformations leads to a conformal supergravity
Lagrangian with up to two derivatives in each term, it is also possible
to include a certain class of higher derivative terms. This elaborates
on the previous observation that one can also construct a Yang-Mills
like action quadratic in the field strength. The field strength
associated with the Weyl multiplet form a reduced chiral tensor
multiplet $\underline{W}_{ab}$, whose lowest component
is the auxiliary tensor field $T^-_{ab}$. The highest component
contains, among other terms, the Lorentz curvature, which 
after superconformal gauge fixing becomes the 
anti-selfdual Weyl tensor ${}^-C^{-}_{\mu \nu \rho
\sigma}$. 
By contraction 
of indices one can form the (unreduced) chiral multiplet
$\underline{W}^2 = \underline{W}_{ab} \underline{W}^{ab}$, 
which is also referred to as
`the' Weyl multiplet. While its lowest component is 
$\hat{A} = (T_{ab}^-)^2$, the highest component contains, among other
terms, the square of the anti-selfdual Weyl tensor.
Higher curvature terms can now be incorporated by allowing
the prepotential to depend {\em explicitly} on the Weyl 
multiplet: $F(X^I) \rightarrow F(X^I, \hat{A})$. Dilatation invariance
requires that this (holomorphic) function must be (graded)
homogenous of degree 2:
\begin{equation}
F(\lambda X^I, \lambda^2 \hat{A}) = \lambda^2 F(X^I, \hat{A}) \;.
\end{equation}
We refrain from writing down the full bosonic Lagrangian. However 
it is instructive to note that the scalar part, which is 
the analogue of  (\ref{ActionConfGrav}) reads
\begin{equation}
8 \pi e^{-1}{\cal L}_{\rm scalar} =
i \left( \overline{F}_I D^a D_a X^I - F_I D^a D_a \overline{X}^I \right)
\;.
\label{Scalar1}
\end{equation}
Here $D_a$ is the covariant derivative with respect to all 
superconformal transformations.

\subsection{$N=2$ Poincar\'e supergravity}

Our goal is to construct the coupling of $n$ 
vector multiplets to $N=2$ Poincar\'e supergravity. 
The gauge equivalent superconformal theory involves the
Weyl multiplet and $n+1$ vector multiplets, one of which
acts a compensator. Moreover, one needs to add a second 
compensating multiplet, which one can take to be a hypermultiplet.
The second compensator does not contribute any physical 
degrees of freedom to the vector multiplet sector. This is 
different for the compensating vector multiplet. The physical
fields in the $N=2$ supergravity multiplet are the graviton 
$e_\mu^a$, the gravitini $\psi_\mu^i$ and the graviphoton
${\cal F}_{\mu \nu}$. While the first two fields come from
the Weyl multiplet, the graviphoton is a linear combination of
the field strength of all the $n+1$ superconformal vector multiplets:
\begin{equation}
{\cal F}^-_{\mu \nu} = X^I G^-_{I|\mu \nu} - F_I F^{I-}_{\mu \nu} \;.
\end{equation}
At the two-derivative level, one obtains $T^-_{\mu \nu} =
{\cal F}^-_{\mu \nu}$ when eliminating the auxiliary tensor by
its equation of motion. Note, however, that once higher derivative
terms have been added, this relation becomes more complicated, and 
can only be solved iteratively in derivatives.

While all $n+1$ gauge fields of the superconformal theory correspond
to physical fields of the Poincar\'e supergravity theory, 
one of the superconformal scalars acts as a compensator for
the complex dilatations. Gauge fixing imposes one complex 
condition on $n+1$ complex scalars, which leaves $n$ physical 
complex scalars. Geometrically, the scalar manifold of the 
Poincar\'e supergravity theory arises by taking the quotient
of the `superconformal' scalar manifold by the action of the complex
dilatations.

To see what happens with the scalars, we split the superconformal
covariant derivative $D_\mu$ into the covariant derivative ${\cal D}_\mu$,
which contains the connections for $M,D,U(1), SU(2)$, and the
remaining connections. Then the scalar term 
(\ref{Scalar1}) becomes
\begin{eqnarray}
8 \pi e^{-1}{\cal L}_{\rm scalar} &=&
i \left( \overline{F}_I {\cal D}^a {\cal D}_a X^I - F_I {\cal D}^a {\cal D}_a 
\overline{X}^I \right) \nonumber \\
&& - i \left(F_I \overline{X}^I - \overline{F}_I X^I 
\right) \left( \frac{1}{6} R -D\right) \;.
\label{Scalar2}
\end{eqnarray}
In absence of higher derivative terms, the only other term 
containing the auxiliary field $D$ is 
\begin{equation}
8 \pi e^{-1}{\cal L}_{\rm comp} = \chi \left( \frac{1}{6} R + \frac{1}{2}
D \right) \;,
\end{equation}
where $\chi$ depends on the compensating hypermultiplet. The
equation of motion for $D$ is solved by\footnote{Thus, at the two-derivative 
level,  $D$ just acts as a Lagrange multiplier. This changes once
higher-derivative terms are added, but we won't discuss the 
implications here.}
\begin{equation}
\frac{1}{2} \chi = i \left(F_I \overline{X}^I - \overline{F}_I X^I 
\right) \;.
\end{equation}
When substituting this back, $D$ cancels out, and we obtain
\begin{eqnarray}
8 \pi e^{-1}({\cal L}_{\rm scalar} + {\cal L}_{\rm comp})
&=&
i \left( \overline{F}_I {\cal D}^a {\cal D}_a X^I - F_I {\cal D}^a {\cal D}_a 
\overline{X}^I \right) \nonumber \\ && +  
\left(i(F_I \overline{X}^I - \overline{F}_I X^I) 
\right) \left( - \frac{1}{2}R \right) \;.
\label{Scalar3}
\end{eqnarray}
The second line gives 
the standard Einstein-Hilbert term, in Planckian
units $G_N=1$, 
\begin{equation}
8 \pi e^{-1}{\cal L} = - \frac{1}{2} R + \cdots \;, 
\end{equation}
once we impose the D-gauge
\begin{equation}
i \left(F_I \overline{X}^I - X^I \overline{F}_I \right) =1 \;.
\end{equation}
Geometrically, imposing the D-gauge amounts to taking the
quotient of the scalar manifold $M$ with
respect to the (real) dilatations $X^I \rightarrow |\lambda| X^I$.
The chiral $U(1)$ transformations act isometrically
on the quotient, and therefore we can take a further quotient 
by imposing a $U(1)$ gauge.
The resulting manifold 
$\overline{M}=M/\mathbbm{C}^*$ is the scalar manifold of the
Poincar\'e supergravity theory. It
is a K\"ahler manifold, whose 
K\"ahler potential can be expressed in terms of the prepotential
$F(X^I)$. The target manifolds of vector multiplets of 
in $N=2$ Poincar\'e supergravity are called `(projective) special
K\"ahler manifolds.'

To see how the geometry of $\overline{M}$ arises, consider
the scalar sigma model given by the first line of (\ref{Scalar3})
\begin{eqnarray}
8 \pi e^{-1} {\cal L}_{\rm sigma} &=&
i ( {\cal D}_\mu F_I {\cal D}^\mu \overline{X}^I -
{\cal D}_\mu X^I {\cal D}^\mu \overline{F}_I ) \\
 &=& - N_{IJ} {\cal D}_\mu X^I {\cal D}^\mu \overline{X}^J \;,
\label{ScLag1}
\end{eqnarray}
where 
\begin{equation}
N_{IJ} = 2 {\rm Im} F_{IJ} = -i (F_{IJ} - \overline{F}_{IJ})  \;,
\label{DefN}
\end{equation}
and
\begin{eqnarray}
{\cal D}_\mu X^I = (\partial_\mu +i A_\mu) X^I \;,\;\;\; && 
{\cal D}_\mu \overline{X}^I = (\partial_\mu -i A_\mu) X^I \;, \\
{\cal D}_\mu F_I = (\partial_\mu +i A_\mu) F_I \;,\;\;\; &&
{\cal D}_\mu \overline{F}_I = (\partial_\mu - i A_\mu) \overline{F}_I \;.
\end{eqnarray}
We imposed the K-gauge $b_\mu =0$, so that only the $U(1)$ 
gauge field $A_\mu$ appears in the covariant derivative.
This gauged non-linear sigma model is the only place where 
$A_\mu$ occurs in the Lagrangian. $A_\mu$
can be eliminated by solving its equation of motion
\begin{equation}
A_\mu = \frac{1}{2} (
\overline{F}_I \stackrel{\leftrightarrow}{\partial}_\mu X^I 
- \overline{X}^I \stackrel{\leftrightarrow} {\partial}_\mu F_I  )\;.
\label{EoMU1GF}
\end{equation}
Substituting this back, we obtain the non-linear sigma model
\begin{eqnarray}
8 \pi e^{-1} {\cal L}_{\rm sigma} &=&
- (N_{IJ} + e^{\cal K} (N \overline{X})_I (N X)_J ) \partial_\mu X^I \partial^{\mu}
\overline{X}^J \nonumber \\ 
&=:&  - M_{IJ} \partial_\mu X^I \partial^\mu \overline{X}^J 
\label{ScLag2}
\end{eqnarray}
Here we suppress indices which are summed over:
\[
(NX)_I := N_{IJ} X^J \;, \;\;\;\mbox{etc} \;.
\]
The scalar metric $M_{IJ}$ has two null directions
\begin{equation}
\label{transversal}
X^I M_{IJ} = 0 = M_{IJ} \overline{X}^J \;.
\end{equation}
This does not imply that that the kinetic term for the physical
scalars is degenerate, because $M_{IJ}$ operates on the 
`conformal scalars' $X^I$, which are subject to dilatations and
$U(1)$-transformations. We have already gauge-fixed the
dilatations by imposing the D-gauge. We could similarly 
impose a gauge condition for the $U(1)$ transformations,
but it is more convenient to introduce the gauge invariant 
scalars 
\begin{equation}
Z^I = \frac{X^I}{X^0} \;.
\end{equation}
One of these scalars is trivial, $Z^0=1$, while the 
others $z^i = Z^i$, $i=1,\ldots, n$ are the physical scalars
of the Poincar\'e supergravity theory. Using the
transversality relations (\ref{transversal}) and the homogenity of 
the prepotential, we can rewrite the Lagrangian in terms of the
gauge-invariant scalars $Z^I$:
\[
8 \pi e^{-1} {\cal L}_{\rm sigma} = 
- g_{IJ} \partial_\mu Z^I \partial^\mu \overline{Z}^J \;,
\]
where 
\begin{equation}
\label{HorizontalMetric}
g_{IJ} = - \frac{N_{IJ}}{ (ZN\overline{Z}) } +
\frac{ (N\overline{Z})_I (NZ)_J }{ (ZN\overline{Z})^2 } \;.
\end{equation}
Note that we have used the homogenity of the prepotential 
to rewrite it and its derivatives in terms of the $Z^I$:
\[
F(X) = (X^0)^2 F(Z) \;,\;\;\;
F_I(X) = X^0 F_I(Z) \;,\;\;\;
F_{IJ}(X) = F_{IJ} (Z) \;,\;\;\;\mbox{etc.}
\]
One can show that $g_{IJ}$ has the following properties:
\begin{enumerate}
\item
$g_{IJ}$ is degenerate along the complex direction $Z^I$,
or, in other words, along the orbits of the  
$\mathbbm{C}^*$-action. We will call this direction 
the vertical direction. As we will see below the
vertical directions correspond to unphysical 
excitations.
\item
$g_{IJ}$ is non-degenerate along the horizontal directions, 
which form the orthogonal complement of the horizontal
direction with respect to the non-degenerate metric $N_{IJ}$.
As we will see below, this implies a non-degenerate kinetic
term for the physical scalars. 
\item
$g_{IJ}$ is positive definite along the horizontal
directions if and only if $N_{IJ}$ has signature $(2,2n)$ or $(2n,2)$. 
This corresponds to the case where 
$N_{IJ}$ has opposite signature along the 
vertical  and horizontal directions. We need to impose this
to have standard kinetic terms for the physical scalars.
\item
$g_{IJ}$ can be obtained from a K\"ahler potential which 
in turn can be expressed by the prepotential of the 
underlying superconformal theory:
\[
g_{IJ} = \frac{\partial^2 K}{\partial Z^I \partial \overline{Z^J}}
\;,\;\;\;
K = - \log \left( i ( F_I \overline{Z}^I - Z^I \overline{F}_I )\right)\;.
\]
Here it is understood that we only set $Z^0=1$ at the end.
\end{enumerate}
Since $Z^0=1$, and, hence, $\partial_\mu Z^0=1$, the 
Lagrangian only depends on in the physical scalars $z^i = Z^i$, 
$i=1,\ldots, n$. Following conventions in the literature,
we distinguish holomorphic indices $i$ and anti-holomorphic
indices $\overline{i}$ when using the physical scalars $z^i$,
despite that we do not make such a distinction for $X^I$, $Z^I$,
etc. Thus the complex conjugate of $z^i = Z^i$ is denoted
$\overline{z}^{\overline{i}} = \overline{Z}^i$.  

To express the Lagrangian in terms of the physical scalars,
we define
\[
{\cal F}(z^1, \ldots, z^n) := F(Z^0, Z^1, \ldots, Z^n) \;.
\]
The Lagrangian only depends on the horizontal  part 
of $g_{IJ}$, which is denoted $g_{i \overline{j}}$, and which
is given by 
\begin{equation}
g_{i\overline{j}} = \frac{\partial^2 K}{\partial z^i \partial 
\overline{z}^{\overline{j}}} \;.
\label{MetricProj}
\end{equation}
with K\"ahler potential 
\begin{equation}
K = - \log \left( 2i ({\cal F} - \overline{\cal F}) -
i (z^i - \overline{z}^{\overline{i}})
({\cal F}_i + \overline{\cal F}_{\overline{i}}) \right)\;,
\label{KPproj}
\end{equation}
where ${\cal F}_i= \frac{\partial {\cal F}}{\partial z^i}$.
The Lagrangian takes the form
\[
8 \pi e^{-1} {\cal L}_{\rm sigma} = - g_{i \overline{j}}
\partial_\mu z^i \partial^\mu \overline{z}^{\overline{j}} 
\;.
\]
Geometrically, we have performed a quotient of the rigid superconformal
scalar manifold $M$ by the $\mathbbm{C}^*$-action and obtained 
the metric $g_{i\overline{j}}$ of the scalar manifold $\overline{M}$ of 
the Poincar\'e supergravity theory in terms of special 
coordinates $z^i$. Metrics and manifolds obtained in this way are
called `projective special K\"ahler metrics' and `projective 
special K\"ahler manifolds,' respectively.
One can reformulate the theory in terms
of general holomorphic coordinates, but we will not persue this
here. The special coordinates are physically distinguished, because they
are the lowest components of Poincar\'e vector multiplets.
They are also natural from the geometrical point of view, because
they can be defined in terms of intrinsic properties of $M$, 
as explained in more details in appendix A.

Since the $z^i$ are not part of a symplectic vector,
the action of the symplectic transformations in the 
scalar sector is complicated.
Therefore it is often more convenient 
to work on the rigid scalar manifold $M$ 
using  the `conformal scalars' $X^I$ and the symplectic 
vector $(X^I, F_I)^T$. As we have seen, the superconformal
and the super Poincar\'e theory are gauge-equivalent, and 
we know how to go back and forth between the two.
The advantage of the superconformal picture is that
there is an equal number of gauge fields and scalars, which
all sit in vector multiplets. Therefore symplectic transformations
act in a simple way on the scalars.

Let us finally have a brief look at the higher derivative terms.
We expand the function $F(X^I,\hat{A})$ in $\hat{A}$:
\begin{equation}
F(X^I,\hat{A}) = \sum_{g=0}^\infty
F^{(g)}(X^I) \hat{A}^g \;.
\end{equation}
While $F^{(0)}(X^I) = F(X^I)$ is the prepotential, the functions
$F^{(g)}(X^I)$ with $g>0$ are coupling functions multiplying
various higher derivative terms. The most prominent class of 
such terms are
\begin{equation}
F^{(g)}(X^I) ( {}^-C^-_{\mu \nu \rho \sigma})^2 
(T^-_{\mu \nu})^{2g-2} + \mbox{c.c.} \;,
\end{equation}
where ${}^-C^-_{\mu \nu \rho \sigma}$ is the antiselfdual
Weyl tensor and $T^-_{\mu \nu}$ is the antiselfdual
auxiliary field in the Weyl multiplet.
To lowest order in derivatives, this field equals the 
anti-selfdual graviphoton field strength ${\cal F}^-_{\mu \nu}$. 
Therefore such terms are related to effective couplings between
two gravitons and $2g-2$ graviphotons.

$N=2$ supergravity coupled to vector multiplets (and hypermultiplets)
arises by dimensional reduction of type-II string theory on 
Calabi Yau threefolds. Terms of the above form arise from 
loop diagrams where the external states are 
two gravitons and $2g-2$ graviphotons, while an infinite
number of massive strings states runs in the loop. It turns out 
that in the corresponding string amplitudes only genus-$g$ 
diagrams contribute, and that only BPS states make a net 
contribution. 
Moreover these amplitudes are 
`topological': upon topological twisting of the world sheet theory
the couplings $F^{(g)}(X^I)$  
turn into the genus-$g$ free energies (logarithms of the partition functions)
of the topolocial type-II string. This means that the couplings
$F^{(g)}(X^I)$  can be computed, at least in principle.

\subsection{Further reading and references}

Besides original papers, my main 
sources for this lecture are the 1984 Trieste lecture notes
of de Wit \cite{deWit:1984hw}, and an (unpublished) 
Utrecht PhD thesis \cite{KleijnPhD}.
Roughly the same material was covered in Chapter 3 of
my review \cite{Mohaupt:2000mj}. 
Readers who would like to study special geometry and $N=2$ supergravity in the
superconformal approach in detail
should definitely look into the original
papers, starting with \cite{deWit:1984pk,deWit:1984px}. 
Electric-magnetic duality in the presence of $R^2$-corrections was 
investigated in \cite{deW:1996/02,deW:1996/03}, and is reviewed in
\cite{Mohaupt:2000mj}. Special geometry
has been reformulated in terms of general (rather than special) holomorphic 
coordinates \cite{Str:1990,CasDauFer:1990,CraRooTrovP:1997}. We will not
discuss this approach in these lectures and refer the reader to 
\cite{Andrianopoli:1996cm} for a review of $N=2$ supergravity within
this framework. The intrinsic definition of special K\"ahler geometry
in terms of the special connection $\nabla$ was proposed in 
\cite{Freed:1997dp}. The equivalent characterisation by a K\"ahlerian
Lagrangian immersion into a complex symplectic vector space 
is described in \cite{Alekseevsky:1999ts}.  Key references
about the topological string and its role in computing 
couplings in the effective action are \cite{Bershadsky:1993cx}
and \cite{Antoniadis:1996qg}. See also \cite{Pioline:2006ni}
for a review of the role of the topological string for 
black holes.

\subsection{Problems}

\begin{Problem}
\label{Prob:StuckelbergGravity}
The St\"uckelberg mechanism for gravity.\\[2ex]
Compute the variation of the Einstein-Hilbert action 
\begin{equation}
S[g] = - \frac{1}{2 \kappa^2} \int d^n x \sqrt{-g} R  
\label{EHactionVar}
\end{equation}
and the variation of the action 
\begin{equation}
\tilde{S}[g,\phi] = - \int d^n x \sqrt{-g}
\left( \phi^2 R - 4 \frac{n-1}{n-2} \partial_\mu \phi 
\partial^\mu \phi \right) 
\label{EHactionDilVar}
\end{equation}
under local dilatations
\begin{equation}
\delta g_{\mu \nu} = - 2 \Lambda(x) g_{\mu \nu} \;,\;\;\;
\delta \phi = \frac{1}{2} (n-2) \Lambda \phi \;.
\end{equation}
You can use that
\begin{eqnarray}
\delta \sqrt{-g} &=& - n \Lambda \sqrt{-g} \;, \nonumber \\
g^{\mu \nu} R_{\mu \nu} &=& -2 (n-1) \nabla^2 \Lambda \;.
\end{eqnarray}
You should find that (\ref{EHactionDilVar}) is invariant while
(\ref{EHactionVar}) is not, as explained in Lecture I. Convince
yourself that you can obtain (\ref{EHactionVar}) from 
(\ref{EHactionDilVar}) by gauge fixing.

If you are not familiar with the St\"uckelberg mechanism, 
use what you have learned to make the action of  a free massive
vector field invariant with respect to local $U(1)$ transformations.
\end{Problem}

\begin{Problem}
Einstein-Hilbert action from conformal matter. \\[2ex]
Show that the Einstein-Hilbert action (\ref{EHactionVar}) can be 
obtained from the conformally invariant matter action 
\begin{equation}
S = - \int d^4 x e \phi D_c^2 \phi \;,
\label{ConfGravVar}
\end{equation}
where $D_c^2 = D_\mu D^\mu$ is the conformal
D'Alambert operator, by gauge fixing the K- and D-transformations.\\[2ex]
{\em Instruction}: the scalar field $\phi$ is neutral under K-transformations
and transforms with weight $w=1$ under $D_\mu$. Its first and second
conformally covariant derivatives are:
\begin{eqnarray}
D_\mu \phi &=& \partial_\mu \phi - b_\mu \phi \;, \\
D_\mu D^a \phi &=& (\partial_\mu - 2 b_\mu) D^a \phi 
- \omega_{\mu}^{ab} D_b \phi + f_\mu^a \phi \;.
\end{eqnarray}
The K-connection $f_\mu^a$ appears in the second line because
the D-connection $b_\mu$ transforms non-trivially under K. 
Note that $b_\mu$ is the only field in the problem which transforms
non-trivially under K, and that 
$D^2 \phi$ is invariant under K. The K-transformations
can be gauged fixed by setting $b_\mu=0$. (In fact, it is clear that
$b_\mu$ will cancel out of (\ref{ConfGravVar}). Why?)
Use this together with the result of Problem
\ref{Prob:StuckelbergGravity} to obtain the Einstein-Hilbert action
(\ref{EHactionVar}) by gauge fixing (\ref{ConfGravVar}).
\end{Problem}

\section{Lecture II: Attractor Mechanism, Variational Principle, and
Black Hole Partition Functions}

We are now ready to look at BPS black holes in $N=2$ supergravity
with vector multiplets. First we review the concept of a BPS state.

\subsection{BPS states}

The $N$-extended four-dimensional supersymmetry algebra has
the following form:
\begin{eqnarray}
\{ Q_\alpha^A, Q^{+B}_{\dot{\beta}} \} &=& 2 \sigma^\mu_{\alpha \dot{\beta}} 
\delta^{AB}  P_\mu \;,\nonumber \\
\{ Q_\alpha^A, Q^{B}_\beta \} &=& \epsilon_{\alpha \beta} Z^{AB} \;. 
\nonumber
\end{eqnarray}
$A,B, \ldots = 1, \ldots, N$ label the supercharges, which we have taken 
to be Weyl spinors. The generators $Z^{AB}=-Z^{BA}$ are central,
i.e. they communte with all generators of the Poincar\'e Lie superalgebra. 
On irreducible representations they are complex multiples of the unit
operators. One can then skew-diagonalise the antisymmetric constant 
matrix $Z^{AB}$, and the skew eigenvalues $Z_1, Z_2, \ldots$ are known
as the central charges carried by the representation. The eigenvalue
of the Casimir operator $P^\mu P_\mu$ is $-M^2$, where $M$ is the
mass. Using the algebra one can derive the BPS inequality
\[
M^2 \geq |Z_1|^2 \geq |Z_2|^2 \geq \cdots \geq 0 \;,
\]
where we have labeled the central charges according to the size
of their absolute values. Thus the mass is bounded from below 
by the central charges.
Whenever a bound on the mass is saturated, some of the supercharges
operate trivially on the representation, and therefore the 
representation is smaller than a generic massive representation.
Such multiplets are called shortened multiplets or BPS multiplets.
The extreme case is reached when all bounds are saturated,
$M=|Z_1| = |Z_2| = \cdots$. In these representations half of the
supercharges operate trivially, and the representation has as many
states as a massless one. These multiplets are called short 
multiplets or $\frac{1}{2}$-BPS multiplets. \\[2ex]
Here are some examples of $N=2$ multiplets.
\begin{enumerate}
\item
$M> |Z|$: these are generic massive multiplets. One example
is the `long' vector multiplet, which has $8+8$ on-shell degrees
of freedom.
\item
$M = |Z|$: these are short or $\frac{1}{2}$-BPS multiplet.
Examples are hypermultiplets and `short' vector multiplets, which
both have $4+4$ on-shell degrees of freedom. The short vector multiplet
is the `Higgsed' version of the massless vector multiplet discussed
earlier in these lectures.\footnote{This has $8+8$ off-shell degrees
of freedom and $4+4$ on-shell degrees of freedom.} The long vector
multiplet combines the degrees of freedom of a hypermultiplet and a
short vector multiplet. This shows that one cannot expect that the
number of BPS multiplets is conserved when deforming the theory 
(by moving through its moduli space of vacua), because BPS multiplets
can combine into non-BPS multiplets. However the 
difference between the number of hypermultiplets and short 
vector multiplets is preserved under multiplet recombination and
has the chance of being an `index'. 
\end{enumerate}

Let us give some examples of $N=4$ multiplets.
\begin{enumerate}
\item
$M > |Z_1| > |Z_2|$: these are generic  massive multiplets.
The number of states is $2^8$.\footnote{We are referring here
to representations of the algebra generated by the supercharges.
Irreducible representations of the full Poincar\'e Lie superalgebra
are obtained by replacing the lowest weight state by any 
irreducible representation of the little group. Their dimension
is therefore a multiple of $2^8$.}
\item
$M = |Z_1| > |Z_2|$: these are called 
intermediate or $\frac{1}{4}$-BPS multiplets.
One quarter of the supercharges operate trivially, and they have
(a multiple of) $2^6$ states.
\item
$M = |Z_1| = |Z_2|$: these are short or $\frac{1}{2}$-BPS multiplets,
with (a multiple of) $2^4$ states.
One example are short $N=4$ vector multiplets which have $8+8$
states, as many as a massless $N=4$ vector multiplet. 
Short or massless multiplets have the same field content as an
large $N=2$ vector multiplet, or, equivalently, as a short or massless
$N=2$ vector multiplet plus a hypermultiplet. 
\end{enumerate}

Finally, there can of course also be singlets under the 
supersymmetry algebra, states which are completely invariant.
Such states are maximally supersymmetric and can therefore be interpreted
as supersymmetric ground states.

\subsubsection*{Further reading and references}

This section summarises basic facts about
the representation theory of Poincar\'e Lie superalgebras,
which can be found in textbooks on supersymmetry, i.p.
in Chapter II of \cite{Wess:1992cp} and Chapter 8 of \cite{West}.

\subsection{BPS solitons and BPS black holes}

One class of BPS states are states in the Hilbert space 
which sit in BPS representations. They correspond to 
fundamental fields
in the Lagrangian, which transform in BPS representations 
of the supersymmetry algebra. Another class of BPS states 
is provided by non-trivial static solutions of the field
equations, which have finite mass and are non-singular. 
Such objects are called solitons and interpreted as extended
particle-like collective excitations of the theory.

Because of the finite mass
condition they have to approach Minkowski space
at infinity\footnote{We only consider theories where Minkowski space
is a superysmmetric ground state.} and can be classified according to their
transformation under the asymptotic Poincar\'e Lie superalgebra
generated by the Noether charges. If this representation is
BPS, the soliton is called a BPS soliton. The corresponding
field configuration admits Killing spinors, i.e. there are
choices of the supersymmetry transformation parameters
$\epsilon(x)$ such that the field configuration is invariant:
\[
\left. \delta_{\epsilon(x)} \Phi(x) \right|_{\Phi_0(x)} = 0\;.
\]
Here $\Phi$ is a collective notation for all fundamental 
fields, and $\Phi_0$ is the invariant field configuration.
The maximal number of linearly independent Killing spinors
equals the number $N$ of supercharges. Solutions with 
$N$ Killing spinors are completely invariant under supersymmetry 
and qualify as supersymmetric ground states.\footnote{Minkowski space
is a trivial example. Here all Killing spinors are constant (in
linear coordinates).}
Generic solitonic solutions of the field equations do not have 
Killing spinors and correspond to generic massive representations.
Solitonic solutions with $\frac{N}{n}$ Killing spinors 
are invariant under $\frac{1}{n}$ of the asymptotic symmetry algebra
and correspond to $\frac{1}{n}$-BPS representations.\footnote{More 
precisely, the collective modes generated by the broken superysmmetries
fall into such representations.}

The particular type of solitons we are interested in are
black hole solutions of $N=2$ supergravity. Black holes 
are asymptotically flat, have a finite mass, and are
`regular' in the sense that they do not have naked singularities.
For static four-dimensional black holes in Einstein-Maxwell type
theories with matter, the BPS bound coincides with the extremality
bound. Therefore BPS black holes are extremal black holes, with 
vanishing Hawking temperature. Since this makes them stable against
decay through Hawking radiation, the interpretation as a particle-like
solitonic excitation appears to be reasonable.

We will restrict ourselves in the following to static, spherically symmetric
$\frac{1}{2}$-BPS solutions of $N=2$ supergravity with $n$ 
vector multiplets. Such solutions describe single black holes.\footnote{
There are also static multi-black hole solutions, which we will not
discuss here.}
As a first step, let us ignore higher derivative
terms and work with a prepotential of the form $F(X)$.

In an asymptotically flat space-time, we can define electric and
magnetic charges by integrating the flux of the gauge fields  over 
an asymptotic two-sphere at infinity:
\begin{equation}
\left( \begin{array}{c}
p^I \\ q_I \\
\end{array} \right) 
=
\left( \begin{array}{c}
\oint F^I_{\mu \nu} d^2\Sigma^{\mu \nu}  \\ \oint G_{I|\mu \nu}
d^2 \Sigma^{\mu \nu}   \\
\end{array} \right)  \;.
\end{equation}
By construction, the charges form a symplectic vector $(p^I, q_I)^T$.
The central charge under the asymptotic Poincar\'e Lie superalgebra
is given by the charge associated with the graviphoton:
\begin{equation}
Z = \oint {\cal F}^-_{\mu \nu} d^2 \Sigma^{\mu \nu} = \oint 
(   F^{I-}_{\mu \nu} F_I - G^-_{I|\mu \nu} X^I ) d^2 \Sigma^{\mu \nu} 
= p^I F_I(\infty)  - q_I X^I(\infty) \;.
\end{equation}
This is manifestly invariant under symplectic transformations. 
By common abuse of terminology, the symplectic function
\[
Z = p^I F_I - q_I X^I 
\]
is also called the central charge, despite that it is actually 
a function of the scalars which are in turn functions
on space-time.

A static, spherically symmetric metric can be brought to the
following form:\footnote{The solution can be constructed without
fixing the coordinate system, but we present it in
this way for pedagogical reasons.}
\begin{equation}
ds^2 = - e^{2g(r)} dt^2 + e^{2 f(r)} ( dr^2 + r^2 d \Omega^2) \;,
\label{BHmetric}
\end{equation}
with two arbitrary functions $f(r)$, $g(r)$ of the radial variable $r$.
We also impose that the solution has four Killing spinors. In this
case one can show that $g(r) = - f(r)$. For the gauge fields and
scalars we impose the same symmetry requirements as for the metric.
Therefore each gauge field has only two independent components,
one electric and one magnetic, which are functions of $r$:
\[
F^I_{\underline{tr}} = F^I_E(r)  \;,\;\;\;
F^I_{\underline{\theta \phi}} = F^I_{M}(r) \;. 
\]
Here $\underline{t}, \underline{r}, \underline{\theta}, 
\underline{\phi}$ are tangent space indices.\footnote{If we use
world indices, $F^I_M$ depends on the angular variables. This dependence
is trivial in the sense that it disappears when the
tensor components are evaluated in an orthonormal frame.}

The physical scalar fields $z^i$ can be functions of the radial variable
$r$, $z^i=z^i(r)$. In order to maintain symplectic covariance, we
work in the gauge-equivalent superconformal theory and use the
conformal scalars $X^I$. It turns out to be convenient to rescale
the scalars and to define
\[
Y^I(r) = \overline{Z}(r) X^I(r) \;,
\]
where $Z(r)$ is the `central charge'. Note that
\[
|Z|^2 = \overline{Z} Z = \overline{Z}
(p^I F_I(X) - q_I X^I) = p^I F_I(Y) - q_I Y^I \;,
\]
where we used that $F_I$ is homogenous of degree one.

In the following we
will focus on the near-horizon limit. In the isotropic coordinates
used in (\ref{BHmetric}), the horizon is located at $r=0$. 
The scalar fields show a very particular behaviour in this limit:
irrespective of their `initial values' $z^i(\infty)$ at spatial
infinity, they approach fixed point values $z^i_*=z^i(p^I, q_I)$
at the horizon. This behaviour was discovered by Ferrara, Kallosh and
Strominger and is called the black hole attractor mechanism.
The fixed point values are determined by the attractor equations,
which can be brought to the following, manifestly symplectic form:
\[
\left( \begin{array}{c}
Y^I - \overline{Y}^I \\
F_I - \overline{F}_I \\
\end{array}
\right)_* = i 
\left( \begin{array}{c}
p^I \\
q_I \\
\end{array}
\right) \;.
\]
Here and in the following `$*$' indicates the evaluation of a 
quantity on the horizon.
Depending on the explicit form of the prepotential it may or may
not be possible to solve this set of algebraic equations to 
obtain explicit formulae for the scalars as functions of the charges.
The remaining data of the near-horizon solution are the metric 
and the gauge fields. The near-horizon metric takes the form
\[
ds^2 = - \frac{r^2}{|Z_*|^2} dt^2 + \frac{|Z_*|^2}{r^2} dr^2 
+ |Z_*|^2 d \Omega^2_{(2)} \;,
\]
where $Z_*$ is the horizon value of the central charge,
\[
|Z_*|^2 = \left( p^I F_I (Y) - q_I Y^I \right)_* \;.
\]
The near horizon geometry is therefore $AdS^2 \times S^2$, with
curvature radius $R=|Z_*|^2$. This is a maximally symmetric space,
or more precisely the product of two maximally symmetric spaces.
The gauge fields become covariantly constant in the near horizon 
limits, i.e., they become fluxes whose strength is characterized
by the charges $(p^I,q_I)$. In suitable coordinates\footnote{Essentially,
$r\rightarrow \frac{1}{r}$ combined with a rescaling of $t$. In these
coordinates it becomes manifest that the metric is conformally flat.}
one simply has
\[
F^I_E = q_I \;,\;\;\;
F^I_M = p^I \;.
\]
$AdS^2 \times S^2$, supported by fluxes and constant scalars
is a generalisation of the Bertotti-Robinson solution of 
Einstein-Maxwell theory. 

This generalised Bertotti-Robinson solution is not only the 
near horizon solution of BPS black holes, but also an interesting
solution in its own right. 
It can be shown that it is the most
general static fully supersymmetric 
solution (8 Killing spinors) of $N=2$ supergravity with vector multiplets. 
Note that the attractor equations follow from imposing full supersymmetry,
or, equivalently, the field equations. Thus in a Bertotti-Robinson 
background the scalars cannot take arbitrary values. This is easily
understood by interpreting the solution as a flux compactification
of four-dimensional supergravity on $S^2$. Since $S^2$ is not
Ricci flat, flux must be switched on to solve the field equations.
The dimensionally reduced theory is a gauged supergravity theory with
a non-trivial scalar potential with a non-degenerate $AdS^2$ ground state
and fixed moduli. 

The BPS black hole solution, which has only four Killing spinor,
interpolates between two supersymmetric ground states with 
eight Killing spinors. At infinity it approaches Minkowski space,
and in this limit the values of the scalars are arbitrary, because
the four-dimensional supergravity theory has no scalar potential 
and a moduli space of vacua, parametrised by the scalars. 
At the horizon we approach another supersymmetric ground state,
but here the scalars have to flow to the fixed point values
dictated by the attractor equations. The black hole solution can 
be viewed as a dynamical system for the radial evolution 
of the scalars\footnote{The
other non-trivial data, namely metric and gauge fields can be expressed
in terms of the scalars.} from arbitrary initial values at $r=\infty$ to 
fixed point values at $r=0$.

For completeness we mention that not all flows correspond
to regular black holes. For non-generic choices of the charges 
(typically when switching off sufficiently many charges) the 
scalar fields can run off to the boundary of moduli space. In these
cases $|Z_*|^2$ becomes zero or infinity, so that there is 
no black hole horizon. The original derivation of the attractor equations
was in fact motivated by this observation: if one imposes that 
the scalars do not run off to infinity at the horizon, this implies
that the solution must approach a supersymmetric ground state, which 
in turn implies that the geometry is Bertotti-Robinson and that 
the scalars take fixed point values. In this context the attractor
equations were called stabilisation equations, because they forbid
that the moduli run off. 

There can also be more complicated phenomena
if the flow crosses, at finite $r$, a line of marginal stability, where the BPS
spectrum changes, or if it runs into a boundary point or other special
point in the moduli space. We will concentrate on regular black hole
solutions here, and make some comments on so-called small black holes
later.

The attractor behaviour of the scalars is important for the concistency
of black hole thermodynamics. The laws of black hole mechanics, combined
with the Hawking effect, suggest that a black hole has a macroscopic
(thermodynamical)
entropy proportional to its area $A$:
\[
S_{\rm macro} = \frac{A}{4} \;.
\]
The corresponding microscopic (statistical) entropy is given by the 
state degeneracy\footnote{The macrostate of a black hole is given 
by its mass, angular momentum and conserved charges.} 
\[
S_{\rm micro} = \log \# \{ \mbox{Microstates corresponding to given 
macrostate} \}\;.
\]
Both entropies should be equal, at least asymptotically in the
semi-classical limit (which, for non-rotating black holes, 
is the limit of large mass and charges). Therefore it should not be 
possible to change the area continuously. This is precisely what
the attractor mechanism guarantees.

From the near horizon geometry we can read off that the area 
of the black hole is $A= 4 \pi |Z_*|^2$. The entropy is given
by the following symplectic function of the charges:
\[
S_{\rm macro} = \frac{A}{4} = \pi |Z|^2_* = 
\pi | p^I F_I(X) - q_I X^I |^2_*  =
\pi  \left( p^I F_I (Y) - q_I Y^I \right)_* \;.
\]

\subsubsection*{Further reading and references}

For a general introduction to solitons (and instantons), see
for example the book by Rajamaran \cite{Rajaraman:1982is}.
The idea to  interprete extremal black holes as supersymmetric
solitons is due to Gibbons \cite{Gibbons:1981ja} (see also
\cite{Gibbons:1982fy}). There are many good reviews on 
BPS solitons in string theory, in particular 
\cite{Townsend:1997wg} and \cite{Stelle:1998xg}.
The black hole attractor mechanism was discovered by
Ferrrara, Kallosh and Strominger \cite{Ferrara:1995ih}.
This section is heavily based on a paper written
jointly with Cardoso, de Wit and K\"appeli \cite{LopesCardoso:2000qm},
where we proved that the attractor mechanism is not only 
sufficient, but also necessary for $\frac{1}{2}$-BPS 
solution, and that the Bertotti-Robinson solution is 
the only static solution preserving full supersymmetry. 
We refer to Sen's recent review \cite{Sen:2007qy} for the
discussion of non-supersymmetric attractors.      

We mentioned that not all attractor flows correspond to 
regular black holes solutions. One phenomenon which can
occur is that the solution becomes singular before the horizon
is reached (i.e. the solution becomes singular at finite values
of $r$.) In string theory such singularities can usually be 
explained by a breakdown of the effective field theory. In 
particular, for domain walls and black holes in five-dimensional
string compactifications it has been shown that one 
always reaches an internal boundary of moduli space before the singularity
forms \cite{Kallosh:2000rn,Mayer:2000zk}. 
When the properties of the internal boundary are taken into
account, the solutions becomes regular.\footnote{At internal boundaries
one typically encounters additional massless states, and this  changes
the flow corresponding to the solution.} In four dimensions 
the variety of phenomena appears to be more complex. There are
so-called split attractor flows, which correspond to situations
where the flow crosses a line of marginal stability \cite{Denef:2000nb}. 
This has the
effect that solutions which look like single-centered black hole
solutions when viewed form infinity, turn out to be complicated
composite objects when viewed from nearby. The role of lines 
of marginal stability has been studied recently in great detail 
in \cite{Denef:2007vg}.

\subsection{The black hole variational principle}

Almost immediately after the black hole attractor mechanism 
was discovered, it was observed that the attractor equations 
follow from a variational principle. More recently it has 
been realized that this variational principle plays an
important role in black hole thermodynamics and can be used
to relate macrophysics (black hole solutions of effective
supergravity) to microphysics (string theory, and in particular
BPS partition functions and the topological string) in 
an unexpectedly direct way. 

To explain the variational principle we start by defining the 
`entropy function'
\[
\Sigma(Y,\overline{Y},p,q) := {\cal F}(Y,\overline{Y})
- q_I (Y^I + \overline{Y}^I) + p^I ( F_I + \overline{F}_I) \;,
\]
where $F(Y,\overline{Y})$ is the 
`free energy'
\[
{\cal F}(Y, \overline{Y}) = -i (F_I \overline{Y}^I - Y^I \overline{F}_I)\;.
\]
The terminology will become clear later. If we extremize
the entropy function with respect to the scalars, the equations
characterising critical points of $\Sigma$ are precisely the
attractor equations:
\[
\frac{\partial \Sigma}{\partial Y^I} = 0 = \frac{\partial \Sigma}{\partial \overline{Y}^I}
\Longleftrightarrow  
\left( \begin{array}{c}
Y^I - \overline{Y}^I \\
F_I - \overline{F}_I \\
\end{array}
\right)_* = i 
\left( \begin{array}{c}
p^I \\
q_I \\
\end{array}
\right) \;.
\]
And if we evaluate the entropy function at its critial point, we 
obtain the entropy, up to a conventional factor:
\[
\pi \Sigma_* = S_{\rm macro}(p,q) \;.
\]
The geometrical meaning of the entropy function becomes clear
if we use the special affine coordinates 
\begin{eqnarray}
x^I &=& \mbox{Re} Y^I \;,\nonumber \\
y_I &=& \mbox{Re} F_I(Y) \;,
\end{eqnarray}
instead of the special coordinates $Y^I=x^I + i u^I$. 
The special affine coordinates $(q^a)=(x^I,y_I)^T$ have
the advantage that they form a symplectic vector. In special
affine coordinates, the special K\"ahler metric can be expressed
in terms of a real K\"ahler potential $H(x^I,y_I)$, called the Hesse 
potential. The Hesse potential is related to the prepotential by a
Legendre transform, which replaces $u^I=\mbox{Im}Y^I$ by $y_I=\mbox{Re}(Y^I)$
as an independent field:
\[
H(x^I, y_I) = 2 \left( \mbox{Im} F(x^I +iu^I(x,y)) - y^I u_I(x,y)  \right)\;,
\]
where 
\[
y_I = \frac{\partial \mbox{Im}F}{\partial u^I} \;.
\]
If we express the entropy function in terms of special affine 
coordinates, we find:
\[
\Sigma(x,y,q,p) = 2 H (x,y) - 2 q_I x^I + 2 p^I y_I  \;,
\]
where
\[
2 H (x,y) = {\cal F}(Y,\overline{Y}) = 
-i (F_I \overline{Y}^I - Y^I \overline{F}_I) \;.
\]
Thus, up to a factor, the Hesse potential is the free energy.
The critical points of the entropy function satisfy the black hole
attractor equations, which in special affine coordinates take 
the following form:
\[
\frac{\partial H}{\partial x^I} = q_I \;,\;\;\;
\frac{\partial H}{\partial y_I} = - p^I  \;.
\]
The black hole entropy is obtained by substituting the critical 
values into the entropy function:
\[
S_{\rm macro}(p,q) = 2 \pi \left(
H - x^I \frac{\partial H}{\partial x^I} - y_I \frac{ \partial H }{\partial y_I}
\right)_* \;.
\]
This shows that, up to a factor, the macroscopic black hole entropy is
Legendre transform of the Hesse potential. Note that at the horizon
the scalar fields are determined by the charges, so that the charges
provide coordinates on the scalar manifold. More precisely, the 
charges are not quite coordinates, because they can only take discrete 
values, but by the attractor equations they are proportional to 
continuous quantities which provide coordinates. The attractor
equations can be rewritten in the form
\begin{equation}
\left( \begin{array}{c}
2 u^I \\ 2 v_I \\
\end{array} \right)
= 
\left(
\begin{array}{c}
p^I \\ q_I \\
\end{array}
\right) \;,
\end{equation}
where $u^I = \mbox{Im} Y^I$ and $v_I = \mbox{Im} F_I$. 
It can be shown that $(u^I, v_I)$ is another system 
of special affine coordinates. Thus the attractor equations
specify a point on the scalar manifold in terms of 
the coordinates $(u^I, v_I)$. The extremisation of the entropy
function can be viewed as a Legendre transform from one set
of special affine coordinates to another. 

The special affine coordinates $(x^I, y_I)$ also have a direct
relation to the gauge fields, which even holds away from the horizon.
By the gauge field equations of motion in a 
static (or stationary) background
the scalars $(x^I,y_I)$ are proportional to the 
electrostatic and magnetostatic 
potentials $(\phi^I, \chi_I)$:
\[
\left( \begin{array}{cc}
2 x^I \\ 2 y_I \\
\end{array} \right)
=
\left( \begin{array}{cc}
\phi^I \\ \chi_I \\
\end{array} \right) \;.
\]
Thermodynamically, the electrostatic and magnetostatic potentials
are the chemical potentials associated with the electric and 
magnetic charges in a grand canonical ensemble.

\subsubsection{Further reading and references}

The black hole variational principle described in this 
section was formulated by Behrndt et al in \cite{Behrndt:1996jn}.
The reformulation in terms of real coordinates is relatively recent
\cite{LopesCardoso:2006bg}. The relation of the black hole variational
principle to the work of Ooguri, Strominger and Vafa \cite{Ooguri:2004zv}
will be explained in the following sections. Sen's entropy function
(see \cite{Sen:2007qy} for a review and references),
which can be used to establish the attractor mechanism for general
extremal black holes, irrespective of supersymmetry and details
of the Lagrangian, can be viewed as a generalisation of the 
entropy function discussed here, in the sense that the two
entropy functions differ by terms which vanish in BPS backgrounds
\cite{Cardoso:2006xz}.\footnote{To be precise, Sen's formalism is
based on an entropy function which is based on the `mixed' rather than
the `canonical' ensemble. This is explained in the next section.}

For completeness, we need to mention that there is yet another
`variational' approach to the black hole entropy. The concept of 
a black hole effective potential was already introduced in 
\cite{Ferrara:1997tw}. The idea is to use the symmetries of 
static, spherically symmetric black holes to reduce the dynamics
to the one of particle moving in an effective potential. This 
does not rely on supersymmetry and has become, besides Sen's entropy
function, the second approach for studying the attractor mechanism
for non-BPS black holes \cite{Tripathy:2005qp}. 
The two  approaches are not completely 
unrelated. Dimensional reduction along Killing vectors plays a
role in both of them, since Sen's entropy function is obtained
by dimenensionally reducing the action, evaluated on the horizon,
along the $S^2$ factor of the horizon geometry $AdS^2 \times S^2$,
and then taking a Legendre transform.\footnote{This is a partial
Legendre transform, because the mixed ensemble is used. See the
next section for explanation.}

\subsection{Canonical, microcanonical and mixed ensemble}

For a grand canonical ensemble, the first law of thermodynamics
takes the following form:
\[
\delta E = T \delta S - p \delta V + \mu_i \delta N_i \;.
\]
Here $E$ is the energy, $T$ the temperatue, $S$ the entropy,
$p$ the pressure, $V$ the volume, $\mu_i$ the chemical potential
and $N_i$ the particle number of the $i$-th species of particles.
In relativistic systems the particle number is replaced by
the conserved charge under a gauge symmetry. For a general 
stationary black hole, the first law of black hole mechanics
has the same structure:
\[
\delta M = \frac{\kappa_S}{2\pi} \delta A + \omega \delta J 
+ \phi^I \delta q_I + \chi_I \delta p^I \;.
\]
Here $M$ is the mass, $\kappa_S$ the surface gravity, $A$ the area,
$\omega$ the rotation velocity, $J$ the angular momentum,
and $\phi^I, \chi_I, p^I, q_I$ are the electric and magnetic 
potentials and charges.
The Hawking effect and the generalized second law of thermodynamics
suggest to take the formal analogy between thermodynamics and 
black hole physics seriously. In particular, the Hawking temperature 
of a black hole is $T=\frac{\kappa_S}{2\pi}$, which fixes the
relation between area and entropy to be $S=\frac{A}{4}$.

In thermodynamics we consider other ensembles as well. 
The canonical ensemble is obtained by freezing the particle number
while the microcanonical ensemble is obtained by freezing 
the energy as well. In general, the result for a thermodynamical
quantity will depend on the ensemble one uses. However, all ensembles
give the same result in the thermodynamical limit. 

We will only discuss non-rotating
black holes, $\omega=0$. The analogous ensemble in thermodynamics
does not seem to have a particular name, but, by common abuse
of terminology, we will call this the canonical ensemble. Moreover,
we only consider extremal black holes, with zero temperature. 
For $\kappa_S=0$ the first law does not give directly a relation between
mass and entropy, but we can interprete extremal black holes 
as limits of non-extremal ones. The independent variables
in the canonical ensemble are the potentials $(\phi^I, \chi_I) 
\propto (x^I, y_I) $. 
This ensemble corresponds to a situation where the electric and 
magnetic charge is allowed to fluctuate, while the corresponding
chemical potentials are prescribed. The ensemble obtained by 
fixing the electric and magnetic charges is called the microcanonical
ensemble. Here the independent variables are $(p^I, q_I) \propto 
(u^I, v_I)$.

At the microscopic (`statistical mechanics') level, all three
ensembles are characterised by a corresponding partition function.
The microcanonical partition function is simply given by the
microscopic state degeneracy:
\[
Z_{\rm micro}(p,q) = d(p,q)\;,
\]
where $d(p,q)$ is the number of microstates of a BPS black hole
with charges $p^I,q_I$. The microscopic (statistical) entropy
of the black hole is
\[
S_{\rm micro}(p,q) = \log d(p,q) \;.
\]

The partition function of the canonical
ensemble is obtained by a formal discrete Laplace transform:
\begin{equation}
Z_{\rm can}(\phi,\chi) = \sum_{p,q} d(p,q) e^{\pi (q\phi - p \chi)}
\;.
\label{Zcan}
\end{equation}
This relation can be inverted (formally):
\[
d(p,q) = \oint d\phi d \chi Z_{\rm can}(\phi, \chi)
e^{-\pi(q\phi -p\phi)} \;.
\]

These partition functions are supposed to provide the microscopic
description of BPS black holes. The macroscopic description
is provided by black hole solutions of the effective supergravity
theory, through the attractor equations, the macroscopic entropy and
the entropy function. The variational principle suggests that 
the Hesse potential should be interpreted as the BPS black hole
free energy with respect to the microscopic ensemble. This leads
to the conjecture
\begin{equation}
\label{ModOSVreal}
e^{ 2 \pi H(\phi,\chi) } \approx  Z_{\rm can} =
\sum_{p,q} d(p,q) e^{ \pi [q_I \phi^I - p^I \chi_I ]} \;,
\end{equation}
or, using special coordinates instead of special affine coordinates:
\begin{equation}
\label{ModOSVcomplex}
e^{\pi {\cal F}(Y,\overline{Y})}  \approx  Z_{\rm can} =
\sum_{p,q} d(p,q) e^{\pi [ q_I (Y^I + \overline{Y}^I)  - p^I
(F_I + \overline{F}_I) ]} \nonumber \;.
\end{equation}
Here `$\approx$' means asymptotic equality in the limit of large
charges, which is the semiclassical and thermodynamic limit. 
Ideally, one would hope to find an exact relation between
macroscopic and microscopic quantities, but so far there is only
good evidence for a weaker, asymptotic relation. We can formally
invert (\ref{ModOSVreal}), (\ref{ModOSVcomplex}) to obtain a
prediction for the state degeneracy in terms of the 
macroscopically defined free energy:
\[
d(p,q) \approx \int dx dy e^{\pi \Sigma(x,y)} 
\approx \int dY d \overline{Y} | \det [\mbox{Im} F_{KL} ]| 
e^{\pi \Sigma (Y, \overline{Y})} 
\]
Observe that this formula is manifestly invariant under 
symplectic transformations, because
\[
dx dy := \prod_{I,J} dx^I dy_J = (dx^I \wedge dy_I)^{\rm{top}}
\]
is the natural volume form on the scalar manifold (the top exterior
power of the symplectic form $dx^I\wedge dy_I$), and 
$\Sigma(x,y)$ is a symplectic function.\footnote{Observe that the
relevant scalar manifold is $M$ rather than $\overline{M}$.} 
Note that there is a 
non-trivial Jacobian if we go to special  coordinates. 

By the variational principle, the saddle point value of $\pi\Sigma$
is the macroscopic entropy. Therefore it is obvious that 
microscopic and macroscopic entropy agree to leading order in a
saddle point evaluation of the integral:
\[
e^{S_{\rm micro}(p,q) }
= d(p,q) \approx 
e^{S_{\rm macro}(p,q) (1 + \cdots) } \;.
\]
However, in general the microscopic entropy (defined through state
counting) and the macroscopic entropy (defined geometrically through
the area law) will be different. The reason is that the
macroscopic entropy is the Legendre transform of the canonical
free energy, while the microcanonical and canonical 
partition functions are  related by the Laplace transform (\ref{Zcan}).
The Legendre tranform between canonical free energy and 
macroscopic entropy provides the  leading order approximation 
of this Laplace transform. In other words, the macroscopic
entropy is not computed in the microcanonical ensemble, and we can only
expect it to agree with the microscopic entropy in the thermodynamical
limit.

\subsection*{The mixed ensemble}

We will now consider the so-called 
mixed ensemble, where the independent variables are 
$p^I$ and $\phi^I$. This corresponds to a situation where 
the magnetic charge is fixed while the electric charge fluctuates
and the electrical potential is prescribed. This ensemble has
the disadvantage that the independent variables do not form 
a symplectic vector, which obscures symplectic covariance. 
However, 
the mixed ensemble is natural in the functional integral
framework, and one obtains a direct relation between black hole
thermodynamics and the topological string.

The partition function of the mixed ensemble is obtained from
the microcanonical partition function 
through a Laplace transform with respect to half of the variables:
\[
Z_{\rm mix}(p,\phi) = \sum_q d(p,q) e^{\pi q \phi} \;,
\]
\[
d(p,q) = \oint Z_{\rm mix}(p,\phi) e^{-\pi q \phi} 
\;.
\]

Let us discuss this ensemble from the macroscopic point of view.
In our previous treatment of the variational principle, we
extremized the entropy function with respect to all scalar 
fields/potentials at once. This extremisation process can
be broken up into several steps. The `magnetic' attractor
equations
\[
Y^I - \overline{Y}^I = i p^I
\]
fix the imaginary parts of the $Y^I$:
\[
Y^I = \frac{1}{2} ( \phi^I + i p^I ) \;.
\]
If we substitute this into $\Sigma$ we obtain a reduced entropy function:
\[
\Sigma (\phi, p,q )_{\rm mix} = {\cal F}_{\rm mix}(p, \phi) - q_I \phi^I\;,
\]
where 
\[
{\cal F}_{\rm mix}(p,\phi) = 4 \mbox{Im} F(Y,\overline{Y}) 
\]
is interpreted as the free energy in the mixed ensemble.
$\Sigma_{\rm mix}$ can be interpreted as the entropy function in
the mixed ensemble, because there is a new, reduced variational
principle in the following sense: if we extremize $\Sigma_{\rm mix}$
with respect to the remaining scalars $\phi^I = \frac{1}{2} \mbox{Re} Y^I$,
then we obtain the remaining `electric' attractor equations:
\[
F_I - \overline{F}_I = q_I \;.
\]
If this is substituted back into the mixed entropy function,
we obtain the macroscopic entropy:
\[
S_{\rm macro}(p,q) = \pi \Sigma_{\rm mix, *}\;.
\]
The extremisation of the mixed entropy function defines a 
Legendre transform between the mixed free energy and the
entropy. Note that the mixed free energy is the imaginary part
of the prepotential. 

The mixed free energy should be related to the mixed partition 
function. One conceivable relation is the original `OSV-conjecture'
\begin{equation}
\label{OSV_leading}
e^{\pi {\cal F}_{\rm mix}(p,\phi)} \approx Z_{\rm mix}(p,\phi) \;.
\end{equation}
To leading order in a saddle point approximation the variational
principle guarantees that macroscopic and microscopic entropy
agree. But one disadvantage of the mixed ensemble is that 
the independent variables $p^I, \phi^I$ do not form a symplectic
vector. Therefore symplectic covariance is obscure. 

Let us then compare (\ref{OSV_leading}) to the symplectically
covariant conjecture (\ref{ModOSVcomplex}) involving the canonical ensemble.
Since the variational principle can be broken up into two steps,
we can perform a partial saddle point approximation of (\ref{ModOSVcomplex})
with respect to the imaginary parts of the scalars and obtain
\[
d(p,q) \approx \int d\phi \sqrt{|\det \mbox{Im} F_{IJ}|} 
e^{\pi [ {\cal F}_{\rm mix}(p,\phi) - q\phi]} \;.
\]
This can be formally inverted with the result:
\begin{equation}
{\sqrt{\Delta^-} e^{\pi {\cal F}_{\rm mix}(p,\phi)}}  \approx 
Z_{\rm mix}= \sum_q d(p,q) e^{\pi q_I \phi^I} \;.
\end{equation}
Thus by imposing symplectic covariance we predict the 
presence of a non-trivial `measure factor' in the mixed
ensemble.

\subsubsection*{Further reading and references}

The idea to interprete the (partial) Legendre transform
of the black hole entropy as a free energy (in the mixed
ensemble) is due to Ooguri, Strominger and Vafa \cite{Ooguri:2004zv}
and has triggered an immense number of publications which
elaborate on their observation. Our presentation, which is
based on \cite{Cardoso:2006xz}, uses the variational
principle of \cite{Behrndt:1996jn} to reformulate the 
`OSV-conjecture' in a manifestly symplectically covariant way.

\subsection{$R^2$-corrections}

Non-trivial tests of conjectures about state counting and
partition functions depend on the ability to compute subleading
corrections to the macroscopic entropy. Such corrections are
due to quantum and stringy corrections to the effective action,
which manifest themselves as higher derivative terms.
Within the superconformal calculus one class of such terms
can be handled by giving the prepotential an explicit dependence
on the lowest component of the Weyl multiplet. Incidentially, 
in type-II Calabi-Yau compactifications  
the same class of terms is controlled by the topolocially
twisted world sheet theory. Therefore these higher derivative
couplings can be computed, at least in principle.

It is possible to find the most general stationary $\frac{1}{2}$-BPS
solution for a general prepotential of the form $F(X^I, \hat{A})$,
at least iteratively. Here we restrict ourselves to the near-horizon
limit of static, spherically symmetric single black hole solutions.
It is convenient to introduce rescaled variables 
$Y^I = \overline{Z} X^I$ and $\Upsilon=\overline{Z}^2 \hat{A}$,
and by homogenity we get a rescaled prepotential $F(Y^I, \Upsilon)=
\overline{Z}^2 F(X^I, \hat{A})$. The near horizon solution 
is completely determined by the generalized attractor equations
\begin{equation}
\label{GenAttrEq}
\left( \begin{array}{c}
Y^I - \overline{Y}^I \\
F_I(Y,\Upsilon ) - \overline{F}_I(\overline{Y}, \overline{\Upsilon}) \\
\end{array}
\right)_* = i 
\left( \begin{array}{c}
p^I \\
q_I \\
\end{array}
\right) \;,\;\;\; \Upsilon_* = -64 \;.
\end{equation}
This is symplectically covariant, because $(Y^I, F_I(Y,\Upsilon))^T$
is a symplectic vector. The variable $\Upsilon$ is invariant and
takes a particular numerical value at the horizon. The geometry
is still $AdS^2 \times S^2$, but the radius and therefore the
area is modified by the higher derivative corrections:
\[
A = 4 \pi |p^I F_I(X,\hat{A}) - q_I X^I|^2_* =
4 \pi \left( p^I F_I(Y,\Upsilon) - q_I Y^I \right)_* \;.
\]
But this is not the only modification of the entropy, because
in theories with higher curvature terms the entropy is not determined
by the area law. Wald has shown by a careful derivation of the first 
law of black hole
mechanics for generally covariant Lagrangians (admitting higher curvature
terms) that the definition of the entropy must be modified, if the
first law is still to be valid. 
Entropy, mass, angular momentum and charges 
can be defined as surface charges, which are the Noether charges 
related to the Killing vectors of the space-time. The entropy is given by the
integral of a Noether two-form over the event horizon:
\[
S = \oint Q  \;.
\]
The symmetry associated with this Noether charge is the one generated by the
so-called horizontal Killing vector field. For static black holes this
is the timelike Killing vector field associated with the time-independence
of the background, while for rotating black holes it is a linear combination 
of the timelike and the axial Killing vector field. In practice the Noether 
charge can be expressed
in terms of variational derivatives of the Lagrangian with respect to 
the Riemann tensor: 
\[
S = \oint \frac{\delta {\cal L}}{\delta R_{\mu \nu \rho \sigma}}
\varepsilon_{\mu \nu} \varepsilon_{\rho \sigma} \sqrt{h} d^2 \Omega_2 \;.
\]
Here $\varepsilon_{\mu \nu}$ is the normal bivector, normalized 
to $\varepsilon_{\mu \nu} \varepsilon^{\mu \nu}=-2$ and 
$\sqrt{h}d^2 \Omega$ is the induced volume element of the horizon.
If one evaluates this formula for $N=2$ supergravity with 
prepotential $F(Y^I, \Upsilon)$, the result is
\begin{equation}
\label{EntropyWald}
S_{\rm macro} = 
\pi \left(  {(p^I F_I (Y,\Upsilon) - q_I Y^I )} - 
{256 \;
\mbox{Im} \left( \frac{\partial F}{\partial \Upsilon} \right) }\right)_* \;.
\end{equation}
This is the sum of two symplectic functions. The first term
corresponds to the area law while the second is an explicit
modification. This modification is crucial for the matching
of subleading contributions to the macroscopic and microscopic
entropy in string theory. 

$R^2$-corrections can be incorporated into the variational principle
in a straightforward way. One defines a generalized Hesse potential
as the Legendre transform of (two times the imaginary part of) 
the prepotential $F(Y^I, \Upsilon)$:
\[
H(x,y,\Upsilon, \overline{\Upsilon}) =2  \left(
\mbox{Im} F(Y^I, \Upsilon) - y_I u^I \right)  \;,
\]
where 
\[
y_I = \mbox{Re} F_I(Y^I, \Upsilon) = \frac{\partial \mbox{Im}
F(Y^I, \Upsilon)}{\partial u^I}\;.
\]
The canonical free energy is 
\[
{\cal F}(Y,\overline{Y}) = 2 H(x,y)  
= -i (\overline{Y}^I F_I - Y^I \overline{F}_I)
-2i (\Upsilon F_{\Upsilon} - \overline{\Upsilon} 
\overline{F}_{\overline{\Upsilon}}) \;.
\]
Here and in the following we adopt a notation where we 
usually surpress the dependence on $\Upsilon$, unless 
where we want to emphasize that $R^2$-corrections have been
taken into account. 
The entropy function takes the form 
\[
\Sigma(x,y,p,q) = 2 (H - q x + p y) \;,
\]
where $H$ is now the generalized Hesse potential. It is 
straightforward to show that the extremization of this entropy
function gives the attractor equations (\ref{GenAttrEq}), 
and that its critical values 
gives the entropy (\ref{EntropyWald}): $S_{\rm macro}=
\pi \Sigma_*$. 

\subsubsection*{Further reading and references}

$R^2$-corrections to BPS solutions of $N=2$ supergravity with
vector (and hyper) multiplets were first obtained in 
\cite{LopesCardoso:1998wt} in the near horizon limit. The comparision with 
subleading corrections to state counting in $N=2$ string 
compactifications 
\cite{Maldacena:1997de,Vafa:1997gr}
showed that is crucial to use Wald's modified definition of 
the black hole entropy \cite{Wald}. This approach assumes a Lagrangian
which is covariant under diffeomorphisms, and identifies the
correct definition of the entropy by imposing the validity of the
first law of black hole mechanics. The entropy is found to be
a Noether surface charge, which can be expressed in terms of variational
derivatives of the Lagrangian \cite{Iyer:1994ys}. 
The full derivation is quite intricate,
and while no concise complete review is available, some elements
of it have been reformulated in \cite{LopesCardoso:1999cv} from a more 
conventional gauge theory perspective. Otherwise, see \cite{Mohaupt:2000mj} 
for a more detailed  account on Wald's entropy formula and its merits in 
string theory.  Sen's entropy function formalism \cite{Sen:2007qy} is 
based on Wald's 
definition of black hole entropy.

The general class of stationary $\frac{1}{2}$-BPS solutions in 
$N=2$ supergravity with $R^2$-terms was described in 
\cite{LopesCardoso:2000qm}.
The generalisation of the black hole variational principle 
to include $R^2$ terms was found in \cite{LopesCardoso:2006bg}.

\subsection{Non-holomorphic corrections \label{NonHolo}}

There is a further type of corrections which need to be taken
into account, the so-called non-holomorphic corrections. One way of
deducing that such corrections must be present is to investigate
the transformation properties of the entropy under string dualities,
specifically under S-duality and T-duality. We will discuss an instructive 
example in section \ref{R2forN=4}. The consequence is that 
the entropy and the attractor equations 
can only be duality invariant, if there are additional
contributions to the entropy and to the symplectic vector
$(Y^I, F_I(Y,\Upsilon))$, which cannot be derived from a holomorphic
prepotential $F(Y,\Upsilon)$. This is related to  a generic feature
of string-effective actions and their couplings. One has to distinguish
between two types of effective actions. The Wilsonian action is 
always local and the corresponding Wilsonian couplings are holomorphic
functions of the moduli (in supersymmetric theories).
The other type of effective action is the
generating functional of the scattering amplitudes. If massless modes
are present this is in general non-local, and the associated physical
couplings have a more complicated, non-holomorphic dependence on 
the moduli. Both types of actions differ by threshold corrections
associated with the massless modes, which can be computed by 
field theoretic methods. The supergravity actions which we have
constructed and discussed so far are based on a holomorphic prepotential 
and have to be interpreted as Wilsonian actions. Their couplings 
are holomorphic, and they are different from the physical couplings, which
can be extracted from string scattering amplitudes. The Wilsonian couplings 
are not necessarily invariant under symmetries, such as string dualities, wheras
the physical couplings are.
The same distinction between holomorphic, but non-covariant quantities
and non-holomorphic, but covariant quanities occurs for the topological
string, which is the tool used to compute the couplings. Here the
non-holomorphicity arises from the integration over the world-sheet 
moduli space, and it is encoded in the holomorphic anomaly equations.

In the following we will describe a general formalism for incorporating
non-holomorphic corrections to the attractor equations and the entropy.
This formalism is model-independent (as such), but we should stress
that it is inspired by the example which we are going to discuss in 
section \ref{R2forN=4}. While it has been shown to work in $N=4$
compacatifications, it is not clear a priori whether the non-holomorphic
modifications that are introduced are general enough to cover generic
$N=2$ compactifications.  Moreover, it should be interesting to 
investigate the relation between this formalism and the holomorphic
anomaly equation of the topological string in more detail. 

The basic assumption underlying the formalism is that 
all non-holomorphic modifications are captured by  a single real-valued
function $\Omega(Y,\overline{Y}, \Upsilon, \overline{\Upsilon})$, which
is required to be (graded) homogenous of degree two:
\[
\Omega(\lambda Y^I, \overline{\lambda} \overline{Y}^I, 
\lambda^2 \Upsilon, \overline{\lambda}^2 \overline{\Upsilon})
= |\lambda|^2 \Omega(Y,\overline{Y}, \Upsilon, \overline{\Upsilon})\;.
\]
We then define a generalized Hesse potential by taking 
the Legendre transform of $\mbox{Im}F + \Omega$:
\begin{equation}
\hat{H}(x,y) = 2 \left( \mbox{Im}F(x+iu, \Upsilon) +
\Omega(x,y,\Upsilon, \overline{\Upsilon}) - qx + p \hat{y} \right) \;,
\end{equation}
where 
\begin{equation}
\hat{y}_I = y_I + i (\Omega_I - \Omega_{\overline{I}}) \;.
\end{equation}
Clearly, this modfication is only non-trivial if $\Omega$ is not
a harmonic function, because otherwise it could be absorbed 
by redefining the holomorphic function $F$.

We now take the generalized Hesse potential as our canonical free
energy and define the entropy function
\begin{equation}
\Sigma = 2 ( \hat{H} -qx +p \hat{y} ) \;.
\end{equation}
By variation of the entropy function with respect to $x,\hat{y}$ we
obtain the attractor equations
\begin{equation}
\frac{\partial \hat{H}}{\partial x}
=q \;,\;\;\;\frac{\partial \hat{H}}{\partial \hat{y}} = - p \;,
\end{equation}
and by substituting the critical values back into the entropy function 
we obtain the macroscopic black hole entropy
\begin{equation}
S_{\rm macro} = \pi \Sigma_* = 2 \pi \left(
\hat{H} - x \frac{\partial \hat{H}}{\partial x} - \hat{y}
\frac{\partial \hat{H}}{\partial \hat{y}} \right)_*  \;.
\end{equation}

In practice, one works with special coordinates rather than special
affine coordinates, because explicit expressions for subleading
contributions to the couplings are 
only known in terms of complex coordinates. In special coordinates
the entropy function has the following form:
\[
\Sigma(Y,\overline{Y},p,q) = {\cal F}(Y,\overline{Y}, 
\Upsilon, \overline{\Upsilon}) -q_I (Y^I + \overline{Y}^I)
+ p^I (F_I + \overline{F}_I +2i (\Omega_I - \Omega_{\overline{I}}))
\;,
\]
with canonical free energy
\[
{\cal F}(Y,\overline{Y}, \Upsilon, \overline{\Upsilon}) =
-i (\overline{Y}^I F_I - Y^I \overline{F}_I) -2i (\Upsilon F_{\Upsilon}
- \overline{\Upsilon} \overline{F}_{\overline{\Upsilon}} )
+  4\Omega -2 (Y^I - \overline{Y}^I) (\Omega_I - \Omega_{\overline{I}})
\;.
\]
The attractor equations are
\[
\left( \begin{array}{c}
Y^I - \overline{Y}^I \\
F_I - \overline{F}_I + 2 i (\Omega_I + \Omega_{\overline{I}})\\
\end{array} \right) =
\left( \begin{array}{c}
p^I \\ q_I \\
\end{array} \right) \;,
\]
and the entropy is
\begin{equation}
S_{\rm macro} = \pi \left( |Z|^2 -256 \mbox{Im}
( F_{\Upsilon} + i \Omega_\Upsilon ) \right)_* \;.
\end{equation}
By inspection, the net effect of the non-holomorphic corrections is
to replace $F\rightarrow F+2i\Omega$ in the entropy function and
in the attractor equations, but $F\rightarrow F+i\Omega$ in the
definition of the Hesse potential and in the entropy.\footnote{As
an exercise, the curious reader is encouraged to verify this statement
by himself, starting from the definition of the generalized 
Hesse potential and re-deriving all the formulae step by step.}

As before we can impose half of the attractor equations and go 
from the canonical to the mixed ensemble. The modified mixed 
free energy is found to be  
\[
{\cal F}_{\rm mix} = 4 ( \mbox{Im}F + 
\Omega) \;.
\]

Since the non-holomorphic modifications are enforced by duality 
invariance, they are relevant for the conjectures about the 
relation between macroscopic quantities (free energy and 
macroscopic entropy) and microscopic quantities (partition functions 
and microscopic entropy). 

Our basic conjecture is that the canonical free energy, including
non-holomorphic modifications, is related to the canonical 
partition function by:
\begin{equation}
e^{ 2 \pi H(x,y) } \approx  Z_{\rm can} =
\sum_{p,q} d(p,q) e^{2 \pi [q_I x^I - p^I \hat{y}_I ]}  \;.
\end{equation}
In special coordinates, this reads 
\begin{equation}
e^{\pi {\cal F}(Y,\overline{Y})} \approx Z_{\rm can} =
\sum_{p,q} d(p,q) e^{\pi [ q_I (Y^I + \overline{Y}^I)  - p^I
(\hat{F}_I + \overline{\hat{F}}_I) ]} \;.
\label{OSVmodNonHol}
\end{equation}
We can formally invert these formulae to get a prediction of 
the state degeneracy in terms of macroscopic quantities:
\begin{equation}
\label{StateDeg}
d(p,q) \approx \int dx d\hat{y} e^{\pi \Sigma(x,\hat{y})} 
\approx \int dY d \overline{Y} \Delta^-(Y, \overline{Y}) 
e^{\pi \Sigma (Y, \overline{Y})} \;,
\end{equation}
where we defined
\begin{equation}
\label{DeltaPM}
\Delta^{\pm}(Y,\overline{Y}) = 
| \det \left[
\mbox{Im} F_{KL} + 2 \mbox{Re}(\Omega_{KL} \pm \Omega_{K\overline{L}}) 
\right] | \;.
\end{equation}
In saddle point approximation, we predict the following 
relation between the microscopic and the macroscopic entropy:
\[
e^{S_{\rm micro} (p,q)} =
d(p,q) \approx e^{S_{\rm macro}(p,q)} \sqrt{ \frac{\Delta^-}{\Delta^+}}
\approx e^{S_{\rm macro}(p,q) (1 + \cdots)} \;.
\]
Here we used 
that both the measure factor $\Delta^-$ and the fluctuation determinant
$\Delta^+$ are subleading in the limit of large charges. 

We can also perform a partial saddle point approximation 
\[
d(p,q) \approx \int d \phi {\sqrt{\Delta^-(p,\phi)}}
e^{\pi [ {{\cal F}_{\rm mix}(\phi,p)} - q_I \phi^I ]}
\]
and get a 
conjecture for the relation between the mixed free energy and 
the mixed partition function:
\begin{equation}
{\sqrt{\Delta^-} e^{\pi {\cal F}_{\rm mix}(p,\phi)}}  \approx 
Z_{\rm mix} = \sum_q d(p,q) e^{\pi q_I \phi^I} \;.
\label{Z_mix_modified}
\end{equation}

The conjecture put forward by Ooguri, Strominger and Vafa is:
\begin{equation}
e^{\pi {\cal F}_{\rm mix}^{\rm hol} (p,\phi)} \approx 
Z^{\rm (mix)}_{\rm BH} = \sum_q d(p,q) e^{\pi q_I \phi^I} \;.
\label{OSV_original}
\end{equation}
This differs from (\ref{Z_mix_modified}) in two ways:
(i) the measure factor $\Delta^-$ is absent, and 
(ii) the mixed free energy does not include contributions
from non-holomorphic terms. Since these modifications are
subleading, the black hole variational principle guarantees
that both formulae agree to leading order for large charges.
As indicated by our presentation, we expect that the
measure factor and the non-holomorphic contributions to 
the free energy are present, because they are needed
for symplectic covariance and duality invariance. In fact,
the presence of subleading modifications in (\ref{Z_mix_modified})
has been verified, and we will review this later.

\subsubsection*{The relation to the topological string}

One nice feature of (\ref{OSV_original}) is that provides
a direct link between the mixed black hole partition function 
and the partition function of the topological string.
The coupling functions $F^{(g)}(X)$ in the effective action of 
type-II strings compactified
on a Calabi-Yau threefold are related to particular set of 
`topological' amplitudes. If one performs a topological twist
of the worldsheet conformal field theory, the function
$F^{(g)}(X)$  becomes the free-energies of the twisted theory 
on a world-sheet of genus $g$. The generalized prepotential $F(X,\hat{A})$ 
is therefore proportional to the all-genus free energy, i.e., to 
the logarithm of the all-genus partition function $Z_{\rm top}$ 
of the topological string. As we have seen, the mixed free energy 
${\cal F}^{\rm hol}_{\rm mix}$
is proportional to the imaginary part of $F(X,\hat{A})$. Taking into
account conventional normalization factors, (\ref{OSV_original})
can be rewritten in the following, suggestive form:
\begin{equation}
Z_{\rm BH}^{\rm mix} \approx | Z_{\rm top}|^2 \;.
\end{equation}
However, general experience with holomorphic quantities
in supersymmetric theories 
suggests that such a relation should not be expected to be 
exact, but should be modified by a non-holomorphic 
factor.\footnote{One example is the mass formula
$M^2 = e^{-K} |{\cal M}|^2$ for orbifold models, where 
$K$ is the K\"ahler potential and ${\cal M}$ is the chiral
mass which depends holomorphically on the moduli. In this case
the presence of the non-holomorphic factor $e^{-K}$ can be
inferred from T-duality.
Another example, which has been pointed out to me by
S. Shatashvili, is the path integral measure for strings. While it
shows holomorphic factorisation for critial strings, this is spoiled
by a correction factor, namely the exponential of the Liouville action,
for the generic, non-critical case.}
And indeed, work done over the last years on state counting
and partition functions in $N=2$ compactifications,
has established that the holomorphic factorisation of the
black hole partition function holds to leading order, but
is spoiled by subleading corrections. The underlying microscopic
picture is that the black hole corresponds, modulo string dualities,
to a system of branes and antibranes. To leading order, when
interactions can be neglected, this leads to the holomorphic
factorisation.

Currently, the  detailed microscopic interpretation of the modified
conjecture (\ref{OSVmodNonHol}), (\ref{Z_mix_modified}) and
its relation to the topological string is still an open question.
In the following two lectures, we will discuss how the general
ideas explained in this lecture can be tested in concrete examples.

\subsubsection*{Further reading and references}

This section is mostly based on \cite{LopesCardoso:2006bg}, where
we used the results of \cite{LopesCardoso:1999ur} to formulate
a modified version of the `OSV conjecture' 
\cite{Ooguri:2004zv}. The relation between Wilsonian and 
physical couplings in string effective actions was worked out 
in \cite{Dixon:1990pc} and is reviewed in \cite{Kaplunovsky:1994fg}. 
Concrete examples for the
failure of physical quantities of supersymmetric theories 
to show holomorphic factorisation are provided by mass formulae
(see e.g. \cite{LopesCardoso:1994ik}) and by the path integral
measure of the non-critical string (see e.g. \cite{Gerasimov:2004yx} for a 
discussion). The topological string can be used to derive the
physical couplings of $N=2$ 
compactifications \cite{Bershadsky:1993cx,Antoniadis:1996qg}.
In this case the non-holomorphic corrections are captured 
by the holomorphic anomaly equations. The relation between
these and symplectic covariance in supergravity have been 
discussed in \cite{deWit:1996wq}, while the relevance of non-holomorphic
corrections for black hole entropy was explained in \cite{LopesCardoso:1999ur}.
The role of non-holomorphic corrections for the microscopic aspects
of the OSV conjecture has been addressed in \cite{Verlinde:2004ck}.
The ramifications of the OSV conjecture for `topological M-theory', 
and the role of non-holomorphic corrections in this context 
have been discussed in \cite{Gerasimov:2004yx,Dijkgraaf:2004te}.

References for tests of the OSV conjecture will be given in Lecture IV.


\section{Lecture III Black holes in $N=4$ supergravity}

\subsection{$N=4$ compactifications}

The dynamics of
string compactifications with $N=4$ supersymmetry is considerably
more restricted than the dynamics of $N=2$ compactifications. 
In particular, the classical S- and T-duality symmetries are exact,
and there are fewer higher derivative terms.
Therefore $N=4$ compactifications can be used to test 
conjectures by precision calculations. We consider the simplest example, 
the compactification
of the heterotic string on a six-torus. This is equivalent to 
the compactification of the type-II string on $K3\times T^2$, 
but we will mostly use the heterotic language. 

The massless spectrum consists of the $N=4$ supergravity multiplet
(graviton, four gravitini, six graviphotons, four fermions, one
complex scalar, which is, 
in heterotic $N=4$ compactifications, the dilaton)
together with 22 $N=4$ vector multiplets
(one gauge boson, four gaugini, six scalars). 
Since the gravity multiplet
contains six graviphotons, the resulting gauge group is $U(1)^{28}$ (at 
generic points of the moduli space). The corresponding electric and
magnetic charges each live on a copy of the Narain lattice
$\Gamma=\Gamma_{22;6}$, which is an even self-dual lattice of
signature $(22,6)$:
\[
(p,q) \in \Gamma \oplus \Gamma \;.
\]
Locally, the moduli space is 
\[
{\cal M}  \simeq \frac{SL(2,\mathbbm{R})}{SO(2)} \otimes 
\frac{SO(22,6)}{SO(22) \otimes SO(6)} \;,
\]
where the first factor is parametrised by the (four-dimensional, heterotic)
dilaton $S$,
\[
S = e^{-2\phi} + i a \;.
\]
The vacuum expectation value of $\phi$ is related to the four-dimensional
heterotic string coupling $g_S$ by $e^{\langle \phi \rangle} = g_S$, 
and $a$ is the universal axion (the dual of the universal 
antisymmetric tensor field).
The global moduli space is obtained by modding out by 
the action of the duality group
\[
SL(2,\mathbbm{Z})_S \otimes SO(22,6,\mathbbm{Z})_T \;.
\]
The T-duality group $SO(22,6,\mathbbm{Z})_T$ is a perturbative
symmetry under which the dilaton $S$ is inert, and which acts
linearly on the Narain lattice $\Gamma$. The S-duality group
$SL(2,\mathbbm{Z})_S$ is a non-perturbative symmetry, which
acts on the dilaton by fractional linear transformations,
\begin{equation}
\label{Sduality}
S \rightarrow \frac{a S + ib}{-icS + d} \;,\;\;\;
\left( \begin{array}{cc}
a & b \\ c& d \\
\end{array} \right) \in SL(2,\mathbbm{Z}) \;,
\end{equation}
while it acts linearly on the charge lattice $\Gamma \oplus \Gamma$
by 
\begin{equation}
\label{SdualityOnCharges}
\left( \begin{array}{c}
p  \\ q \\
\end{array} 
\right) \longrightarrow 
\left( \begin{array}{cc}
a\; \mathbbm{I}_{28} & b \;\mathbbm{I}_{28} \\ 
c\; \mathbbm{I}_{28} & d \;\mathbbm{I}_{28}  \\
\end{array} \right) 
\left( \begin{array}{c}
p \\ q \\
\end{array} 
\right)
\;.
\end{equation}
Using the Narain scalar product, we can form three 
quadratic T-duality invariants out of the charges:
$p^2, q^2, p \cdot q$. 
Under S-duality these quantities
form a `vector', i.e., they transform in the ${\bf 3}$-representation,
which is the fundamental representation of $SO(2,1)\simeq SL(2)$.
The scalar product of two such S-duality vectors is an S-duality
singlet. One particularly important example is the S- and T-duality
invariant combination of charges
\[
p^2 q^2 - (p\cdot q)^2 \;,
\]
which discriminates between different types of BPS multiplets.
Recall that the $N=4$ algebra has two complex central charges.
Short ($\frac{1}{2}$-BPS) multiplets satisfy
\[
M = |Z_1 | = |Z_2| \Leftrightarrow p^2 q^2 - (p\cdot q)^2 =0 \;,
\]
whereas intermediate ($\frac{1}{4}$-BPS) multiplets satisfy
\[
M = |Z_1 | > |Z_2| \Leftrightarrow p^2 q^2 - (p\cdot q)^2 \not=0 \;.
\]

\subsection{$N=4$ supergravity in the $N=2$ formalism}

In constructing BPS black hole solutions, we can make use of the
$N=2$ formalism. The $N=4$ gravity multiplet decomposes into
the $N=2$ gravity multiplet, one vector multiplet (which contains
the dilaton), and 2 gravitino multiplets (each consisting of 
a gravitino, two graviphotons, and one fermion). Each $N=4$ 
vector multiplet decomposes into an $N=2$ vector multiplet plus
a hypermultiplet. We will truncate out the gravitino and hypermultiplets
and work with the resulting $N=2$ vector multiplets. This means that
we `loose' four electric and four magnetic charges, corrsponding the
the four gauge fields in the gravitino multiplets. But as we will see
we can use T-duality to obtain the entropy formula for the full
$N=4$ theory.

At the two-derivative level, the effective action is an $N=2$ 
vector multiplet action with prepotential 
\begin{equation}
\label{N=4prepotential}
F(Y) = - \frac{ Y^1 Y^a \eta_{ab} Y^b}{Y^0}  \;,
\end{equation}
where 
\[
Y^a \eta_{ab} Y^b = Y^2 Y^3 - (Y^4)^2 - (Y^5)^2 - \cdots \;.
\]
The dilaton is given by
\[
S=-i\frac{Y^1}{Y^0} \;.
\]
The corresponding scalar manifold is (locally)
\[
\overline{M} \simeq \frac{SL(2,\mathbbm{R})}{SO(2)} \otimes
\frac{SO(22,2)}{SO(22) \otimes SO(2)} \;,
\]
with duality group $SL(2,\mathbbm{Z})_S \otimes SO(22,2,\mathbbm{Z})_T$.

The prepotential (\ref{N=4prepotential})
corresponds to a choice of the symplectic frame 
where the symplectic vector of the scalars is
$(Y^I, F_I(Y))^T$. The magnetic and electric charges corresponding
to this frame are denoted $(p^I, q_I)$. This symplectic frame is 
called the supergravity frame in the following.
Heterotic string
perturbation theory distinguishes a different symplectic frame, 
called the heterotic frame, which is defined by imposing that all
gauge coupling go to zero in the limit of weak string coupling
$g_S \rightarrow 0$ (equivalent to $S\rightarrow \infty$).
In this frame 
$p^1$ is an electric charges while $q_1$ is a magnetic charge.
An alternative way of defining the heterotic frame is to 
impose that the electric charges are those which are carried
by heterotic strings, while magnetic and dyonic charges are carried
by solitons (wrapped five-branes). The heterotic frame has the
particular property that `there is no prepotential' (see also appendix
\ref{AppA}). 
The symplectic transformation relating the heterotic frame and the
supergravity frame is $p^1 \rightarrow q_1$, $q_1 \rightarrow - p^0$.
If one applies this transformation to 
$(Y^I, F_I)^T$, then the transformed $Y^I$ are dependent and do not
form a coordinate system on $M$ (the complex cone over $\overline{M}$), 
while the transformed $F_I$
do not form the components of a gradient.

Since one frame is not adapted to string perturbation theory 
while the other is inconvenient, 
one uses a hybrid formalism, where calculations are performed in the
supergravity frame but interpreted in the heterotic frame. 
The vectors of physical electric and magnetic charges are:
\begin{eqnarray}
q &=& (q_0, p^1, q_a) \in \Gamma \;, \nonumber \\
p &=& (p_0, -q_1, p^a) \in \Gamma \;,
\end{eqnarray}
where $a,b=2, \ldots$. In this parametrisation, the explicit expressions
for the T-duality invariant scalar products are:
\begin{eqnarray}
q^2 &=& 2 ( q_0 p^1 - \frac{1}{4} q_a \eta^{ab} q_b )\nonumber \\
p^2 &=& 2 (-p^0 q_1 - p^a \eta_{ab} p^b ) \nonumber \\
p\cdot q &=& q_0 p^0 - q_1 p^1 + q_2 p^2 + q_3 p^3 + \cdots  \;,
\end{eqnarray}
where 
\begin{eqnarray}
p^a \eta_{ab} p^b &=& p^2 p^3 - (p^4)^2 - (p^5)^2 - \cdots \;,\nonumber \\
q_a \eta^{ab} q_b &=& 4 q_2 q_3 - (q_4)^2 - (q_5)^2 - \cdots \;.
\end{eqnarray}
In the heterotic frame, S-duality acts according to (\ref{SdualityOnCharges}),
and the three quadratic T-duality invariants transform in the vector
representation of $SO(2,1) \simeq SL(2)$, where $SO(2,1)$ is
realised as the invariance group of the indefinite bilinear form
$a_1 a_2 - a_3^2$.  The scalar product of two S-duality vectors is 
a scalar, and the quartic S- and T-duality invariant of the charges is
\[
q^2 p^2 - (p \cdot q)^2 = 
\left( q^2, p^2, p \cdot q \right)
\left( \begin{array}{ccc}
0& \frac{1}{2} & 0 \\
\frac{1}{2} & 0 & 0 \\
0& 0& -1 \\
\end{array} \right) 
\left( \begin{array}{c}
q^2 \\ p^2 \\ p \cdot q \\
\end{array} \right) \;.
\]

For a prepotential of the form (\ref{N=4prepotential})
the attractor equations can be solved
in closed form, and the resulting formula for the entropy is
\begin{equation}
\label{N=4Entropy}
S_{\rm macro} = \pi \sqrt{ p^2 q^2 - (p \cdot q)^2 } \;.
\end{equation}
This formula is manifestly invariant under 
$SL(2,\mathbbm{Z})_S \otimes SO(22,2,\mathbbm{Z})_T$,
and we can reconstruct the eight missing charges by passing to the
corresponding invariant of the full duality group
$SL(2,\mathbbm{Z})_S \otimes SO(22,6,\mathbbm{Z})_T$.
This result agrees with the direct derivation of the 
solution within $N=4$ supergravity.

When using the prepotential (\ref{N=4prepotential}) we neglect
higher derivative corrections to the effective action. 
Therefore the solution is only valid if both the string coupling 
and the curvature are small at the event horizon. This is the case
if the charges are uniformly large in the following sense:
\[
q^2 p^2 \gg (p\cdot q)^2 \gg 1 \;.
\]
Note that if the scalars take values inside the moduli
space\footnote{The moduli space is realised as 
an open domain in $\mathbbm{R}^n$, which is given by a set
of inequalities. In our parametrisation one of these inequalities
is $\mbox{Re}S = e^{-2\phi} >0$,
which implies that the dilaton lives in a half plane (the 
right half plane). Solutions where $\mbox{Re}S <0$ at the horizon 
are therefore unphysical. Similar remarks apply to the other moduli.} 
then $q^2 < 0$ and $p^2 < 0$ in our parametrisation.

From the entropy formula (\ref{N=4Entropy}) it is obvious that 
there are two different
types of BPS black holes in $N=4$ theories.
\begin{itemize}
\item
If $p^2 q^2 - (p \cdot q)^2 \not=0$ the black hole is $\frac{1}{4}$-BPS
and has a finite horizon. These are called large black holes.
\item
If $p^2 q^2 - (p \cdot q)^2 =0$ the black hole is $\frac{1}{2}$-BPS
and has a vanishing horizon. These are called small black holes.
They are null singular, which means that the event horizon 
coincides with the singularity. 
\end{itemize}

\subsubsection*{Further reading and references}

The conventions used in this section are those of \cite{LopesCardoso:1999ur}. 
See
there for more information and references about the relation between $N=4$
and $N=2$ compactifications. The entropy for large black holes in
$N=4$ compactifications was computed in \cite{Cvetic:1995uj,Bergshoeff:1996gg} 
and rederived using 
the $N=2$ formalism in \cite{LopesCardoso:1999ur}.

\subsection{$R^2$-corrections for $N=4$ black holes \label{R2forN=4}}

Let us now incorporate higher derivative corrections. Since no
treatment within $N=4$ supergravity is available, it is
essential that we can fall back onto 
the $N=2$ formalism. One simplifying feature of
$N=4$ compactifications is that all higher coupling functions 
$F^{(g)}(Y)$ with $g>1$ vanish. The only higher derivative coupling
is $F^{(1)}(Y)$, which, moreover, only depends on the dilaton $S$.
The generalized prepotential takes the following form:
\[
F(Y, \Upsilon) = - \frac{ Y^1 Y^a \eta_{ab} Y^b}{Y^0} 
+ F^{(1)}(S) \Upsilon  \;.
\]
In order to find duality covariant attractor equations and
a duality invariant entropy, we must incorporate the non-holomorphic
corrections to the Wilsonian coupling $F^{(1)}(Y)$, which are
encoded in a homogenous, real valued, non-harmonic function 
$\Omega(Y,\overline{Y}, \Upsilon, \overline{\Upsilon})$.

One way to find this function is to compute the physical coupling
of the curvature-squared term in string theory. Since this
coupling depends on the dilaton (but not on the other moduli), it can receive 
non-perturbative corrections (though no perturbative ones). 
At this point one has to invoke the duality between the heterotic string 
on $T^6$ and the type-IIA string on $K3 \times T^2$. Since the
heterotic dilaton corresponds to a geometric type-IIA modulus, 
the exact result can be found by a perturbative calculation in the
IIA theory. This calculation is one-loop, and can be done exactly in 
$\alpha'$, because there is no dependence on the K3-moduli.

Alternatively, one can start with the perturbative heterotic
coupling and infer the necessary modifications of the attractor
equations and of the entropy by imposing S-duality invariance.
It turns out that there is a minimal S-duality invariant completion, which
in principle could differ from the full result by further subleading
S-duality invariant terms. But for the case at hand the minimal
S-duality completion turns out to give complete result. 


At tree level, the coupling function $F^{(1})$ is given by 
\[
F^{(1)}_{\rm tree}(S) = c_1 i S \;,\;\;\;
\mbox{where} \;\;\; c_1 = - \frac{1}{64} \;.
\]
We know a priori that there can be instanton corrections 
${\cal O}(e^{-S})$. The function $F^{(1)}(S)$ determines the
`$R^2$-couplings'
\begin{equation}
{\cal L}_{R^2} \simeq \frac{1}{g^2} C_{\mu \nu \rho \sigma}
C^{\mu \nu \rho \sigma} + \Theta C_{\mu \nu \rho \sigma}
\tilde{C}^{\mu \nu \rho \sigma} \;,
\label{R2Lagrangian}
\end{equation}
where $C_{\mu \nu \rho \sigma}$ is the Weyl tensor, 
through $g^{-2} \simeq \mbox{Im} F^{(1)}$ and 
$\theta \simeq \mbox{Re} F^{(1)}$. Therefore $\mbox{Im}F^{(1)}$
must be an S-duality invariant function, whereas $\mbox{Re}F^{(1)}$ must
only be invariant up to discrete shifts. Accoring to  (\ref{Sduality}),
the linear tree-level piece is not invariant. Restrictions on 
the functional dependence of $F^{(1)}$ on $S$ result from the 
requirement that the S-duality transformation (\ref{Sduality}) of the dilaton
induces the symplectic transformation (\ref{SdualityOnCharges}) of 
the symplectic vector
$(Y^I, F_I)^T$. This implies that 
\[
f(S) := -i \frac{\partial F^{(1)}}{\partial S}
\]
must transform with weight 2:
\[
f \left( \frac{aS -ib}{icS + d} \right) =
(icS + d)^2 f(S) \;.
\]
A classical result in the theory of modular forms\footnote{
We refer the reader to appendix \ref{AppB} for a brief
review of modular forms and references.} implies 
that $f(S)$, (and, hence, $F^{(1)}$) cannot be holomorphic. The 
holomorphic object which comes closest to transforming with weight
2 is the holomorphic second Eisenstein series
\[
G_2 (S) = - 4 \pi \partial_S \eta(S) \;,
\]
where $\eta(S)$ is the Dedekind $\eta$-function.\footnote{Here
$G_2(S)$ is short for
$G_2(iS)$, etc.} 
To obtain a function which transforms with weight 2 one needs to
add a non-holomorphic term and obtains the non-holomorphic second 
Eisenstein series:
\[
G_2 (S,\overline{S}) = G_2(S) - \frac{2\pi}{S + \overline{S}} \;.
\]
This is the only candidate for $f(S)$. We will write
$f(S,\overline{S})$ in the following, to emphasize that this function is
non-holomorphic. We need to check that we get 
the correct asymptotics in the weak coupling limit 
$S\rightarrow \infty$. Since $F^{(1)} \rightarrow 
c_1 i S$, we know that $f(S, \overline{S})$ must go to a constant.
This is indeed true for the second Eisenstein series (the non-holomorphic
term is subleading):
\[
G_2(S,\overline{S}) \rightarrow \frac{\pi^2}{3}\;,
\]
and therefore the minimal choice for $f(S,\overline{S})$ is
\[
f(S,\overline{S}) = c_1 \frac{3}{\pi^2} G_2(S,\overline{S}) \;.
\]
This can be integrated, and we obtain the non-holomorphic function 
\begin{equation}
F^{(1)}(S,\overline{S}) = - i c_1 \frac{6}{\pi} 
\left( \log \eta^2 (S) + \log (S+\overline{S}) \right) \;.
\end{equation}
This function generates a symplectic vector 
$(Y^I, F_I(Y,\overline{Y}))^T$ with the 
correct behaviour under S-duality.
Moreover, the function $p^I F_I(Y,\overline{Y})
- q_I Y^I$, which is proportional to the area, is S-duality 
invariant.
However $F^{(1)}(S,\overline{S})$ is not S-duality invariant, but
transforms as follows:
\[
F^{(1)}(S,\overline{S}) \rightarrow 
F^{(1)}(S,\overline{S}) + i c_1 \frac{6}{\pi} 
\log ( -ic \overline{S} + d )  \;.
\]
This was to be expected, because derivatives (and, hence, integrals)
of modular forms are not modular forms but transform with additional
terms. The function $F^{(1)}(S,\overline{S})$
was constructed by requiring that its derivative is a 
modular form of weight 2. Therefore it does not quite transform 
as a modular form of weith zero (modular function). 
In order to get an S-duality invariant function,
we need to add a further non-holomorphic piece:
\[
F^{(1)}_{\rm phys}(S,\overline{S}) = 
F^{(1)}(S,\overline{S})  + i c_1 \frac{3}{\pi} \log (S+\overline{S}) =
F^{(1)}(S)_{\rm hol} + i c_1 \frac{6}{\pi} \log ( S + \overline{S}) \;,
\]
where 
\[
F^{(1)}_{\rm hol}(S) = - i c_1 \frac{6}{\pi} \log \eta^2 (S) \;.
\]
The invariant function $F^{(1)}_{\rm phys}$ is the minimal
S-duality completion of the $R^2$-coupling (\ref{R2Lagrangian}).
An explicit caculation of this coupling in string theory 
shows that this is in fact the full $R^2$-coupling.

Since the entropy must be S-duality invariant, it is also clear
that the correct way of generalizing the holomorphic function $F^{(1)}(S)$
in the entropy formula is:\footnote{Remember $\Upsilon_*= -64$.}
\[
S_{\rm macro} = \pi \left[
(p^I F_I(Y,\overline{Y}) - q_I Y^I) 
+ 4 \mbox{Im} \left( \Upsilon F^{(1)}_{\rm phys}(S,
\overline{S} ) \right) \right]_* \;.
\]

Note that the non-holomorphic modifications are purely imaginary.
Therefore they only modify the $R^2$-coupling $g^{-2} \simeq
\mbox{Im} F^{(1)}$ and 
reside in a real-valued, non-harmonic function $\Omega$.
In the following we find it convenient to absorbe the holomorphic
function $\Upsilon F^{(1)}(S)$ into $\Omega$:
\begin{eqnarray}
\Omega(S,\overline{S},\Upsilon,\overline{\Upsilon})
&=& \mbox{Im} \left( \Upsilon F^{(1)}(S,\overline{S}) 
+ \Upsilon i c_1 \frac{3}{\pi} \log (S + \overline{S}) \right)
\nonumber \\
&=&
\mbox{Im} \left( \Upsilon F^{(1)}(S) - \Upsilon i c_1 \frac{3}{\pi}
\log(S + \overline{S}) \right) \;.
\end{eqnarray}
This function encodes all higher derivative corrections to the 
tree-level prepotential. 

We already mentioned that the holomorphic $R^2$-corrections
correspond to instantons. To make this explicit we expand
$F^{(1)}_{\rm hol}(S)$ for large $S$:
\[
F^{(1)}_{\rm hol} (S) \simeq 
\log \eta^{24}(S) = - 2 \pi S - 
{24 e^{-2\pi S} + {\cal O}
(e^{-4 \pi S})}\;.
\]
This shows that the $R^2$-coupling has a classical piece proportional to
$S$,  followed by an infinite series of instanton corrections, which
correspond to wrapped five-branes.

\subsubsection*{Further reading and references}

This section is based on \cite{LopesCardoso:1999ur}. 
The treatment of the non-holomorphic
corrections illustrates the general formalism introduced in 
\cite{LopesCardoso:2006bg}.
In fact, the formalism is modelled on this example, and it is not
excluded that generic $N=2$ compactifications need more general
modifications. The $R^2$-term in the effective action for
$N=4$ compactifications was computed in \cite{Harvey:1996ir}.

\subsection{The reduced variational principle for $N=4$ theories }

It is possible and in fact 
instructive to analyse the attractor equations and entropy
without using the explicit form of $\Omega$. Using that  $\Omega$
depends on the dilaton $S$, but not on the other moduli
$T^a \simeq \frac{Y^a}{Y^0}$, 
one can solve all but
two of the attractor equations explicitly.
The remaining two 
`dilaton attractor equations' are the only ones which involve
$\Omega$, and they 
determine the dilaton as a function
of the charges.
Substituting the solved attractor equations into the 
entropy function, we obtain the following, reduced entropy function:
\begin{equation}
\label{DilatonEntropyFunction}
\Sigma(S, \overline{S},p,q) =
- \frac{q^2 - i p\cdot q (S - \overline{S}) + p^2 |S|^2}{S+\overline{S}}
+ 4 \Omega(S, \overline{S},\Upsilon, \overline{\Upsilon}) \;.
\end{equation}
Extremisation of this function yields the remaining
dilatonic attractor equations
\[
\partial_S \Sigma = 0 = \partial_{\overline{S}} \Sigma 
\;\;\;\Leftrightarrow \;\;\;\mbox{Dilaton attractor equations,}
\]
and its critical value gives the entropy:
\begin{eqnarray}
S_{\rm macro}(p,q) &=& \pi \Sigma_*(p,q) \nonumber \\
&=&  \left( 
- \frac{q^2 - i p\cdot q (S - \overline{S}) + p^2 |S|^2}{S+\overline{S}}
+ 4 \Omega(S, \overline{S},\Upsilon, \overline{\Upsilon})
\right)_{| \partial_S \Sigma =0} \;.
\end{eqnarray}
The entropy function is manifestly 
S-duality and T-duality invariant, provided that $\Omega$ is
an S-duality invariant function.\footnote{
$\frac{1}{S+\overline{S}} ( 1, |S|^2, -i (S-\overline{S}))$
transforms as an $SO(2,1)$ vector under S-duality, and therefore the 
contraction with the vector $(q^2, p^2, p \cdot q)$ gives an 
invariant.}

\subsubsection*{Further reading and references}

The observation that all but two of the $N=4$ attractor
equations can be solved, even  in presence of $R^2$-terms,
was already made in \cite{LopesCardoso:1999ur} and exploited in 
\cite{LopesCardoso:2004xf} and \cite{LopesCardoso:2006bg}.

\subsection{Small $N=4$ black holes}

Let us now have a second look at small black holes. For convenience
we take them to be electric black holes, $p=0$. By this explicit
choice, S-duality is no longer manifest, but T-duality remains manifest.
As we saw above, as long as $\Omega=0$ the area of a $\frac{1}{2}$-BPS
black hole
vanishes, $A=0$, and therefore the Bekenstein-Hawking entropy
is zero, too. In fact, the moduli also show singular behaviour, and, 
in particular, the dilaton runs of to infinity at the horizon $S_*=\infty$. 
Thus small black holes live on the boundary of moduli space.

The lowest order approximation to the $R^2$-coupling is to
take its classical part,
\[
F^{(1)} \simeq
\log \eta^{24}(S) = -2 \pi S + {\cal O}(e^{-2\pi S})\;,
\]
and to neglect all instanton and non-holomorphic corrections.
In this approximation one can solve the dilatonic attractor
equations explicitly. This results in the following, non-vanishing
and T-duality invariant area:
\[
A = 8 \pi \sqrt{ \frac{1}{2} |q^2|} \not=0\;.
\]
Thus the $R^2$-corrections smooth out the null-singularity and 
create a finite horizon. We need to impose that $|q^2| \gg 1$\footnote{In 
our parametrisation 
$q^2 < 0$, if the horizon values of the scalars are inside the 
moduli space.} in order that the dilaton $S$ is large, which we
need to impose because 
we neglect subleading corrections to the $R^2$-coupling. Note that
in contrast to the two-derivative approximation the dilaton is 
now finite at the horizon. Thus not only the metric but also 
the moduli are smoothed by the higher derivative corrections.
The horizon area is small in string units, even though it
is large in Planck units. This motivates the terminology 
`small black holes.'

The resulting Bekenstein-Hawking entropy is: 
\[
S_{\rm Bekenstein-Hawking} = \frac{A}{4} = 2 \pi 
\sqrt{ \frac{1}{2} |q^2|} \;. 
\]
However, since the area law does not apply to theories with
higher curvature terms, the correct way to compute the 
macroscopic black hole entropy is (\ref{EntropyWald}). 
Evaluating this for the
case at hand gives
\[
S_{\rm macro} = \frac{A}{4} + {\mbox{correction}}
= \frac{A}{4} {+ \frac{A}{4}}
 = \frac{A}{2} = 4 \pi \sqrt{ \frac{1}{2} |q^2| } \;.
\]
In this particular case the 
correction is as large as the area term itself. Later we will have
the opportunity to confront both formulae with string microstate
counting.

In the limit of large $S$ the next subleading correction
comes from the non-holomorphic corrections $\propto \log (S + \overline{S})$.
We can still find an explicit formula for the entropy:
\[
S_{\rm macro} = 4 \pi \sqrt{ \frac{1}{2} |q^2| } - 6 \log |q^2| \;,
\]
which we will compare to microstate counting later.

If we include further corrections, ultimately the full series of 
instanton corrections encoded in $\log \eta^{24}(S)$, we cannot
find an explicit formula for the entropy anymore. However, we know 
that the exact macrocsopic entropy is given as the solution
of the extremisation problem for the dilatonic entropy function 
(\ref{DilatonEntropyFunction}). This can be used for a 
comparison with state counting. 

\subsubsection*{Further reading and references}

The observation that $R^2$-corrections smooth or `cloak'
the null singularity of small black holes was made 
in \cite{Dabholkar:2004dq}.
This result follows immediately from \cite{LopesCardoso:1999ur}.


\section{Lecture IV: $N=4$ state counting and black hole partition functions}

The BPS spectrum of the heterotic string on $T^6$ consists of the
excited modes of the heterotic string itself, and solitons. Heterotic string
states are labeled by 28 quantum numbers:
six winding numbers, six discrete momenta around $T^6$ and
16 charges of the unbroken $U(1)^{16} \subset E_8 \otimes E_8$
gauge group. They combine into 
22 left- and 6 right-moving momenta, which 
take values in the Narain lattice:
\[
(p_L;p_R) \in \Gamma  \;.
\]
Modular invariance of the world sheet conformal field theory implies
that the lattice $\Gamma$ must be even and selfdual with repect to the bilinear
form $p_L^2 - p_R^2$, which has signature $(+)^{22}(-)^6$.
From the four-dimensional point of view,
the 28 left- and rightmoving momenta are the 28 electric charges
with respect to the generic gauge group $U(1)^{16+6+6}$: 
$q=(p_L;p_R) \in \Gamma$.

Similarly, the
winding states of heterotic five-branes carry magnetic charges
$p\in \Gamma^*=\Gamma$. If purely electric or purely magnetic 
states satisfy a BPS bound, they must be $\frac{1}{2}$-BPS states,
because $p^2 q^2 - p\cdot q =0$ if either $p=0$ or $q=0$.
However, there are
also dyonic solitonic states with $q^2 p^2 - p\cdot q
\not=0$, which are $\frac{1}{4}$-BPS. By string--black hole 
complementarity, the BPS states with charges 
$(p,q) \in \Gamma \oplus \Gamma$ should be
the microstates of $N=4$ black holes with the same charges. 
We will now discuss how these states are counted and compare
our results to the macroscopic black hole entropy and free
energy.

\subsection{Counting $\frac{1}{2}$-BPS states}

Without loss of generality, we take the $\frac{1}{2}$-BPS
states to be electric, $p=0$. Such states correspond to
excitations of the heterotic string, and are called
Dabholkar-Harvey states. Recall that the world-sheet theory
of the heterotic string has two different sectors. The left-moving sector
consists of 24 world sheet bosons (using the light cone
gauge), namely the left-moving projections of the eight
coordinates transverse to the world sheet, and 16 bosons 
with values in the maximal torus of $E_8 \otimes E_8$. 
The right-moving sector consists of the right-moving projections
of the eight transvserse coordinates, together with 
eight right-moving fermions. This sector is supersymmetric
in the two-dimensional, world-sheet sense. World-sheet supersymmetry
combined with a condition on the spectrum of charges implies
the existence of an extended chiral algebra on the world-sheet,
which is equivalent to $N=4$ supersymmetry in the ten-dimensional,
space-time sense.
The generators of the space-time supersymmetry algebra are
build exclusively out of right-moving degrees of freedom.
To obtain BPS states one needs to put the right-moving sector
into its ground state, but still has the freedom to 
excite the left-moving sector. A basis of such states is 
\begin{equation}
\alpha^{i_1}_{-m_1} \alpha^{i_2}_{-m_2} \cdots | q \rangle 
\otimes {\bf 16}  \;,
\end{equation}
where $\alpha^{i_k}_{-m_l}$ are creation operators for the
oscillation modes of the string.
The indices $i_k=1,\ldots, 24$ label the directions transverse
to the world-sheet of the string, while 
$m_k=1,2,3,\ldots$ label the oscillation modes. $q = (p_L;p_R)
= \Gamma$ are the electric charges, which 
correspond to the winding modes, momentum modes and 
$U(1)^{16}$ charges. ${\bf 16}$ denotes the ground state
of the right-moving sector, which carries the degrees of freedom 
of an $N=4$ vector multiplet (with 16 on-shell degrees of freedom). 
States of this form are invariant
under as many supercharges as the right-moving groundstate, and 
therefore they belong to $\frac{1}{2}$-BPS multiplets. To be
physical, the state must satisfy the 
level matching condition,
\begin{equation}
N - 1 + \frac{1}{2} p_L^2 = \tilde{N} + \frac{1}{2} p^2_R \;,
\end{equation}
where $N,\tilde{N}$ are the total left-moving and right-moving
excitation numbers. BPS states have
$\tilde{N}=0$, and therefore the excitation level is fixed 
by the charges:
\begin{equation}
N - 1 + \frac{1}{2} p_L^2 = \tilde{N} + \frac{1}{2} p^2_R \;
\Rightarrow N = \frac{1}{2} p^2_R - \frac{1}{2} p^2_L -1 
= -q^2 -1 = |q^2| - 1  \;.
\end{equation}
This is equivalent to the statement that the mass saturates
the BPS bound. Note that $q^2 <0$ for physical BPS states.
For large charges we can use $N\approx |q^2|$.

The problem of counting $\frac{1}{2}$-BPS states amounts to 
counting in how many ways a given total excitation number 
$N \approx |q^2|$ can be distributed among the creation 
operators $\alpha^i_{-m}$. If we ignore the additional
space-time index $i=1,\ldots, 24$, 
this becomes the classical problem of
counting the partitions of an integer $N$, which was
studied by Hardy and Ramanujan. The space-time index $i$
adds an additional 24-fold degeneracy, and we  might say
that we have to count partitions of $N$ into integers with 
24 different `colours'. Incidentially exactly the same 
problems arises (up to the overall factor 16 from the degeneracy 
of the right-moving ground state)
when counting the physical states of the open 
bosonic string. From the world-sheet perspective, both problems
amount to finding the partition function of 24 free bosons,
which is a standard  problem in quantum statistics and conformal
field theory. 

The reader is encouraged to solve Problem 3, which 
is to derive the following formula for the state degeneracy:
\begin{equation}
\label{1/2BPSstates}
d(q) = d (q^2) = 16 \oint  d \tau
\frac{ e^{i \pi \tau q^2}}{\eta^{24}(\tau) }  \;,
\end{equation}
where $\tau = \tau_1 + i \tau_2 \in {\cal H}$,
where ${\cal H}=\{ \tau \in \mathbbm{C} | \tau_2 > 0 \}$ is
the upper half plane and where 
$\eta(\tau)$ is the Dedekind $\eta$-function. The integration
contour runs through a strip of width one in the upper half plane, i.e.,
it connects two points $\tau^{(1)}$ and $\tau^{(2)} = 
\tau^{(1)} + 1$. Since the integrand
is periodic under $\tau \rightarrow \tau + 1$
(which is a general property of modular forms), this integration 
contour is effectively closed. (It becomes a closed contour when 
going to the new variable $e^{2 \pi i \tau}$, which takes
values in the interior of the unit disc.) 

In its present form this expression is not very useful, because we
want to know $d(q)$ explicitly, at least asymptotically for large
values of $|q^2|$. This type of problem was studied already by Hardy
and Ramanujan, and a method for solving it exactly was found
by Rademacher. For our specific problem with 24 `colours'
the Rademacher expansion takes the following form:
\begin{equation}
\label{Rademacher}
d(q^2) = 16 \sum_{c=1}^\infty c^{-14} \mbox{Kl}\left(\frac{1}{2} |q^2|, -1 ;
c \right)
\hat{I}_{13}\left( \frac{4 \pi}{c} \sqrt{\frac{1}{2} |q^2|} \right) \;, 
\end{equation}
where $\hat{I}_{13}=$ is the modified Bessel function of index 13, 
and $Kl(l,m;c)$ are the so-called Kloosterman sums.

Modified Bessel functions have the following integral representation:
\[
\hat{I}_\nu (z) = -i (2\pi)^\nu \int_{\epsilon -i\infty}^{\epsilon + 
i \infty} \frac{dt}{t^{\nu+1}} e^{t + \frac{z^2}{4t}} \;,
\]
and their asymptotics for $\mbox{Re}(z) \rightarrow \infty$ is:
\[
\hat{I}_\nu(z) \approx \frac{e^z}{\sqrt{2}} \left( 
\frac{z}{4\pi} \right)^{-\nu - \frac{1}{2}} \left( 
1 - \frac{2 \nu^2 -1 }{8 z} + {\cal O}(z^{-2}) \right) \;.
\]
We will not need the values of the Kloostermann sums, except 
that $Kl(l,m;1)=1$.

In the limit of large $|q^2|$ the term with $c=1$ is leading, while
the terms with $c>1$ are exponentially surpressed:
\[
d(q^2) = 16 \; \hat{I}_{13} ( 4 \pi \sqrt{ \frac{1}{2} |q^2|}) 
+ {\cal O}(e^{-|q^2|})\;.
\]
Using the asymptotics of Bessel functions, this
can be expanded in inverse powers of $|q^2|$:
\[
S_{\rm micro}(q^2) = \log d(q^2) \approx 
{4 \pi \sqrt{\frac{1}{2} |q^2|} - \frac{27}{4} \log |q^2| +} 
\frac{15}{2} \log(2) 
- \frac{675}{32 \pi |q^2|} + \cdots
\]
The first two terms correspond to a saddle point evaluation of the
integral representation (\ref{1/2BPSstates}):
The first term is the value of integrand at its
saddle point, while the second term is the `fluctuation determinant'.
The derivation of the first two terms using a saddle point
approximation of (\ref{1/2BPSstates}) is left to the reader 
as Problem 4. 
A derivation of the full Rademacher
expansion (\ref{Rademacher})
can be found in the literature.

\subsubsection*{Further reading and references}

An excellent and accessible account on the Rademacher 
expansion can be found in \cite{Dijkgraaf:2000fq}. See in particular
the appendix of this paper for two versions of the proof
of the Rademacher expansion. We have also borrowed
some formulae from \cite{Dabholkar:2005by,Dabholkar:2005dt}, who
have studied the state counting for $\frac{1}{2}$-BPS states in 
great detail, including various $N=4$ and $N=2$ orbifolds of the
toroidal $N=4$ compactification considered in this lecture.

\subsection{State counting for $\frac{1}{4}$-BPS states}

For the problem of counting $\frac{1}{4}$-BPS states the dual
type-II picture of the $N=4$ compactification is useful. 
Here all the heterotic $\frac{1}{2}$- and $\frac{1}{4}$-BPS
states arise as winding states of the NS-five-brane. It is believed
that the dynamics of an NS-five-brane is described
by a string field theory whose target space is the world volume
of the five-brane. If one assumes that the counting of BPS states
is not modified by interactions, the problem of state counting 
reduces to counting states in a multi-string Fock space. For 
$\frac{1}{2}$-BPS states the resulting 
counting problem is found to be
equivalent to the one described in the last section, as required
by consistency. For $\frac{1}{4}$-BPS states the counting problem
is more complicated, but one can derive the following integral
representation:
\begin{equation}
\label{1/4BPSstates}
d(p,q) = \oint d\rho d \sigma d v
\frac{ e ^{i \pi  [ \rho p^2 + \sigma q^2 + (2v-1) pq)]}}
{ \Phi_{10}(\rho, \sigma, v) } \;.
\end{equation}
This formula requires some explanation. Essentially 
it is a generalisation of (\ref{1/2BPSstates}), where the 
single complex variable $\tau$ has been replaced by three
complex variables $\rho, \sigma, \tau$, which live in the
so called rank-2 Siegel upper half space ${\cal S}_2$. 
In general the rank-$n$ Siegel upper half space is the
space of all symmetric $(n\times n)$-matrices
with positive definit imaginary part. This is a symmetric 
space,
\[
{\cal S}_n \simeq \frac{Sp(2n)}{U(n)} \;,
\]
which can be viewed as a generalisation of the upper half plane
\[
{\cal H} = \frac{Sp(2)}{U(1)} = {\cal S}_1 \;.
\] 
The group $Sp(2n,\mathbbm{Z})$ 
acts by fractional linear transformations on the $(n\times n)$
matrices $\Omega \in {\cal S}_n$,
\[
\Omega \rightarrow (A \Omega + B ) (C \Omega +D)^{-1} \;,\;\;\;
\mbox{where}\;\;\;
\left( \begin{array}{cc}
A & B \\ C & D \\
\end{array} \right) \in Sp(2n) \;.
\]
The discrete subgroup $Sp(2n,\mathbbm{Z})$ is a generalisation 
of the modular group $Sp(2,\mathbbm{Z}) \simeq SL(2,\mathbbm{Z})$,
and there is a corresponding theory of Siegel modular forms.
A Siegel modular form is said to have weight $2k$, if it 
transforms as
\[
\Phi(\Omega) \rightarrow \Phi \left( 
(A\Omega +B)(C \Omega +D)^{-1} \right) =
(\det(C\Omega +D))^k  \Phi(\Omega) \;.
\]
In the rank-2 case, we can parametrise the matrix $\Omega$ as
\[
\Omega = \left( \begin{array}{cc}
\rho & v \\ v & \sigma \\
\end{array} 
\right) \;,
\]
and positive definitness 
of the imaginary part implies that 
\[
\rho_2 > 0\;,\;\;\;
\sigma_2 > 0 \;,\;\;\;
\rho_2 \sigma_2 - v^2 > 0 \;,
\]
where $\rho = \rho_1 + i \rho_2$, etc.

In the theory of rank-2 Siegel modular forms, the analogon 
of the weight-12 cusp form $\eta^{24}(\tau)$ is the 
weight-10 Siegel 
cusp form $\Phi_{10}(\rho, \sigma, v)$,
which enters into the 
state counting formula (\ref{1/4BPSstates}).
Like modular forms, Siegel modular forms are periodic in the real 
parts of the variables $\rho, \sigma, v$. The integration contour in the
Siegel half space is along a path of the form 
$\rho \rightarrow \rho + 1$, $\sigma \rightarrow \sigma +1$,
$v \rightarrow v+1$,  
which is effectively a closed contour since the integrand is 
periodical.\footnote{In  the numerator one has to use that 
the Narain lattice is even selfdual.} The state counting 
formula (\ref{1/4BPSstates}) is manifestly T-duality
invariant. It is also formally S-duality invariant, in the sense
that S-duality transformations can be compensated by 
$Sp(4,\mathbbm{Z})$ transformations of the integration variables.

As in the $\frac{1}{2}$-BPS case one would like to evaluate 
(\ref{1/4BPSstates})
asymptotically, in the limit of large charges 
$q^2 p^2 - (p\cdot q)^2 \gg 1$.
One important difference between $\Phi_{10}$ and $\eta^{24}$ is
that the Siegel cusp form has zeros in the interior of
the Siegel half space ${\cal S}_2$, namely 
at $v=0$ and its images under $Sp(4,\mathbbm{Z})$. The 
$v$-integral therefore evaluates the residues of the integrand. At $v=0$,
the asymptotics of $\Phi_{10}$ is
\[
\Phi \simeq_{v=0} v^2 \eta^{24}(\rho) \eta^{24}(\sigma) \;.
\]
The asymptotics at the other zeros can be found by applying 
$Sp(4,\mathbbm{Z})$ transformations.

If one sets the magnetic charges to zero, the residue at $v=0$
is the only one which contributes to (\ref{1/4BPSstates}). 
This can be used to  derive the $\frac{1}{2}$-BPS formula 
(\ref{1/2BPSstates}) as a special case 
of (\ref{1/4BPSstates}).\footnote{Incidentially, the problem
is equivalent to the factorisation of a genus-2 string 
partition function into two genus-1 string partition functions.}

For $\frac{1}{4}$-BPS states
it can be shown that for large charges 
the dominant contribution comes from the
residue at 
\[
D = v + \rho \sigma - v^2 =0 \;,
\]
while all other residues are exponentially surpressed. 
Neglecting the subleading residues, one
can perform the  $v$-integral. The
remaining integral has the following structure:
\begin{equation}
d(p,q) = \oint  d\rho d\sigma 
e^{i \pi (X_0+X_1)(\rho, \sigma)} \Delta(\rho, \sigma) \;.
\end{equation}
The parametrisation has been chosen such that $X_1$ and $\Delta$
are subleading for large charges.

This integral can be evaluated in a saddle point approximation, 
analogous to (\ref{1/2BPSstates}).
The leading term for large charges is given by
the approximate saddle point value of the integrand,
\begin{equation}
d(p,q) \approx e^{i \pi \left. X_0 \right|_*} =
e^{\pi \sqrt{ p^2 q^2 - (pq)^2}} \;.
\end{equation}
This result is manifestly T- and S-duality invariant. 

A refined approximation can be obtained as follows: one 
identifies the exact critical point of $e^{i\pi X} = e^{i \pi (X_0 + X_1)}$,
expands the integrand $e^{i\pi X} \Delta$ 
to second order and performs a Gaussian integral. This is different
from a standard saddle point approximation, where one would 
expand around the critical point of the full integrand $e^{i\pi X} \Delta$.
This modification is motivated by the observation that 
the critical point of $i\pi X$ agrees exactly with the 
the critical point of the reduced dilatonic entropy function
(\ref{DilatonEntropyFunction}),
which gives the exact macroscopic entropy:
\[
i \pi X_* = \pi \Sigma_* = S_{\rm macro}(p,q) \;.
\]
At the critical point one has 
the following relation between the critical values of
$\rho, \sigma$ and the fixed point value of the dilaton:
\[
\rho_* = \frac{i |S_*|^2}{S_*+\overline{S}_*} \;,\;\;\;
\sigma_* = \frac{i}{S_*+\overline{S}_*} \;.
\]

One might think that the subleading contributions from $\Delta$ 
spoil the resulting equality between microscopic and macroscopic
entropy. However, these cancel against the contributions from 
the Gaussian integration (the `fluctuation determinant'), at
least to leading order in an expansion in inverse powers of the charges:
\begin{equation}
e^{S_{\rm micro}(p,q)} =
d(p,q) \approx e^{\pi \Sigma_* + \cdots} = e^{S_{\rm macro}(p,q) + \cdots} \;.
\end{equation}
This shows that the  modified saddle point approximation is compatible with 
a systematic expansion in large charges. Moreover, there
is an intriguing direct relation between the saddle point 
approximation of the exact microscopic state degeneracy 
(\ref{1/4BPSstates}) and the black hole variational principle.

\subsubsection{Further reading and references}

The state counting formula for $\frac{1}{4}$-BPS
states in $N=4$ compactifications was proposed 
in \cite{Dijkgraaf:1996it}. There several ways of deriving it
were discussed, which provide very strong evidence for
the formula.
Further evidence was obtained more recently in 
\cite{Shih:2005uc}, by using the relation between four-dimensional
and five-dimensional black holes \cite{Gaiotto:2005gf}. While
the leading order matching between state counting and black
hole entropy was already observed in \cite{Dijkgraaf:1996it},
the subleading corrections were obtained in \cite{LopesCardoso:2004xf} 
by using the modified saddle point evaluation explained above.

Another line
of development is the generalisation from toroidal
compactifications to a class of $N=4$ orbifolds, 
the so-called CHL-models \cite{Jatkar:2005bh,David:2006yn}. The issue
of choosing integration countours is actually more
subtle than apparent from our review, see \cite{Cheng:2007ch, Cheng:2008fc}
for a detailed account. For a comprehensive account of Siegel
modular forms, see \cite{Freitag}.

\subsection{Partition functions for  large black holes}

The strength of this result becomes even more obvious when we 
use it to compare the (microscopically defined) 
black hole partition function  to the (macroscopically defined)
free energy. 

One way of doing this is to the
evaluate mixed partition function 
$
Z_{\rm mix}(p,\phi) = \sum_q d(p,q) e^{\pi q_I \phi^I}
$
using integral representation (\ref{1/4BPSstates}) 
of $d(p,q)$. The result can be brought to the following form
\begin{equation}
Z_{\rm mix}(p,\phi) = {\sum_{\rm shifts}} 
{\sqrt{ \tilde{\Delta}(p,\phi)}}
e^{\pi {\cal F}_{\rm mix}(p,\phi)}\;.
\label{Zmix1/4}
\end{equation}
${\cal F}_{\rm mix}$ is the black hole free energy, including
all, both the holomorphic and the non-holomorphic corrections:
\[
{\cal F}_{\rm mix}(p,\phi) = \frac{1}{2} ( S + \overline{S}) \left(
p^a \eta_{ab} p^b - \phi^a \eta_{ab} \phi^b \right)
-i (S - \overline{S}) p^a \eta_{ab} \phi^b + 
{4 \Omega(S, \overline{S}, 
\Upsilon, \overline{\Upsilon})}\;.
\]
By imposing the magnetic attractor equations in the transition to
the mixed ensemble, 
the dilaton has become a function of the electric potentials and 
the magnetic charges:
\[
S = \frac{-i \phi^1 + p^1}{\phi^0 + i p^0} \;.
\]

The mixed partition functions is by construction invariant 
under shifts $\phi \rightarrow \phi + 2i$.
The mixed free energy is found to have a different periodicity,
and this manifests itself by the appearance of 
a finite sum over additional shifts of $\phi$
in (\ref{Zmix1/4}).
As predicted on the basis of symplectic covariance, the relation 
between the partition function and the free energy is modified by a 
`measure factor' $\tilde{\Delta}^-$, which we do not need to 
display explicitly. This factor agrees with the
measure  factor $\Delta^-$ in (\ref{Z_mix_modified}), 
which we found by imposing symplectic covariance
in the limit of large charges:
\[
{\tilde{\Delta}^- \approx \Delta^-}\;.
\]
Since we already made a partial saddle point approximation 
when going from the canonical to the mixed ensemble, we could 
not expect an exact agreement. It is highly non-trivial
that we can match the full mixed free energy, including 
the infinite series of instanton corrections. Moreover,
we have established that there is a non-trivial
measure factor, which agrees to leading order with the one
constructed by symmetry considerations.

\subsubsection*{Further reading, references, and some comments}

The idea to evaluate the mixed partition function using
microscopic state counting in order to check the OSV conjecture
for $N=4$ compactifiactions was first used in \cite{Shih:2005he}.
This confirmed the expectation that the OSV conjecture needs to be
modified by a measure factor once subleading corrections are taken
into account. This result was generalized in \cite{LopesCardoso:2006bg},
where we showed that the measure factor agrees asymptotically with
our conjecture which is based on imposing symplectic covariance.
Above, we pointed out that in (\ref{Zmix1/4}) we obtain the full
mixed ${\cal F}_{\rm mix}$, including the non-holomorphic corrections.
Of course, this way of organising the result is motivated by 
our approach to non-holomorphic corrections, and it is consistent
to regard these contributions as part of the measure factor,
as other authors appear to do. Further work is needed, in 
particular on the role played by the non-holomorphic corrections
in the microscopic description, before we can decide which way
interpreting the partition function is more adaequate.
Let us also mention that while we specifically considered
toroidal $N=4$ compactifications in this section, all results
generalise to CHL models.


There has also been much activity on $N=2$ compactifications
over the last years. Much of this work has focussed on establishing
and explaining the asymptotic factorisation 
\[
Z_{\rm mix} \simeq | Z_{\rm top}|^2
\]
predicted by the OSV conjecture 
\cite{Gaiotto:2006ns,Kraus:2006nb,Beasley:2006us,deBoer:2006vg}. 
The strategy persued
in these papers is to use string-dualities, in particular the
$AdS_3/CFT_2$-correspondence, to reformulate the proplem in terms
of two-dimensional conformal field theory. In comparison to the
simpler $N=4$ models, the relevant microscopic partition functions
are  related to the so-called elliptic genus of the underlying CFT.
Roughly, the elliptic genus is a `BPS partition function', i.e.
a partition function which has been modified by operator insertions
such that it only counts BPS states. The main problem is
to find a suitable generalisation of the Rademacher expansion which
allows to evaluate these BPS partition functions asymptotically 
for large charges. The picture emerging from this treatment is
that the black hole can be described microscopically (modulo 
string dualities) as a non-interacting state of branes and
anti-branes. This explains the asymptotic factorisation.

But as we have stressed throughout, non-holomorphic corrections
are expected to manifest themselves at the subleading level, which
microscopically correspond to interactions between branes and antibranes.
And indeed, a more recent refined analysis \cite{Denef:2007vg} has revealed the
presence of a measure factor, which agrees with the one 
found in  \cite{Shih:2005he} and \cite{LopesCardoso:2006bg}
in the limit of large charges. 

There is one further point which we need to comment on. During
this lecture we have tentatively assumed that `state counting'
means literally to count all the BPS states. But, as we have
mentioned previously, the BPS spectrum changes when crossing
a line of marginal stability. This is a possible cause for
discrepancies between state counting and thermodynamical entropy,
because they are computed in different regions of the parameter
space. In their original work \cite{Ooguri:2004zv} therefore 
conjectured that the microscopical entropy entering into the
OSV conjecture is an `index', i.e. a weighted sum over states 
which remains invariant when crossing lines of marginal stability.
The detailed study of \cite{Dabholkar:2005by, Dabholkar:2005dt} 
showed that it is very hard in practice to discriminate 
between absolute versus weighted state counting. While one 
example appeared to support absolute state counting, it was
pointed out later that there are several candidates for 
the weighted counting \cite{Denef:2007vg}. One intriguing 
proposal is that the correct absolute state counting is in
fact captured by an index, once it is taken into account that
states which are stable in the free limit become unstable once
interactions are taken into account \cite{Denef:2007vg}.

In conclusion, the OSV conjecture appears to work well in the
semi-classical approximation, if supplemented by a mearsure
factor. The concrete proposal discussed in these lectures
works correctly in this limit. It is less clear what is the
status of the original, more ambitious goal of finding an 
exact relation \cite{Ooguri:2004zv}, which would have various ramifications,
such as helping to find a non-perturbative definition 
of the topolocial string \cite{Ooguri:2004zv}, formulating a mini-superspace
approximation of stringy quantum cosmology \cite{Dijkgraaf:2006ab}, studying
$N=1$ compactifications via `topological M-theory' 
\cite{Gerasimov:2004yx,Dijkgraaf:2004te},
and approaching the vacuum selection problem of string 
theory by invoking an `entropic principle' 
\cite{Gukov:2005bg,Cardoso:2006nt,Cardoso:2006cb}.

\subsection{Partition functions for small black holes}

The counting of $\frac{1}{2}$-BPS states gave rise to the
following microscopic entropy:
\begin{equation}
S_{\rm micro} \approx \log \hat{I}_{13} 
\left( 4\pi \sqrt{\frac{1}{2} |q^2|} \right) 
\approx 4 \pi \sqrt{ \frac{1}{2} |q^2| } - 
{\frac{27}{4}} \log |q^2| 
 + \cdots
\end{equation}
This is to be compared with the macroscopic entropy. Including
the classical part of the $R^2$-coupling and the non-holomorphic
corrections, but neglecting instantons, this is:
\begin{equation}
S_{\rm macro} = 4 \pi \sqrt{ \frac{1}{2} |q^2| } - 
{6} \log |q^2| + \cdots
\end{equation}
While the leading terms agree, the first subleading term
comes with a slightly different coefficient. However,
we have seen that both entropies belong to different
ensembles, so that we can only expect that they agree in 
the thermodynamical limit. Since we have a conjecture about
the exact (or at least assymptotically exact) relation between
both entropies, we can check whether the shift in the coefficient
of the subleading term is predicted correctly.
Our conjecture about the relation between the canonical
free energy and the canonical partition function predicts the
following relation (see section \ref{NonHolo}):
\[
S_{\rm micro} = S_{\rm macro} + \log \sqrt{\frac{\Delta^-}{\Delta^+}}\;.
\]
This shows that both entropy are indeed different
if the measure factor $\Delta^-$ and the fluctuation determinant
$\Delta^+$ are different. For dyonic black holes we found that
both were equal, up to subleading contribution. Unfortunately 
our relation is not useful for small black holes, because
\begin{eqnarray}
\Delta^- &=& 0 \;,\;\;\;\mbox{up to non-holomorphic terms and 
instantons,} \nonumber \\
\Delta^+ &=& 0 \;,\;\;\;\,\mbox{up to instantons} \nonumber \;.
\end{eqnarray}
Since the measure factor and the fluctuation determinant are
degenerate (up to subleading contributions) the saddle 
point approximation is not well defined. This reflect that
small black hole live on the boundary of moduli space.

We can still test our conjecture about the relation between
the mixed partation and the mixed free energy, in particular
the presence of a measure factor and the role of non-holomorphic
contributions.
This requires to evaluate 
\[
\exp (S_{\rm micro}) = d(p^1,q) \approx \int d \phi \sqrt{\Delta^-(p^1, \phi)}
e^{\pi [{\cal F}_{\rm mix} (p^1, \phi) - q_I \phi^I ]} \;,
\]
where a non-vanishing $\Delta^-$ is obtained by including the non-holomorphic 
correstions.\footnote{Remember that $p^1$ is an electric charge for the 
heterotic string. We take $q_1=0$, because this is a magnetic charge.} 
We still neglect the contributions of the instantons. 

The integral can be evaluated, with the result:
\begin{equation}
d(p^1,q) \approx \int \frac{ d S d \overline{S}}{(S + \overline{S})^{14}}
\sqrt{ S + \overline{S} - \frac{12}{2\pi}}
\exp \left[ - \frac{\pi q^2}{S + \overline{S}} + 2 \pi (S + \overline{S})
\right] \;.
\label{OSV_modified}
\end{equation}
Here the integrals over $\phi^a =\phi^2, \phi^3, \ldots, \phi^{27}$
have been performed and the remaining integrals over 
$\phi^0$ and $\phi^1$ have been expressed in terms of the dilaton 
$S$. If we approximate
\[
\sqrt{ S + \overline{S} - \frac{12}{2\pi}} \approx
\sqrt{ S + \overline{S}} \;,
\]
this becomes the integral representation of a modified Bessel function.

Then our conjecture predicts 
\begin{equation}
S_{\rm micro}^{\rm (predicted)} \approx  
\log \hat{I}_{{13 - \frac{1}{2}}} ( 4 \pi \sqrt{ \frac{1}{2} |q^2| }) \approx
 4 \pi \sqrt{ \frac{1}{2} |q^2| } - {\frac{13}{2}} 
\log |q^2|  + \cdots \;,
\end{equation}
while the entropy obtained from state counting is:
\begin{equation}
S_{\rm micro} \approx \log  \hat{I}_{{13}} 
( 4 \pi \sqrt{ \frac{1}{2} |q^2| }) \approx
 4 \pi \sqrt{ \frac{1}{2} |q^2| } - {\frac{27}{4}} 
\log |q^2|  + \cdots
\end{equation}
Thus there is a systematic mismatch in the index of the Bessel function,
and while the leading terms agree, the coefficients of the $\log$-terms 
and all the following inverse-power terms mismatch.

This result can be compared with the original OSV-conjecture, where
one does not have a measure factor, and where only holomorphic
contributions to the free energy are taken into account:
\begin{eqnarray}
d(p^1,q) &\approx &\int d \phi e^{\pi [ {\cal F}_{OSV}(p^1,\phi) - q_I \phi^I]}
\approx  {(p^1)^2} 
\hat{I}_{{15}} ( 4 \pi \sqrt{\frac{1}{2} |q^2|}) \nonumber \\
S_{\rm micro}^{\rm (predicted)} &=& 4 \pi \sqrt{ \frac{1}{2} |q^2| }
- \frac{31}{4} \log |q^2| + \log (p^1)^2 + \cdots
\label{OSV_unmodified}
\end{eqnarray}
In this case the index of the Bessel function deviates even
more, and in addition there is an explicit factor $(p^1)^2$ 
which spoils T-duality. This clearly shows that the 
OSV conjecture needs to be modified by a measure factor.

When deriving (\ref{OSV_unmodified}), we have integrated over
28 potentials $\phi^I$, as we have done in our discussion of
large black holes, and in (\ref{OSV_modified}). There is
one subtlety to be discussed here. The full $N=4$ theory
has 28 gauge fields, but we have used the $N=2$ formalism.
Since we disregard the gravitini multiplets (and the hypermultiplets),
we work with a truncation to a subsector consisting of 
the $N=2$ gravity multiplet and 23 vector multiplets. This theory
only has 24 gauge fields, and therefore it only has 24 
electrostatic potentials $\phi^I$. However, at the end we
should reconstruct the missing 4 gauge potentials, and as we
have seen when recovering the $N=4$ entropy formula using  the
$N=2$ formalism, this extension is uniquely determined by
T-duality. As we have seen this prescription works for 
large black holes, but for small black holes we do not quite obtain
the right index for the Bessel function. 

However, the correct index for the Bessel function is obtained 
when using the unmodified OSV conjecture, but integrating only
over 24 instead of 28 electrostatic potentials:
\begin{eqnarray}
d(p^1,q) &\approx & \int d \phi e^{\pi [ {\cal F}_{OSV}(p^1,\phi) - q_I \phi^I]}
\approx {(p^1)^2} 
\hat{I}_{{13}} ( 4 \pi \sqrt{\frac{1}{2} |q^2|})\;, \\
S_{\rm micro}^{\rm (predicted)} &=& 4 \pi \sqrt{ \frac{1}{2} |q^2| }
- \frac{27}{4} \log |q^2| + \log (p^1)^2 + \cdots
\label{OSV_24}
\end{eqnarray}
Note that this does not cure the problem with the prefactor 
$(p^1)^2$, which is incompatible with T-duality. It is intriguing,
but at the same time puzzling that the correct value for the
index is obtained by reducing the number of integrations. However,
it is not clear how to interprete this restriction. Moreover, 
it is unavoidable to include a measure factor to implement T-duality,
and this is very likely to have an effect on the index.

\subsubsection*{Further reading and references}

In this section, we followed \cite{LopesCardoso:2006bg}, and compared the 
result with the calculation based on the original OSV conjecture 
\cite{Dabholkar:2005by,Dabholkar:2005dt}. Both approaches find
agreement for the leading term, but disagreement for the
subleading terms. Moreover, when sticking to the original
OSV conjecture, the result is not compatible with T-duality. 
Further problems and subtleties with the OSV conjecture 
for $\frac{1}{2}$-BPS black holes have been discussed in 
detail in \cite{Dabholkar:2005by,Dabholkar:2005dt}. One obvious
explanation for these difficulties is that in the `would-be leading'
order approximation small black holes are singular: they
have a vanishing horizon area and the moduli take values 
at the boundary of the moduli space. While the higher curvature
smooth the null singularity, leading to agreement between
macroscopic and microscopic entropy to leading order in the
charges, the semi-classical expansion is still ill defined,
since one attempts to expand around a singular configuration.
Apparently one needs to find a different way of organising
the expansion, if some version of the  OSV conjecture 
is to hold at the semi-classical level. A more drastic 
alternative is that the OSV conjecture simply does not apply
to small black holes. But since the mismatch of the 
subleading corrections appears to follow some systematics,
there is room for hope. The situation is less encouraging
for the non-perturbative corrections coming form instantons. As observed 
both in \cite{Dabholkar:2005by,Dabholkar:2005dt} and in 
\cite{LopesCardoso:2006bg} the analytical structure of the
terms observed in microscopic state counting is different 
from the one expected on the basis of the OSV conjecture.

\subsection{Problems}

\begin{Problem} \label{Prob:statecounting}
Counting states of the open bosonic string.\\[2ex]
In the light cone gauge, a basis for the Hilbert space of the
open bosonic string (neglecting the center of mass momentum)
is given by
\begin{equation}
\alpha^{i_1}_{-m_1} \alpha^{i_2}_{-m_2}  \cdots |0 \rangle \;,
\end{equation}
where $i_k=1,\ldots, 24$ and $m_k = 1,2,3,\ldots$. States with
the same (total) excitation number $n=m_1+m_2+\ldots$ have the
same mass. Incidentially, the problem of counting states of the
open bosonic string with given mass, is the same as counting
the number of $\frac{1}{2}$-BPS states for the heterotic
string, compactified on $T^6$, with given charges $q \in 
\Gamma_{\rm Narain}$. \\[2ex]

The number of states with given excitation number $n$ is
encoded in the partition function 
\begin{equation}
Z(q) = \mbox{Tr} \,q^N \;,
\label{Definition}
\end{equation}
where the trace is over the Hilbert space of physical 
string states (light cone gauge), $q\in \mathbbm{C}$, and
$N$ is the number operator with eigenvalues $n=0,1,2,3,\ldots$.
Evaluation of the trace gives
\begin{equation}
Z(q) = \left( \prod_{l=1}^\infty (1-q^l) \right)^{-c} \;,\;\;\;
|q| < 1 \;,
\label{ProductRepresentation}
\end{equation}
where $c=D-2=24$ is the number of space-time dimensions
transverse to the string world sheet (the physical excitations).
The number $d_n$ of string states at level $n$ is encoded in the 
Taylor expansion 
\begin{equation}
Z(q) = \sum_{n=0}^\infty d_n q^n \;.
\label{TaylorSeries}
\end{equation}
Verify that $d_n$ counts string states, for small $n=0,1,2,\ldots$
Do this either for the critical open bosonic string, $c=24$, or for just
one string coordinate, $c=1$. The latter is the classical problem
of counting partitions of an integer. It is instructive to 
evaluate $d_n$ both directly, by reorganising the 
product representation (\ref{ProductRepresentation}) into a
Taylor series, and by the integral representation of $d_n$
obtained by inverting (\ref{TaylorSeries}). \\[2ex]
Hints:
Note that 
\begin{equation}
Z(q) = q \Delta^{-1}(q) \;,
\end{equation}
where $\Delta(q) = \eta^{24}(q)$ is the cusp form ($\eta$ is the
Dedekind eta-function). $\Delta(q)$ is a modular form of weight 12 
and has the following expansion around the cusp $q=0$:
\begin{equation}
\Delta(q) = q - 24 q^2 + 252 q^3 - 1472 q^4 + 4830 q^5 +
{\cal O}(q^6) \;.
\end{equation}
$\Delta(q)$ has no zeros for $0<|q|<1$. \\
$d_n$ can be computed by a contour integral in the unit
disc $|q|<1$.
\end{Problem}

\begin{Problem} The asymptotic state density of the open bosonic
string. \\[2ex]
Given the information provided in Problem \ref{Prob:statecounting},
compute the asymptotic number of open bosonic string states $d_n$ for
$n\rightarrow \infty$. (You may restrict yourself to the case $c=24$,
which corresponds to the critical dimension.)\\[2ex]
Instructions:
\begin{enumerate}
\item
The unit disc $|q|<1$ can be mapped to the semi-infinite
strip $-\frac{1}{2} < \tau_1 < \frac{1}{2}$, $\tau_2 >0$
in the complex $\tau$-plane, $\tau =\tau_1 + i \tau_2$ by
\begin{equation}
q = e^{2 \pi i \tau} \;.
\end{equation}
(Like other modular forms, $\Delta$ extends to a holomorphic
function on the whole upper half plane by periodicity in $\tau_1$.)\\
Rewrite the contour integral for $d_n$ as a contour integral 
over $\tau$. 
\item
Use the modular properties of $\Delta(\tau)$ to find the behaviour
of the integrand close to $\tau =0$ from the known behaviour
of $\Delta(\tau)$ at $\tau = i \infty \Leftrightarrow q=0$. 
Show that for $n\rightarrow \infty$ the integrand has a sharp
saddle point. Use this to evaluate the contour integral 
in saddle point approximation. (Expand the integrand to second
order around the saddle point and perform the resulting 
Gaussian integral.)
\item
The correct result is
\begin{equation}
d_n \approx \mbox{Const.} \; e^{4 \pi \sqrt{n}} n^{-\frac{27}{4}} \;.
\end{equation}

\end{enumerate}

\end{Problem}

\subsection*{Acknowledgements}

These notes are based on lectures given at the 
School on Attractor Mechanism 2007 in Frascati,
and they make use of the material prepared for  
a one semester course taught at the Erwin
Schr\"odinger Insitute as part of a senior research
fellowhsip in the academic year 2006/07. I would 
like to thank Stefano Belluci for inviting me to 
lecture in Frascati and to publish these notes as part of
the proceedings. I would also like to 
thank the ESI for its great hospitality during
my two stays in Vienna. The participants of the
lecture series in both Frascati and Vienna 
have helped me to shape these notes.
A special thanks goes to Maximilian Kreuzer and his
group for their active participation. 

The material presented in these notes is to a large 
extent based on work done in various collaborations with
Gabriel Cardoso, Vicente Cort\'es, Bernard de Wit, J\"urg K\"appeli, 
Christoph Mayer and Frank Saueressig.

\begin{appendix}

\section{K\"ahler manifolds and special K\"ahler manifolds \label{AppA}}

In this appendix we review K\"ahler manifolds and special K\"ahler manifolds
from the mathematical perspective. The first part is devoted to
the basic definitions and properties of complex, hermitean and
K\"ahler manifolds. For a more extensive review we recommend
the book by Nakahara \cite{Nakahara}, and, for readers with a
stronger mathematical inclination, the concise lecture notes
by Ballmann \cite{Ballmann}. The characterisation of complex
and K\"ahler manifolds in terms of holonomy groups
can be found in \cite{Joyce}.
The second part reviews special
K\"ahler manifolds and is mostly based on  
\cite{Freed:1997dp,Alekseevsky:1999ts} with 
supplements from \cite{Cortes:2003zd,Cortes:2005uq,Cortes:2007}.
A review of special geometry from a modern perspective can also be found 
in \cite{Mohaupt:2007md}.

A para-complex variant of special 
geometry, which applies to the target manifolds of 
Euclidean $N=2$ theories has been developed in 
\cite{Cortes:2003zd,Cortes:2005uq}. The framework of 
$\epsilon$-K\"ahler manifolds, which will be employed in 
\cite{Cortes:2007}, is particularly suitable for treating
Euclidean supersymmetry and 
standard (Lorentzian) supersymmetry in parallel.

\subsection{Complex and almost complex manifolds}

Let $M$ be a differentiable manifold of dimension $2n$. 

\begin{Definition}
An {\em almost complex structure} $I$ on $M$ is tensor field of 
type $(1,1)$ with the property that (pointwise)
\[
I^2 = - \mbox{Id} \;.
\]
\end{Definition}
In components, using real coordinates $\{x^m|m=1, \ldots, 2n\}$, 
this condition reads:
\begin{equation}
\label{IisMinusOne}
I^m_{\;p} I^p_{\;n} = - \delta^m_n  \;.
\end{equation}
\begin{Definition}
An almost complex structure is called {\em integrable} if 
the associated Nijenhuis tensor $N_I$ vanishes for all 
vector fields $X,Y$ on $M$:
\[
N_I(X,Y) := [IX, IY] + [X,Y] - I[X,IY] - I[IX,Y] = 0 \;.
\]
\end{Definition}
The expression for $N_I$ in terms of local coordinates $\{x^m \}$ can
be found by substituting the coordinate expressions $X=X^m \partial_m$,
$Y=Y^m\partial_m$ for the vector fields.\footnote{In fact,
it is sufficient
to substitute a basis of coordinate vector fields $\{\partial_m \}$
to obtain the components
$N_{mn} = N(\partial_m, \partial_n)$.}
We will not need this
explicitly.\\[2ex]
{\bf Remark:}
The integrability of an almost complex structure
is equivalent to the existence of local 
complex coordinates $\{z^i | i=1,\ldots, n \}$.
An integrable almost complex structure
is therefore also simply called a {\em complex structure}.  
\begin{Definition}
A manifold which is equipped with 
an (almost) complex structure is called an (almost) complex manifold.
\end{Definition}

\noindent
{\bf Remark:} The existence of an (almost) complex structure can
be rephrased in terms of holonomy. An almost complex structure is 
a $GL(n,\mathbbm{C})$ structure, and a complex structure is 
a torsion-free $GL(n,\mathbbm{C})$ structure.

\subsection{Hermitean manifolds}

Let $(M,I)$ be a complex manifold and let $g$ be a 
(pseudo-)Riemannian metric on $M$. 

\begin{Definition}
$(M,g,I)$
is called a {\em hermitean} manifold, if $I$ generates isometries of $g$:
\begin{equation}
\label{hermitean}
I^* g= g  \;.
\end{equation}
\end{Definition}
{\bf Remark:}
Condition (\ref{hermitean}) is equivalent to saying that
\[
g(IX, IY) = g(X,Y) \;,
\]
for all vector fields $X,Y$ on $M$. In local coordinates
the condition reads:
\begin{equation}
g_{pq} I^p_{\;m} I^q_{\;n} = g_{mn} \;.
\label{IisIsometry}
\end{equation}
{\bf Remark:} If the metric is indefinite, $(M,g,I)$ is
called pseudo-hermitean, but we will usually drop the
prefix `pseudo-'. \\[2ex]
On a hermitean manifold one can define the so-called {\em 
fundamental two-form}:
\[
\omega(X,Y) := g(IX,Y)\;,
\]
or, in coordinates,
\begin{equation}
\label{OfromIP}
\omega_{mn} = -g_{mp} I^p_{\;n} \;.
\end{equation}
Note that $\omega_{mn}=-\omega_{nm}$, because $g_{mn}$ is symmetric,
while $I$ satisfies (\ref{IisMinusOne}) and (\ref{IisIsometry}). 
Moreover the two-form $\omega$ is non-degenerate, because $g$ is.

Equation (\ref{OfromIP}) can be solved for the metric $g$ or for
the complex structure $I$:
\begin{eqnarray}
g_{mn} &=& \omega_{mk}I^k_{\;n} \;, \nonumber \\
I^m_{\;n} &=& -g^{mk} \omega_{kn} \;.
\end{eqnarray}
Thus, if any two of the three data $g$ (metric), $I$ (complex structure)
or $\omega$ (fundamental two-form) are given on a hermitean manifold,
the third is already determined.

When we use complex coordinates $\{z^i \}$, the complex structure
only has `pure' components:
\[
I^i_{\;j} = i \delta^i_j \;,\;\;\;
I^{\overline{i}}_{\; \overline{j}}= -i \delta^{\overline{i}}_{\overline{j}}\;.
\]
For a hermitean metric the pure components vanish,
$g_{ij}=0$ and $g_{\overline{i} \overline{j}} =0$. Only the
`mixed' components $g_{i\overline{j}}$ and $g_{\overline{i}j} =
\overline{ g_{i \overline{j}}}$ remain. Note that 
the matrix $g_{i\overline{j}}$ is hermitean. The fundamental two-form 
also only has mixed 
components, and $\omega_{i\overline{j}} = i g_{i \overline{j}}$. 
Thus in complex coordinates the matrices representing the metric
and the fundamental two-form are hermitean and anti-hermitean, respectively,
while in real coordinates they are symmetric and antisymmetric,
respectively.

On a hermitean manifold the metric 
\[
g = g_{i\overline{j}} (dz^i \otimes \overline{z}^{\overline{j}}
+ d \overline{z}^{\overline{j}} \otimes dz^i )
\]
and the fundamental two-form 
\[
\omega = i g_{i\overline{j}} dz^i \wedge d \overline{z}^{\overline{j}}
= i  g_{i\overline{j}} (dz^i \otimes d \overline{z}^{\overline{j}}
-  d \overline{z}^{\overline{j}} \otimes dz^i ) 
\]
can be combined into the hermitean form
\[
\tau = g_{i \overline{j}} dz^i \otimes d\overline{z}^{\overline{j}}
= \frac{1}{2} (g -i \omega) \;.
\]
The hermitean form defines a hermitean metric on the complexified
tangent bundle $TM_{\mathbbm{C}}$ of $M$. All statements
and formulae in this section apply irrespective of  $g$ being
positive definit or indefinit (but non-degenerate).


\subsection{K\"ahler manifolds}

\begin{Definition}
A {\em K\"ahler manifold} $(M,g,I)$ is a hermitean manifold
where the fundamental form is closed:
\[
d \omega =0 \;.
\]
\end{Definition}
{\bf Remark:} 
Equivalently, one can impose that 
the complex structure is parallel with respect to the 
Levi-Civita connection,
\[
\nabla^{(g)} I =0 \;.
\]
{\bf Comment:}
Hermitean manifolds are characterised by `pointwise' compatibility 
conditions between
metric, complex structure and fundamental form. For K\"ahler
manifolds one imposes a stronger compatibility condition:
the complex structure $I$ must be parallel (covariantly constant) with
respect to the Levi-Civita connection $\nabla^{(g)}$. 
Since the metric $g$ itself
is parallel by definition of $\nabla^{(g)}$, parallelity of 
$I$ is equivalent to the parallelity of the fundamental form $\omega$. 
Moreover, it can be shown that if $\omega$ is closed, it is
automatically parallel with respect to $\nabla^{(g)}$.\\[2ex]
The fundamental form of a K\"ahler manifold is called its {\em K\"ahler 
form.} It can be shown that a K\"ahler metric can be expressed in 
terms of a real-analytic function, the K\"ahler potential, 
by\footnote{In fact, this might serve as yet another equivalent definition 
of a K\"ahler manifold.}
\[
g_{i \overline{j}} = \frac{\partial^2 K (z,\overline{z})}
{\partial z^i \partial 
\overline{z}^{\overline{j}}} \;.
\]
The K\"ahler form can also be expressed as the second derivative
of the K\"ahler potential:
\[
\omega = i \partial \overline{\partial} K 
= i
g_{i \overline{j}} dz^i \wedge d \overline{z}^{\overline{j}} 
\;,\;\;\;\mbox{where}\;\;\;
\partial = dz^i \partial_i \;,\;\;\;
\overline{\partial} = d \overline{z}^{\overline{j}} 
\partial_{\overline{j}} 
\]
are the Dolbeault operators (holomorphic exterior derivatives). \\[2ex]
{\bf Remark:} 
If the metric $g$ is positive definit, 
a K\"ahler manifold can be defined equivalently 
as a $2n$-dimensional manifold with a torsion-free $U(n)$ 
structure. Note that $U(n) \simeq GL(n,\mathbbm{C}) \cap 
SO(2n) \subset GL(2n,\mathbbm{R})$, which shows that 
$U(n)$ holonomy implies that there is a connection such that 
both the metric and the complex structure are parallel. \\[2ex]
{\bf Remark:}
If the metric is not positive definit, $U(n)$ is replaced 
by a suitable non-compact form. Pseudo-hermitean manifolds
with closed fundamental form are called pseudo-K\"ahler manifolds.
We have seen in the main text that the (conical affine special)
K\"ahler manifolds occuring in the construction of supergravity
theories within the superconformal calculus always have
indefinite signature, because the compensator of complex dilatations
has a kinetic term with an inverted sign. We usually omit the prefix
`pseudo-' in the following and in the main text.

\subsection{Affine special K\"ahler manifolds}

Special K\"ahler manifolds are distinguished by the fact that
the K\"ahler potential $K(z,\overline{z})$
can itself be expressed in terms of
a holomorphic prepotential $F(z)$. 
The intrinsic definition of such manifolds is as follows \cite{Freed:1997dp}.
\begin{Definition}
An {\em affine special K\"ahler manifold} $(M,g,I,\nabla)$
is a K\"ahler manifold $(M,g,I)$ equipped with 
a flat, torsion-free connection $\nabla$,  which has the following properties:
\begin{enumerate}
\item
The connection is symplectic, i.e., the K\"ahler form is parallel
\[
\nabla \omega = 0 \;.
\]
\item
The complex structure satisfies:
\[
d^\nabla I = 0 \;,
\]
which means, in local coordinates, that
\[
\nabla_{[m} I_{n]}^{\;p} = 0 \;.
\]
\end{enumerate}
\end{Definition}
{\bf Remark:}
The complex structure is not parallel with respect 
to the special connection $\nabla$, but only `closed' (regarding
$I$ as a vector-valued one-form). This, together with the fact
that $\nabla$ is flat shows that  the connections 
$\nabla$ and $\nabla^{(g)}$ are different,
except for the trivial case of a flat Levi-Civita connection.

It can be shown that the existence of a special connection 
$\nabla$ is equivalent to 
the existence of a K\"ahlerian Lagrangian immersion of 
$M$ into a model vector space, namely the standard complex
vector space of doubled dimension \cite{Alekseevsky:1999ts}. 
Let us review this construction
in some detail. 

The standard complex symplectic vector space of complex dimension 
$2n$ is $V= T^* \mathbbm{C}^n$. As a vector space, this is isomorphic
to $\mathbbm{C}^{2n}$. Let $z^i$ be linear  
coordinates on $\mathbbm{C}^n$ and $w_i$ be coordinates 
on $T_z \mathbbm{C}^n$. Then we can take $(z^i, w_i)$ as coordinates
on $T^*\mathbbm{C}^n$, and the symplectic form is 
\[
\Omega_V = dz^i \wedge dw_i \;.
\]
If we interpret $V$ as a phase space, then
the $z^i$ are the coordinates and the $w_i$ are the associated 
momenta. Symplectic rotations of $(z^i, w_i)$ give rise
to different `polarisations' (choices of coordinates vs momenta)
of $V$. 

The vector space $V$ can be made a K\"ahler manifold in 
the following way: starting form 
the antisymmetric complex bilinear form $\Omega_V$ one can 
define an hermitean sesquilinear form $\gamma_V$ 
by applying complex conjugation in the second argument of
$\Omega$, plus multiplication by $i$:
\[
\gamma_V = i \left( dz^i \otimes d \overline{w}_i - dw_i \otimes
d\overline{z}^i \right) \;.
\]
The real part of $\gamma_V$ is a flat K\"ahler metric of 
signature $(2n,2n)$:
\[
g_V = \mbox{Re} (\gamma_V) = i \left( dz^i \otimes_{\rm sym} d\overline{w}_i
- dw_i \otimes_{\rm sym} d\overline{z}^i \right)\;,
\]
while the imaginary part is the associated K\"ahler form:
\[
\omega_V = \mbox{Im} (\gamma_V) = dz^i \wedge d \overline{w}_i
- dw_i \wedge d \overline{z}^i  \;.
\]

Now consider the immersion of a manifold $M$ into $V$. 
An immersion is a map with invertible
differential. An immersion need not
be an invertible map, but it can be made invertible by restriction.
Invertible immersion are called embeddings. (Intuitively, the
difference between immersions and embeddings is that embeddings
are not allowed to have self-intersections, or points where two image
points come arbitrarily close.)
\begin{Definition}
An immersion $\Phi$ of a complex manifold $M$ into a K\"ahler manifold is called
{\em K\"ahlerian}, if it is holomorphic and if
the pullback $g=\Phi^* g_V$ of the K\"ahler metric is nondegenerate.
\end{Definition}
{\bf Remark:} Equivalently, one can require that the pullback of
the hermitean form or of the K\"ahler form is non-degenerate.
\begin{Definition}
An immersion $\Phi$ of a complex manifold $M$ into a complex symplectic 
manifold is called {\em Lagrangian} if the pullback of the complex
symplectic form vanishes, $\Phi^* \Omega_V=0.$ 
\end{Definition}
{\bf Remark:} For generic choices of coordinates, a 
Lagrangian immersion $\Phi$ is generated by a holomorphic
function $F$ on $M$, i.e. $\Phi=dF$. \\ \\
It has been shown that for any affine special K\"ahler manifold
of complex dimensions $n$
there exists\footnote{locally, and if the manifold is simply 
connected even globally,} 
a K\"ahlerian Lagrangian immersion into $V=T^*\mathbbm{C}^n$.
Moreover every K\"ahlerian Langrangian 
immersion of an $n$-dimensional  complex manifold $M$ into $V$ induces 
on it the structure of an affine special K\"ahler manifold.

By the immersion $\Phi$, the special K\"ahler manifold 
$M$ is mapped into $V$ as the graph\footnote{More precisely, the image
is {\em generically} the graph of map. We comment on non-generic 
immersions below.}
of a map $z^i \rightarrow w_i = \frac{\partial F}{\partial z^i}$,
where $F$ is the prepotential of the special K\"ahler metric,
which is the generating function of the immersion: $\Phi=dF$.
Using the immersion, 
one obtains `special' coordinates on $M$ by 
picking half of the coordinates $(z^i, w_i)$ of $V$ (say, the $z^i$). 
Along the graph,
the other half of the coordinates of $V$ are dependent 
quantities, and can be expressed through the prepotential:
$w_i = w_i (z) = \frac{\partial F}{\partial z^i}$.
The special K\"ahler metric 
$g$, the K\"ahler form $\omega$ and the hermitean form $\gamma$ 
on $M$ are the pullbacks of the corresponding data
$g_V, \omega_V, \tau_V$ of $V$ under the immersion.\\[2ex]
{\bf Remark:} For non-generic choices of $\Phi$ 
the immersed $M$ may be not a graph. Then the 
$z^i$ do not provide local coordinates, the $w_i$ are not
the components of a gradient, and $\Phi$ does not have a 
generating function, i.e., `there is no prepotential'.\footnote{In 
the physics literature, this phenomenon and its consequences
have been discussed in  detail in \cite{Ceresole:1995jg,Craps:1997gp}.}
This is not a problem, since one can work perfectly well
by using only the symplectic vector $(z^i, w_i)$. Moreover,
by a symplectic transformation one can always make the 
situation generic and go to a symplectic basis (polarisation of 
$V$) which admits a prepotential. \\[2ex]
{\bf Remark:}
In the main text we denoted the component expression for the
affine special K\"ahler metric on $M$ by
$N_{IJ}$ instead of $g_{i\overline{j}}$. The scalar fields
$X^I$ correspond to the special coordinates $z^i$. More prescisely, 
the scalar fields can be interpreted as compositions of maps from space-time
into $M$ with coordinate maps $M \supset U \rightarrow \mathbbm{C}^n$.
The key formulae which express the K\"ahler potential and 
the metric in terms of the prepotential are 
(\ref{KaehlerPotentialRigid}) and (\ref{RigidMetric}).


\subsubsection{Special affine coordinates and the Hesse potential}

K\"ahler manifolds are in particular symplectic manifolds,
because the fundamental form is both non-degenerate and
closed. The additional 
structure on affine special K\"ahler manifolds is the special
connection $\nabla$, which is both flat and symplectic
(i.e. the symplectic form $\omega$ is parallel with respect
to $\nabla$.)\footnote{It is of course also parallel with respect
to the Levi-Civita connection $\nabla^{(g)}$, but the Levi-Civita
connection is not flat (except in trivial cases).}
As a consequence, there exist $\nabla$-affine (real) coordinates
$x^i, y_i$, $i=1,\ldots, n$ on  
$M$, 
\[
\nabla dx^i =0 \;,\;\;\;
\nabla dy_i = 0 \;,
\]
which are adapted to the symplectic structure,
\[
\omega = 2 dx^i \wedge dy_i \;.
\]
The relation between these special affine coordinates 
and the special coordinates $z^i$ can be elucidated by
using the immersion of $M$ into $V$. We can decompose
$z^i,w_i$ into their real and imaginary parts:
\[
z^i = x^i + i u^i \;,\;\;\;
w_i = y_i + i v_i  \;.
\]
Then the K\"ahler form $\omega_V$ takes the form
\[
\omega_V = dx^i \wedge dy_i + du^i \wedge dv_i \;.
\]
Using that the pullback of the complex symplectic 
form $\Omega_V$ vanishes, one finds that the pullback 
of $\omega_V$ is\footnote{For notational simplicity, we denote
the pulled back coordinates $\Phi^* x^i, \Phi^*y^i$ by $x^i, y_i$.}
\[
\omega = \Phi^* \omega_V = 2 dx^i \wedge dy_i \;.
\]
Thus the special real coordinates form the real part of the
symplectic vector $(z^i, w_i)$. 
The real and imaginary parts of $z^i = x^i + u^i$ also form
a system of real coordinates on $M$, which is 
induced by the complex coordinate system $z^i$, but not adapted
to the symplectic structure (since $x^i, u^i$ do not form a 
symplectic vector).  
The change of coordinates
\[
(x^i, u^i) \rightarrow (x^i, y_i)
\]
can be viewed as a Legendre transform, because
\begin{equation}
y_i = \mbox{Re} \left( \frac{\partial F}{\partial z^i}
\right) = 
\frac{\partial \mbox{Im} F}{\partial \mbox{Im} z^i} =
\frac{\partial \mbox{Im} F}{\partial u_i} \;.
\end{equation}
The Legendre transform maps the imaginary part  
of the prepotential to the Hesse potential 
\[
H (x,y) = 2 \left( \mbox{Im} F(x+iu(x,y)) - u_i y^i \right) \;.
\]
A Hesse potential is a real K\"ahler potential, i.e., a potential
for the metric, but based on real rather than complex coordinates. 
Denoting the affine special coordinates
by $\{ q^a | a=1, \ldots, 2n \} = \{ x^i, y_i | i=1,\ldots, n\}$,
the special K\"ahler metric on $M$ is given by
\[
g = \frac{\partial^2 H}{\partial q^a \partial q^b} dq^a \otimes_{\rm sym} 
dq^b \;.
\]

The special connection present on an affine special K\"ahler
manifold is not unique. The $U(1)$ action generated by the
complex structure generates a one-parameter family of such 
connections. Each of these comes with its corresponding
special affine coordinates. The imaginary part $(u^i, v_i)$
of the symplectic vector $(z^i, w_i)$ provides one of these
special affine coordinate systmes. The coordinate systems
$(x^i, y_i)$ and $(u^i,v_i)$ both occur naturally in the
construction of BPS black hole solutions.

\subsection{Conical affine special K\"ahler manifolds and projective special
K\"ahler manifolds}

\begin{Definition}
A {\em conical} affine special K\"ahler manifold
$(M, g, I, \nabla, \xi)$ is an affine special K\"ahler
manifold endowed with a vector field $\xi$ such that
\begin{equation}
\label{Xi}
\nabla^{(g)} \xi = \nabla \xi = \mbox{Id} \;.
\end{equation}
\end{Definition}
The condition $\nabla^{(g)} \xi = \mbox{Id}$ implies that $\xi$ is a 
homothetic Killing vector field, and that it is hypersurface
orthogonal. Then one can introduce adapted coordinates 
$\{r,v^a\}$ such that 
\[
\xi = r \frac{\partial}{\partial r}
\]
and
\[
g = dr^2 + r^2 g_{ab}(v) dv^a dv^b \;.
\]
Thus $M$ is a real cone. However, in our case $M$ carries
additional structures, and $\xi$ satisfies the additional 
condition $\nabla \xi = \mbox{Id}$. It can be shown that this implies that
$M$ has a freely acting $U(1)$ isometry, with Killing
vector field $I\xi$. The surfaces $r =\mbox{const.}$ are the
level surfaces of the moment
map of this isometry. Therefore the isometry preserves the
level surfaces, and $M\subset T^*\mathbbm{C}^{n+1}$ has the
structure of a complex cone, with $\mathbbm{C}^*$-action 
generated by $\{\xi, I \xi \}$. 

One can choose special affine coordinates such that 
$\xi$ has the form\footnote{These are called conical 
special affine coordinates, but we will usually drop `conical'.}
\begin{equation}
\label{Xireal}
\xi = q^a \frac{\partial}{\partial q^a} = 
x^i \frac{\partial}{\partial x^i} + 
y_i \frac{\partial}{\partial y_i} \;.
\end{equation}

Moreover, it can be shown that that the existence
of a vector field $\xi$ which satisfies (\ref{Xi}) is equivalent to the
condition that the prepotential is homogenous of degree 2:
\[
F(\lambda z^i) = \lambda^2 F(z^i) \;,
\]
where $z^i \rightarrow \lambda z^i$ is the action 
of $\mathbbm{C}^*$ on the (conical) special coordinates $\{ z^i \}$
associated with the (conical) special affine coordinates
$\{x^i, y_i\}$. 
In special coordinates, $\xi$ takes the form\footnote{Note 
that this is equivalent to (\ref{Xireal}) if and only if 
the prepotential is homogenous of degree 2.}
\[
\xi = z^i \frac{\partial}{\partial z^i} \;.
\]
The quotient 
$\overline{M} = M / \mathbbm{C}^*$ is a K\"ahler manifold
which inherits its metric from $M$.
Manifolds which are 
obtained from conical affine special K\"ahler manifolds in this way
are called {\em projective special K\"ahler manifolds}. 
These are the scalar manifolds of vector multiplets in
$N=2$ Poincar\'e supergravity. The corresponding conical
affine special K\"ahler manifold is the target space of 
a gauge equivalent theory of superconformal vector multiplets.
As we have seen from the physical perspective one can go
back and forth between $M$ and $\overline{M}$. Geometrically,
$M$ can be regarded as a $\mathbbm{C}^*$-bundle over
$\overline{M}$. 
In turn $M$ itself is embedded into $V=T^*\mathbbm{C}^{n+1}$,
where $n+1$ is the complex dimensions of $M$. In the main text
the D-gauge is fixed by imposing
\[
-i(X^I \overline{F}_I - F_I \overline{X}^I) = 1
\]
on the symplectic  vector $(X^I,F_I)$. Geometrically, this means
that $(X^I,F_I)$ is required to be a
unitary section of the so-called universal
line bundle over $\overline{M}$. Instead of using unitary sections,
one can also reformulate the
theory in terms of holomorphic sections of the universal bundle.
This is frequently done when working with general (in contrast
to special) coordinates, see \cite{Andrianopoli:1996cm}.
For a more detailed account on the universal bundle, 
see \cite{Cortes:2007}. 

In the main text we gave explicit formulae for various quantities
defined on  projective special K\"ahler manifolds
in the notation used in the supergravity literature. In particular,
(\ref{MetricProj}) and (\ref{KPproj}) are the expressions for 
the metric and K\"ahler potential in terms of special coordinates
on $\overline{M}$. There we also discussed the relation between
the signatures of the special K\"ahler metrics on $M$ and $\overline{M}$
The `horizontal' metric $g_{IJ}$ (\ref{HorizontalMetric}) 
vanishes along the vertical directions (the directions 
orthogonal to $\overline{M}$
under the natural projection with respect to the special K\"ahler metric
of $M$), but it is
non-degenerate
along the horizontal
directions (the directions which project orthogonally onto $\overline{M}$). 
If the metric of $M$ is complex Lorentzian
$(\mp, \mp, \pm, \ldots , \pm)$, then the metric defined on 
$\overline{M}$ by projection is even positive definit.
This defines a projective special K\"ahler metric on $\overline{M}$,
for which an explicit formula in terms of special coordinates 
is given by (\ref{MetricProj}),
(\ref{KPproj}).

\section{Modular forms \label{AppB}}

Here we summarize some standard results on modular forms.
See \cite{Zagier} for a more detailed account. As we mentioned
in the main text, the theory of Siegel modular forms is a
generalisation of the theory of `standard' modular forms
reviewed here. Some facts are stated in the main text. 
For a detailed account on Siegel modular forms see for 
example \cite{Freitag}.

The action of the modular group $PSL(2,\mathbbm{Z}) \simeq 
SL(2,\mathbbm{Z})/\mathbbm{Z}_2$ on the upper half plane
${\cal H} = \{ \tau \in \mathbbm{C} | \mbox{Im} \tau > 0 \}$ is:
\[
\tau \rightarrow \tau' = \frac{a \tau +b}{c\tau +d} 
\;,\;\;\;
\mbox{where}\;\;\;
\left( \begin{array}{cc}
a & b \\ c & d \\
\end{array} \right) \in SL(2,\mathbbm{Z}) \;.
\]
The modular group is generated by the two transformations\footnote{The
notation $T$ and $S$ is standard in the mathematical literature, and 
does not refer to T- or S-duality. However, there are several examples
where either T-duality or S-duality acts by $PSL(2,\mathbbm{Z})$ transformations
on complex fields.}
\[
T: \tau \rightarrow \tau +1 \;,\;\;\;
S: \tau \rightarrow - \frac{1}{\tau} \;.
\]
The interior of the standard fundamental domain for this group action is
\[
{\cal F} = \{ \tau \in {\cal H} | 
-\frac{1}{2} < \mbox{Re} \tau < \frac{1}{2} \;,
|\tau| > 1 \} \;.
\]
The full domain is obtained by 
adding a point at infinity, denoted $i\infty$, and identifying
points on the boundary which are related by the group action.
The point $i\infty$ is called the cusp point.

A function on ${\cal H}$ is said to transform 
with (modular) weight $k$:
\[
\phi(\tau') = (c\tau + d)^k \phi(\tau)
\]
A function on ${\cal H}$ is called a modular function,
a modular form, a cusp form, if it is meromorphic, holomorphic,
vanishing at the cusp, respectively. 

The ring of modular forms is generated by the 
Eisenstein series $G_4, G_6$, which have weights $4$ and $6$
respectively. The (normalized\footnote{With these prefactors, the
coefficients of an expansion in $q=e^{2 \pi i \tau}$ are rational
numbers. In fact, they are related to the Bernoulli numbers.}) 
Eisenstein series of weigth 
$k$ is defined by 
\[
G_k(\tau) = \frac{(k-1)!}{2 (2\pi i)^k} \sum'_{m,n} 
\frac{1}{(m\tau + n)^k} \;,
\]
where the sum is over all pairs of integers $(m,n)$ except
$(0,0)$. The sum converges absolutely for $k>2$ and vanishes
identically for odd $k$. 
For $k=2$ the sum is only conditionally convergent, and one can
define two functions with interesting properties. The holomorphic
second Eisenstein series is defined by 
\[
G_k(\tau) = \frac{(k-1)!}{(2 \pi i)^k} \sum_{n=1}^\infty 
\frac{1}{n^k} + \sum_{m=1}^\infty \left(
\frac{(k-1)!}{(2 \pi i)^k} \sum_{n\in \mathbbm{Z}}
\frac{1}{(m\tau +n)^k} \right) \;,
\]
with $k=2$ (the same organisation of the sum can be used for $k>2$).
The non-holomorphic second Eisenstein series is defined by 
\[
\overline{G}_2 (\tau, \overline{\tau}) 
= - \frac{1}{8 \pi^2} \lim_{\epsilon \rightarrow 0+}
\left( \sum'_{m,n} \frac{1}{(m\tau +n)|m\tau +n|^\epsilon} 
\right) \;.
\]
Both are related by
\[
\overline{G}_2(\tau, \overline{\tau}) = G_2(\tau) + \frac{1}{8 \pi \tau_2} \;.
\]
While the non-holomorphic $\overline{G}_2(\tau, \overline{\tau})$
transforms with weight two, the holomorphic function $G_2(\tau)$
transforms with an extra term:
\[
G_2 \left( \frac{a \tau +b}{c\tau +d} \right) =
(c \tau + d)^2 G_2(\tau) - \frac{c(c\tau +d)}{4 \pi i} \;.
\]
There is no modular form of weight two: $G_2(\tau)$ is holomorphic
but does not strictly transform with weight two, while 
$G_2(\tau, \overline{\tau})$ transforms with weight two but is not holomorphic.

There is a unique cusp form $\Delta_{12}$ of weigth 12,
which can be expressed in terms of the Dedekind $\eta$-function by
\[
\Delta(\tau) = \eta^{24} (\tau) \;,
\]
where
\begin{eqnarray}
\Delta(\tau) = \eta^{24} (\tau) &=& q \prod_{l=1}^\infty (1 - q^l)^{-24} \;,
\nonumber \\
\eta(\tau) &=& q^{\frac{1}{24}} \prod_{l=1}^\infty (1 - q^l)^{-1} \;.
\end{eqnarray}
The Dedekind $\eta$-function is a modular form of weight $\frac{1}{2}$
with multiplier system, i.e. a `modular form up to phase':
\[
\eta(\tau+1) = e^{\frac{2\pi i}{24}} \eta(\tau) \;,\;\;\;
\eta \left( - \frac{1}{\tau} \right) = \sqrt{-i\tau} \eta(\tau) \;.
\]

Modular forms are periodic under $\tau \rightarrow \tau +1$ 
and therefore they have a Fourier expansion in $\tau_1 = \mbox{Re} \tau$.
It is convenient to introduce the variable
\[
q = e^{2i\pi \tau} \;.
\]
In the main text we avoid using the variable $q$, because 
it might be confused with the electric charge vector 
$q \in \Gamma$. 
The transformation $\tau \rightarrow q$ maps the 
the semi-infinite strip 
$\{ \tau \in \mathbbm{C}  | \; |\tau_1| \leq 1, \tau_2 >0 \}
\subset  {\cal H}$ onto the unit disc 
$\{ q \in \mathbbm{C} | \; |q| < 1 \} \subset \mathbbm{C}$.
In particular, the cusp $\tau =i\infty$ is mapped to the
origin $q=0$. The Fourier expansion in $\tau_1$ maps to 
a Laurent expansion in $q$, known as the $q$-expansion.

The $q$-expansion of the cusp form $\Delta_{12} = \eta^{24}$
is 
\[
\eta^{24}(q) = q - 24 q^2 + 252 q^3 + \cdots
\]
In the main text we express modular forms in terms
of variables which live in right half plane rather than 
in the upper half plane, e.g., the heterotic dilaton $S$, 
where $\tau = iS$. For notational simplicity we then
write $\eta(S)$ instead of $\eta(iS)$.



\end{appendix}



\end{document}